\documentclass[webpdf,traditional,mediumone]{oup-authoring-template}

\theoremstyle{thmstyleone}%
\theoremstyle{thmstyletwo}%
\theoremstyle{thmstylethree}%

\usepackage[utf8]{inputenc} 
\usepackage[T1]{fontenc}    
\usepackage{hyperref}       
\usepackage{url}            
\usepackage{booktabs}       
\usepackage{amsfonts}       
\usepackage{nicefrac}       
\usepackage{microtype}      
\usepackage{lipsum}
\usepackage{graphicx}
\graphicspath{ {./images/} }
\usepackage{bm}
\usepackage{amssymb}
\usepackage{graphicx}
\usepackage{xcolor}
\usepackage{bigints}
\usepackage{amsmath}

\newcommand{\inv}[1]{#1^{-1}}

\begin{document}

\journaltitle{}
\DOI{DOI added during production}
\copyrightyear{YEAR}
\pubyear{YEAR}
\vol{XX}
\issue{x}
\access{Published: Date added during production}
\appnotes{Paper}

\firstpage{1}


\title[Spatially continuous modelling of aggregated outcome data]{Spatially continuous modelling of aggregated outcome data}

\author[1,$\ast$]{Stephen Jun Villejo\ORCID{0000-0002-0510-3143}}
\author[2]{Peter J Diggle\ORCID{0000-0003-3521-5020}}
\author[3]{Finn Lindgren\ORCID{0000-0002-5833-2011}}
\author[4]{H{\aa}vard Rue\ORCID{0000-0002-0222-1881}}
\author[5]{Guangquan Li\ORCID{0000-0002-8736-5349}}
\author[1]{Ella White\ORCID{0009-0002-0739-1094}}
\author[6]{Matthew Wade\ORCID{0000-0001-9824-7121}}
\author[1]{Marta Blangiardo\ORCID{0000-0002-1621-704X}}

\address[1]{\orgdiv{Faculty of Medicine}, \orgname{Imperial College London}, \orgaddress{\postcode{W12 0BZ}, \state{London}, \country{UK}}}
\address[2]{\orgdiv{Faculty of Health and Medicine}, \orgname{Lancaster University}, \orgaddress{ \postcode{LA1 4AT},\country{UK}}}
\address[3]{\orgdiv{School of Mathematics}, \orgname{University of Edinburgh}, \orgaddress{\postcode{EH9 3FD},\country{UK}}}
\address[4]{\orgdiv{Computer, Electrical and Mathematical Sciences \& Engineering Division}, \orgname{King Abdullah University of Science and Technology}, \orgaddress{\country{Saudi Arabia}}}
\address[5]{\orgdiv{School of Engineering, Physics and Mathematics}, \orgname{Northumbria University}, \orgaddress{\postcode{NE1 8SU},\country{UK}}}
\address[6]{\orgdiv{Chief Data Officer Group}, \orgname{UK Health Security Agency}, \orgaddress{\postcode{E14 4PU},\country{UK}}}

\corresp[$\ast$]{Corresponding author. \href{email:s.villejo@imperial.ac.uk}{s.villejo@imperial.ac.uk}}

\received{Date}{0}{Year}
\revised{Date}{0}{Year}
\accepted{Date}{0}{Year}



\abstract{This work develops a block aggregation approach to spatial estimation and prediction
 when the response is observed at a coarse spatial scale, for example
 as counts of events in administrative areas, or blocks, while covariates are available at a finer spatial resolution, typically as raster images. Our approach specifies
 a linear predictor at the finer resolution as a combination of covariate effects and a latent,
 spatially continuous Gaussian process. This linear predictor then determines
 the  distribution of the response through an inverse link function and
 spatial integration. We use a simulation study to evaluate the performance of the proposed approach in comparison to two industry standard approaches: a traditional geostatistical model that associates each response with the centroid of its block; and a Markov random field (MRF) approach that aggregates covariate data to block-level. As expected, the differences in performance among the three approaches are small with respect to block-level prediction. The rationale for,
 and advantage of, the
 block aggregation approach lies in its delivery of reliable inferences
 at whatever spatial resolution is required in a particular application.
 We describe two applications: 
  a linear Gaussian sampling model of wastewater virus concentrations in England, using population density as covariate; and log-linear Poisson  model of cardiovascular hospitalisations in England using socio-demographic variables at fine-scale administrative units as covariates.} 


\maketitle

\section{Introduction}

Spatial misalignment arises when there is a mismatch between the spatial resolution of the response and covariate data, a situation common in spatial epidemiology \citep{gryparis2009measurement, lee2017rigorous, rutten2025bayesian, alahmadi2025bayesian,cameletti2019bayesian}. In this paper, we consider the following scenario.
The response is aggregated over a set of
 administrative units, or {\it blocks}, $B_i: i=1,...,n$ that partition
the whole or part of the study area, $A$.
Each $B_i$ is sub-divided into {\it cells},
$b_{ij}: j=1,...,m_i$,
 on each of which
covariate data are available. Typically, 
the $b_{ij}$ correspond to a raster image over $A$,
whose resolution is sufficiently fine that we can
treat the $b_{ij}$ as points in continuous space. Our inferential objectives are: understanding the relationship between response and covariates;
prediction of the response at the
resolution of the covariates for high-resolution risk mapping \citep{pittiglio2018wild, brus2018geostatistical, li2012log, rutten2025bayesian};  prediction over
a partition of $A$ that may or may not coincide
with the $B_i$. 

When predictions are required only over the
$B_i$, a widely used approach is to pre-process the covariate data
by spatial averaging over each $B_i$, in order to obtain values that are spatially aligned with the $B_i$ \citep{cameletti2019bayesian, blangiardo2016two, lee2017rigorous}. An
example where prediction is required for blocks other than the $B_i$
arises in  wastewater epidemiology, where concentrations of pathogens in wastewater are measured 
for sewage treatment works catchment areas, whereas the spatial units for prediction of
disease risk are  administrative areas \citep{morvan2022analysis, li2023spatio, mills2024utility}; we return to
this example in Section \ref{sec:dataapplication}. 

Modelling strategies that involve pre-processing of covariate and/or response data to a common set of blocks have
two limitations. Firstly, in the case of non-linear models, estimates of the relationship between response and
covariate suffer from aggregation bias. Secondly, where there is a choice in how blocks are defined,
this choice can
dramatically affect prediction, a phenomenon known as the Modifiable Areal Unit Problem 
\citep{wong2004modifiable,manley2021scale}. 
To address these limitations, we argue
that most natural processes are inherently continuous in space and should be modelled as such, regardless of the spatial structure of the observed data. We therefore propose 
a {\it block aggregation model} that
combines a latent, spatially continuous {\it process model}, ${\mathcal S}$, which is the object of scientific interest,
with a {\it sampling model} for the distribution
of the response, ${\mathcal Y}$, conditional on ${\mathcal S}$.  To fit the model, we use a Bayesian inferential framework,
implemented with a linearised Integrated Nested Laplace Approximation (INLA) \citep{rue2009approximate, suen2026coherent, serafini2022approximation}. 

Related work includes the following.
In \cite{diggle2013spatial}, the process model is the exponential of a spatially continuous Gaussian process and the sampling model is either
an inhomogeneous Poisson process of individual events or a set of conditionally 
independent Poisson-distributed counts of the number of events in each block.  \cite{nandi2023disaggregation} proposed a method and
associated R package
for disaggregating block-level data to
raster-level. Composite link models,
 first proposed by \cite{thompson1981composite} and further
 developed in \cite{eilers2007ill}, assume more generally that the block-level mean is a function of a linear combination of
 the values of a latent process. 
Several 
authors have considered an {\it area-to-point kriging} approach \citep{kyriakidis2004geostatistical, truong2014bayesian,nagle2011geostatistical,goovaerts2006geostatistical,pardo2007modelling} which, however, cannot estimate covariate effects. \cite{kelsall2002modeling} modelled the spatial variation in disease risk based on aggregated data as a Gaussian random field. \cite{suen2026coherent} developed a disaggregation approach for spatial point process data and covariate fields, motivated  by landslide susceptibility modelling and considering scenarios where the covariate field is only incompletely
observed. \cite{rutten2025bayesian} proposed a Bayesian method for spatial disaggregation of count data, using spline-based low-rank kriging to model spatial correlation, and Laplace approximations for inference. 

The remainder of the paper is structured as follows. Section \ref{sec:model} describes our proposed {\it block aggregation methodology} in
more detail. We present
specific results for Gaussian and Poisson sampling
models, but the methodology is applicable for other sampling models under certain conditions. In Section \ref{sec:inference} we describe
the iterative linearised INLA method for 
model-fitting. Section \ref{sec:simstudy} gives the
results of a simulation study in which we compare the performance of our proposal against two
comparators, a geostatistical analysis that 
treats block-level data as  point-level
data at the block centroids, and a
spatially discrete Markov Random Field approach that uses covariate data 
aggregated to block-level.
Section \ref{sec:dataapplication} describes
two applications. 
The first models SARS-CoV-2 virus concentrations at
wastewater treatment works catchment areas in England using population density as covariate and a Mat\'ern field \citep{matern1960spatial} as the latent process model,
with the dual objectives of estimating the
process model parameters and making predictions of  wastewater virus concentration over the partition of England into Lower Tier Local Authority (LTLA) areas, this being the set of spatial units for which disease prevalence data are routinely recorded. The second application is an analysis
of LTLA-level counts of cardiovascular-related hospitalisations in relation to two socio-demographic covariates that are available at smaller administrative units, the Lower-layer Super Output areas (LSOAs). 
Section \ref{sec:discussion} is
a concluding discussion.

\section{Model}
\label{sec:model}

We denote by $B_i: i=1,\ldots, n$ the set of blocks on which the response variable is observed,
and by 
$b_{ij}, j=1,\ldots,m_i$  the spatial locations of  high-resolution covariates. The extension
to scenarios in which the covariate spatial units are not nested within the blocks is straightforward.

The response data are ${\mathcal Y}=\{Y_i: i=1,...,n\}$. 
The latent process is ${\mathcal S}=\{S(x):x \in A\}$, where
\begin{equation}\label{eq:latentmodel}
        S(b_{ij}) = \bm{\beta}^\intercal \mathbf{z}(b_{ij}) + R(b_{ij}).
    \end{equation}
In Equation \eqref{eq:latentmodel},  $\mathbf{z}(b_{ij})$ is the set of covariates (including an intercept), $\bm{\beta}$ is a vector of fixed effects, and $R(b_{ij})$ is a spatially correlated random effect.
Conditional on ${\mathcal S}$, 
the $Y_i$ are independent,
with
expectations
$\mu_i$ and probability distribution ${\mathcal F}$.  
Finally, ${\mathcal S}(\cdot)$ determines the conditional
expectations through the relationship
\begin{align}\label{eq:aggregation}
    \mu_i = \int_{B_i} w(x) f\Big(S(x)\Big)dx \approx \sum_{j=1}^{m_i} w(b_{ij}) f\Big(S(b_{ij})\Big),
\end{align}
for suitable inverse link function $f(\cdot)$ and weighting function $w(x)$. 

In summary, the ingredients of our block aggregation model are as follows: a set of predictor expressions $S(b_{ij})$;
an inverse link function $f(\cdot)$;
a weighting function $w(\cdot)$ that determines a discrete set of weights $w_{ij} = w(b_{ij})$;
a sampling model $\mathcal{F}$.


In Sections \ref{sec:model_Gaussian}
and \ref{sec:model_Poisson} we give a more detailed
description for sampling models: Gaussian with 
identity link; and Poisson with log link.  

\subsection{Gaussian sampling model}
\label{sec:model_Gaussian}

Here, we specify the sampling model
$Y_i\Big|\mu_i \overset{\text{iid}}{\sim} \mathcal{N}\Big(\mu_i, \sigma^2_e\Big)$.
Using the identity link and equal weights,
$w_{ij}=m_i^{-1}$, gives
    \begin{align}\label{eq:gaussian_functional1}
        Y_i&=\mu_i + e_i = m_i^{-1}\sum_{j=1}^{m_i} \Big[\mathbf{z}_{ij}^\intercal\boldsymbol{\beta} + R(b_{ij})\Big] + e_i, \;\;\; \sim\mathcal{N}(0,\sigma^2_e)
        \end{align}
    which is equivalent to
    \begin{align}\label{eq:gaussian_model_equiv}
        Y_i = \Bigg\{m_i^{-1} \sum_{j=1}^{m_i} \mathbf{z}_{ij}\Bigg\}^\intercal \boldsymbol{\beta} +m_i^{-1}\sum_{j=1}^{m_i}  R(b_{ij}) + e_i.
    \end{align}
    This shows that using the unweighted average of the covariates $\mathbf{z}_{ij}$ 
    to specify $\mu_i$ is equivalent to using the fine resolution $\mathbf{z}_{ij}$. Similarly, setting all $w_{ij}=1$ is equivalent to using the totals, $\sum_{j=1}^{m_i}  \mathbf{z}_{ij}$ to
    determine the $\mu_i$.

For arbitrary weights $w_{ij}$,
 we can express
  the sampling model as
\begin{align*}
       Y_i&=  \Bigg\{m_i^{-1}\sum_{j=1}^{m_i} w_{ij}\mathbf{z}_{ij}\Bigg\}^\intercal\boldsymbol{\beta}  +  m_i^{-1} w_{ij}\sum_{j=1}^{m_i}  w_{ij}R(b_{ij}) + e_i 
        \end{align*}

 \subsection{Poisson sampling model}\label{sec:model_Poisson}
 
 Here we specify the sampling model
 $Y_i\Big|\mu_i \overset{\text{iid}}{\sim}  \text{Poisson}\Big(\mu_i\Big)$ and exponential
 inverse link function. For
 arbitrary weights, this gives
\begin{align}\label{eq:logsumexp}
       \log\mu_i = \log\int w(x) \mu(x)dx \approx \log \sum_{j=1}^{m_i} w_{ij}f\Big(S(b_{ij})\Big) = \log \sum_{j=1}^{m_i} w_{ij}\exp\Big\{ \mathbf{z}_{ij}^\intercal\boldsymbol{\beta} + R(b_{ij}) \Big\},
\end{align}
The non-linear link function results in a set
of conditional expectations that can 
be markedly different from 
the widely used approach of simply aggregating the covariates and specifying
$\mu_i \approx \exp\Big\{m_i^{-1}\sum_{j=1}^{m_i}S({b_{ij}})\Big\}$ in a spatially discrete model \citep{alahmadi2025bayesian,lee2017rigorous,cameletti2019bayesian}.  Equation \eqref{eq:logsumexp} with
constant weights is the
analogue for count data
of the point process model proposed in \cite{diggle2013spatial}, where $f\Big(S(b_{ij})\Big)$ was the
stochastic intensity  of a log-Gaussian Cox process \citep{moller1998log}. 

\section{Inference}\label{sec:inference}
For Bayesian inference, we 
consider the following hierarchical form of the model:
\begin{equation}\label{eq:bayesianmodel}
        \begin{aligned}
    &{\mathcal Y}\big|\bm{\mu},\bm{\theta} \overset{\text{iid}}{\sim} \prod_{i}^n \mathcal{F}\Big(Y(B_i)\big|\mu_i,\bm{\theta}\Big), \;\;\; \mu_i=\mathbb{E}\big[Y(B_i)\big] \\ 
    & g\big(\mu_i \big) = 
     g \Big( \sum_{j=1}^{m_i} w(b_{ij})f\Big(S(b_{ij})\Big) \Big) =
    g \Big( \sum_{j=1}^{m_i} f\Big(\mathbf{a}_{j(i)}^\intercal\bm{u}\Big) \Big)
   \\
        & \bm{u}|\bm{\theta} \sim \mathcal{N}\Big(\bm{\mu}_{u},\mathbf{Q}(\bm{\theta})\inv{}\Big) \\
        &\bm{\theta} \sim p(\bm{\theta})
\end{aligned}
    \end{equation}

 In Equation \eqref{eq:bayesianmodel}, $\bm{u}$ includes all latent quantities in Equation \eqref{eq:latentmodel}, i.e., the fixed effects $\bm{\beta}$ and the spatial effect $R(\cdot)$, while $\mathbf{a}_{ij}$ is a known vector, which includes the covariate information $\mathbf{z}(b_{ij})$ and other mapping vectors depending on the model specification for $R(b_{ij})$. Also, Equation \eqref{eq:bayesianmodel} specifies a Gaussian prior for $\bm{u}$, with mean $\bm{\mu}_{u}$ (fixed) and precision matrix $\mathbf{Q}(\bm{\theta})$, where $\bm{\theta}$ are model hyperparameters. Finally, $p(\bm{\theta})$ is the prior distribution of $\bm{\theta}$.

For the associated computations,
we use the Integrated Nested Laplace Approximation method (INLA)  \citep{rue2009approximate}, a fast deterministic approach for Bayesian inference. For non-linear functionals $f(\cdot)$, \eqref{eq:bayesianmodel} cannot be fitted directly using INLA since the data-level predictor expression is a non-linear function of $\bm{u}$.
To overcome this limitation, \cite{lindgren2024inlabru} proposed a linearisation approach,
using 
a 1st order Taylor approximation of
$f\Big(\mathbf{a}_{j(i)}^\intercal\bm{u}\Big)$ at $\bm{u}_0$. The linearised predictor is:
\begin{eqnarray}
    g \Big( \sum_{j=1}^{m_i} f\Big(\mathbf{a}_{j(i)}^\intercal\bm{u}\Big) \Big) & \approx & g \Big( \sum_{j=1}^{m_i} f\Big(\mathbf{a}_{j(i)}^\intercal\bm{u}_0\Big) \Big) + \nabla g \Big( \sum_{j=1}^{m_i} f\Big(\mathbf{a}_{j(i)}^\intercal\bm{u}\Big) \Big)  (\bm{u} - \bm{u}_0) \\
& = & \bm{\delta} + \nabla \bm{g}^\intercal \bm{u},
\end{eqnarray}
where $\nabla\bm{g}$ is the vector of derivatives evaluated at
the {\it linearisation point}, $\bm{u}_0$, and $\bm{\delta}$ is a constant term, which depends on $\bm u_0$. 

The resulting linearised Bayesian hierarchical model is
\begin{equation}\label{eq:bayesianmodel_linear}
        \begin{aligned}
    &\mathbf{y}\big|\bm{\mu},\bm{\theta} \overset{\text{iid}}{\sim} \prod_{i}^n \mathcal{F}\Big(y(B_i)\big|\mu(B_i),\bm{\theta}\Big), \;\;\; \mu(B_i)=\mathbb{E}\big[y(B_i)\big] \\ 
    & g\big(\mu(B_i) \big) \approx \bm{\delta} + \nabla \bm{g}^\intercal \bm{u}\\
        & \bm{u}|\bm{\theta} \sim \mathcal{N}\Big(\bm{\mu}_{u},\mathbf{Q}(\bm{\theta})\inv{}\Big) \\
        &\bm{\theta} \sim p(\bm{\theta})
\end{aligned}
    \end{equation}
    
Equation \eqref{eq:bayesianmodel_linear} defines a latent Gaussian model whose latent parameters $\bm{u}$ are linear with respect to the predictor expression. Hence, we can use INLA to fit the model. The joint posterior obtained from \eqref{eq:bayesianmodel_linear} is
\begin{equation}
    \pi(\bm{u},\bm{\theta}) \propto \pi(\bm{\theta})\pi(\bm{u}|\bm{\theta})\prod_{i=1}^n\pi(y_i|\bm{u},\bm{\theta}).
\end{equation}
The posterior distribution of $\bm{\theta}$ is approximated as
\begin{equation}
    \pi(\bm{\theta}|\mathbf{y}) \propto \dfrac{\pi(\bm{u},\bm{\theta},\mathbf{y})}{\pi_G(\bm{u}|\bm{\theta},\mathbf{y})}\Bigg|_{\bm{u}=\bm{u}^*(\bm{\theta})},
\end{equation}
where $\pi_G(\bm{u}|\bm{\theta},\mathbf{y})$ is the Gaussian approximation to  $\pi(\bm{u}|\bm{\theta},\mathbf{y})$
obtained from a second order Taylor expansion of the likelihood around the mode $\bm{u}^*(\bm{\theta})$. The Gaussian approximation is used to approximate the marginal posteriors $\pi(u_i|\bm{\theta},\bm{y})$ \citep{van2023new}, with a variational Bayes correction to the mean \citep{van2024low}. The  marginal posteriors $\pi(u_i|\bm{y})$ are then obtained by numerical integration over $\bm{\theta}$.

This linearised model in \eqref{eq:bayesianmodel_linear} has to be fitted iteratively from an initial choice of linearisation point ${\bm u}_0^{(0)}$ until the  conditional posterior mode of the linearised model is the same as the linearisation point, ${\bm u}_0^{(k)}$. The
iteration scheme is
\begin{equation}\label{eq:iterativemethod}
    \begin{aligned}
        &\hat{\bm{\theta}}^{(k+1)} = \texttt{arg max}_{\bm{\theta}}\; \hat{\pi}\big(\bm{\theta}|\mathbf{y}, \bm{u}_0^{(k)}\big) \\
        &\hat{\bm{u}}^{(k+1)} = \texttt{arg max}_{\bm{u}}\; \hat{\pi}\big(\bm{u}|\mathbf{y}, \bm{\theta}^{(k+1)},\bm{u}_0^{(k)}\big) \\
        &\bm{u}_0^{(k+1)} = (1-\alpha)\bm{u}_0^{(k)} +\alpha \hat{\bm{u}}^{(k+1)}
    \end{aligned}
\end{equation}
where $\alpha$ is the value that minimizes $\Big\| g\Big(\sum_{j=1}^{m_i} f\Big(\mathbf{a}_{j(i)}^\intercal\bm{u}\Big)\Big) - (\bm{\delta} + \nabla \bm{g}^\intercal \bm{u})\Big\|$, i.e., the difference between the non-linear and linearized predictors.  \cite{lindgren2024inlabru} proposed the following convergence criteria: (i) the relative maximum change in the linearisation point relative to the posterior standard deviation is below a threshold and (ii) $\alpha \approx 1$. The approximation accuracy depends on the degree of nonlinearity of the model. The properties of these approximations have been investigated in \cite{suen2026coherent}.

\section{Simulation study}\label{sec:simstudy}

\subsection{Study area and model specification}\label{subsec:studydomain_modelspec}

For all simulations, the study area is the unit square,  $A =[0,1] \times [0,1]$, which we divide into 100 square blocks $B_i$. The observed response data are aggregations to block-level of a latent spatial process at the resolution of nested grids, $b_{ij}, j = 1,\ldots, 25$, 
within each $B_i$ (see Figure 1 of Supplementary Material) .
    
We consider Gaussian and Poisson sampling models, as discussed in Section \ref{sec:model}, with the latent model as presented in Equation \eqref{eq:latentmodel}. The block aggregation scheme for the Gaussian and Poisson cases follow Equations \eqref{eq:gaussian_functional1} and \eqref{eq:logsumexp}, respectively.
We consider a single covariate $z_{ij}$ and a latent Mat\'ern Gaussian field $R(\cdot)$ with mean square differentiability parameter fixed at 1, hence
\begin{equation*}
        S(b_{ij}) = \beta_0+\beta_1 z_{ij} + R(b_{ij}).
    \end{equation*}
The data generating model follows the aggregation scheme defined by Equation \eqref{eq:aggregation}. We use the identity functional $f(\cdot)$ for the Gaussian case, and the exponential functional for the Poisson model as discussed in Section \ref{sec:model}. For the Gaussian case, we use $\beta_0=10$, $\beta_1=1.5$, and $\sigma^2_e=1$. For the Poisson case, we set $\beta_0=-2$ and $\beta_1=0.15$.  Covariate values $z(b_{ij})$ are 
an independent random sample
from $\mathcal{U}(0,20)$ and $\mathcal{U}(0,40)$ for the Gaussian  and Poisson cases, respectively. Parameter values for the Mat\'ern field are given in Section \ref{subsec:sim_scenarios}.

\subsection{Simulation scenarios}\label{subsec:sim_scenarios}

We consider all combinations of the following three
factors: (i) the sampling proportion, which determines the number of blocks for which response data are observed; (ii) the practical range parameter for $R(\cdot)$, the distance at which the correlation reaches the value of approximately 0.10; (iii) the marginal variance of the spatial field
(Table \ref{tab:simscenarios}).

\begin{table}[h]
\centering
\renewcommand{\arraystretch}{1.1}
\scalebox{1}{\begin{tabular}{|l|l|}
  \hline
 \textbf{Factors} & \textbf{Values} \\ 
  \hline
Sampling proportion & \{$30\%, 60\%, 100\%$\} \\ \hline
Range parameter for $R(\cdot)$ & \{$0.05, 0.1, 0.4$\} \\ \hline
\multirow{3}{*}{Marginal variance for $R(\cdot)$} 
    & Gaussian case: \{$2,4$\} \\ 
    & Poisson case: \{$0.05,0.15$\} \\
   \hline
\end{tabular}}
\caption{Simulation scenarios}
\label{tab:simscenarios}
\end{table}

Note that the blocks $B_i$ have length 0.1 units. Hence,
when the range parameter value
is 0.05,
correlation decays over a range smaller than the observed data can potentially resolve.

\begin{figure}[t]
\centering

\begin{minipage}{0.65\textwidth}
    \centering

    \begin{minipage}{0.46\linewidth}
        \centering
        \includegraphics[width=\linewidth]{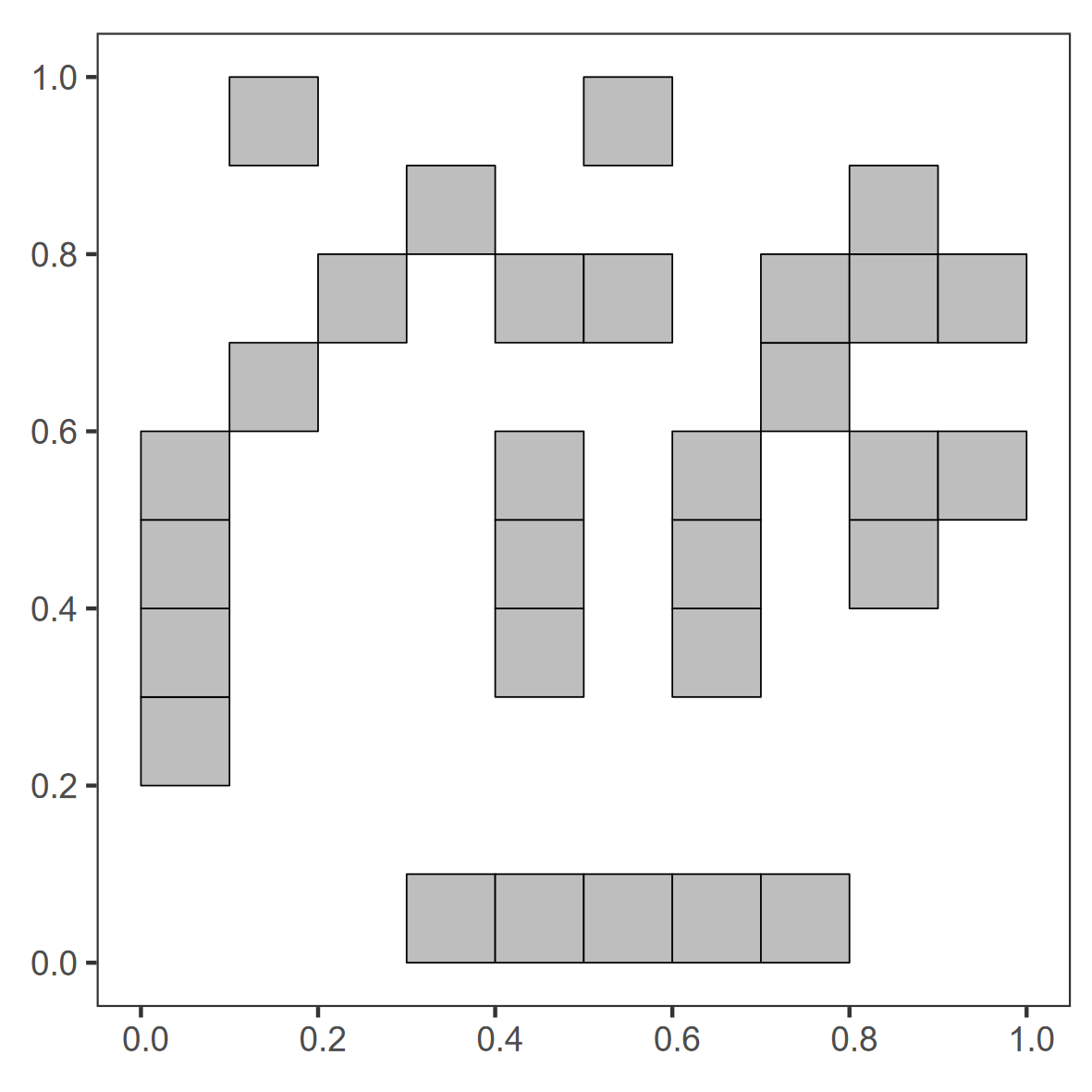}
        \vspace{0.5ex}
        (a) 10\% sampling proportion
    \end{minipage}
    \hspace{2mm}
    \begin{minipage}{0.45\linewidth}
        \centering
        \includegraphics[width=\linewidth]{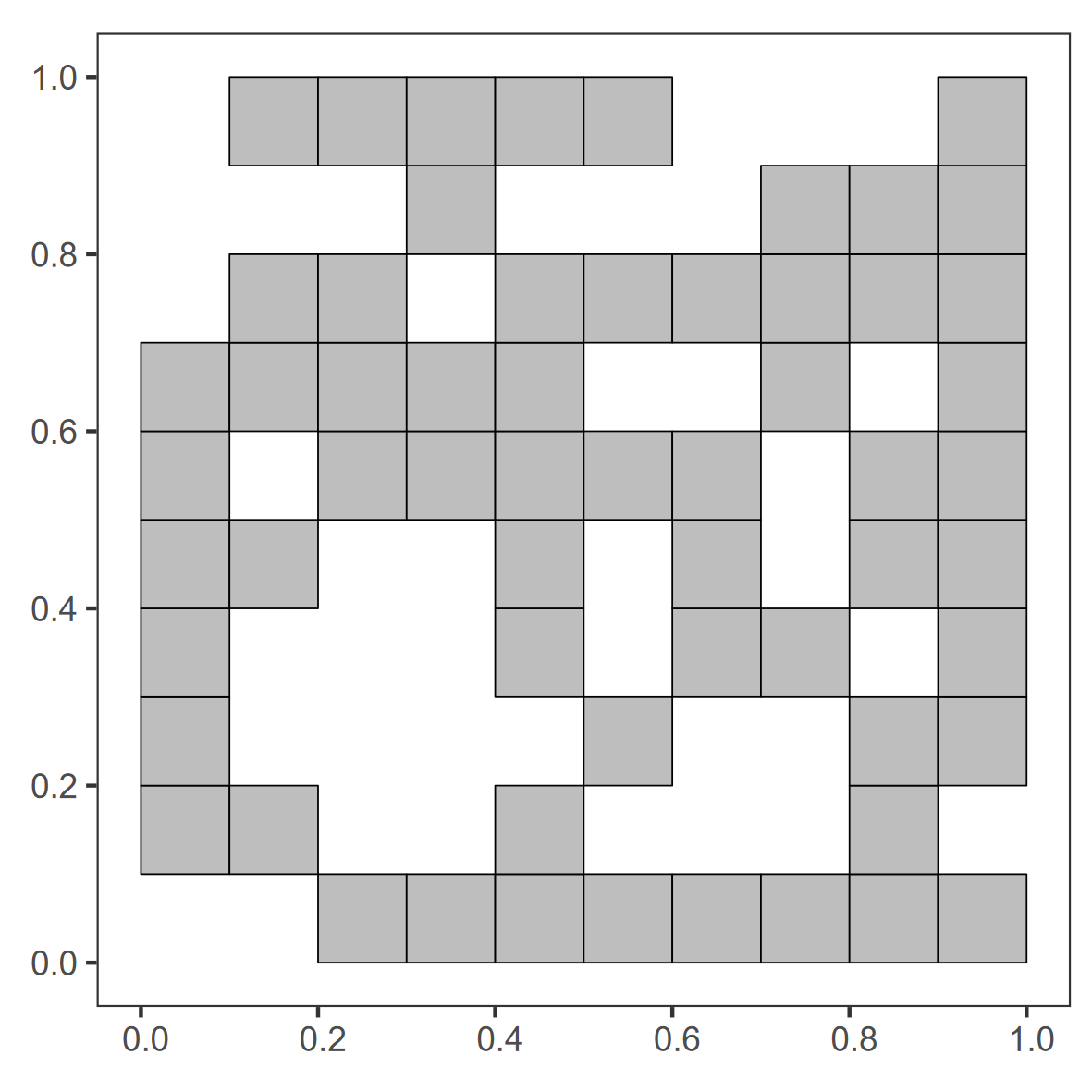}
        \vspace{0.5ex}
        60\% sampling proportion
    \end{minipage}

    \caption{Spatial locations of data $B_i$}
    \label{fig:samplingprop}
\end{minipage}
\hfill
\begin{minipage}{0.28\textwidth}
    \centering
    \includegraphics[width=\linewidth]{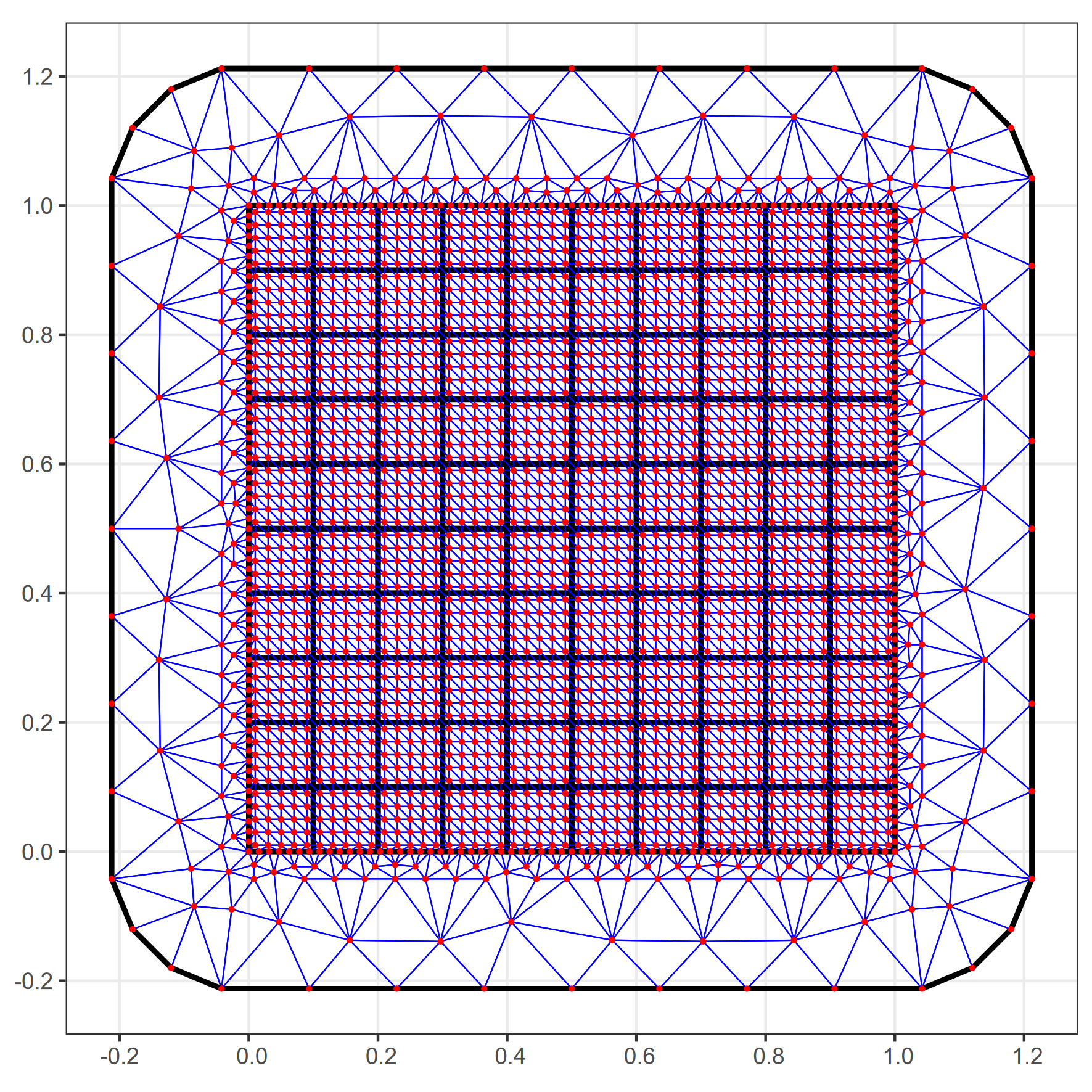}
    \vspace{0.5ex}
    \textbf{(c)}

    \caption{Mesh}
    \label{fig:mesh}
\end{minipage}

\end{figure}

Figure \ref{fig:samplingprop} shows
the spatial locations for which response data are observed under the 30\% and 60\% sampling proportions considered.

We use the stochastic partial differential equation (SPDE) approach \citep{lindgren2011explicit} to fit the Gaussian field $R(\cdot)$. Figure \ref{fig:mesh} shows the mesh used for model fitting. The maximum triangle edge lengths for the study area and its outer extension are set at 0.05 units and 0.4 units, respectively. The choice of 0.05 units for the inner triangle is based on the minimum spatial range we considered in the simulation settings. 

We assign diffuse Normal priors for the fixed effects: $\beta_0
\sim \mathcal{N}(0,\infty)$ and $\beta_1 \sim \mathcal{N}(0,1000)$. The prior
for the error variance $\sigma^2_e$ in the Gaussian sampling model is a penalized complexity (PC) prior \citep{fuglstad2019constructing},
such that $\mathbb{P}(\sigma_{e} > \sqrt{0.1} = 0.5)$. The Mat\'ern range and marginal variance parameters, $\rho_{R}$ and $\sigma^2_{R}$, are also given the PC priors. For the Gaussian sampling model, we specify $\mathbb{P}(\sigma_{R} > 1.73) = 0.5$ and $\mathbb{P}(\rho_{R} < 0.1 = 0.5)$. For the Poisson sampling model, we 
again specify
$\mathbb{P}(\rho_{R} < 0.1 = 0.5)$, but for $\sigma_{R}$
we specify $\mathbb{P}(\sigma_{R} > 0.1) = 0.5$. The values of $R(\cdot)$ are on the data scale for the Gaussian model, but on the log scale for the Poisson model; hence the difference in the PC priors.


\subsection{Comparator models}\label{subsec:baselineapproaches}

In each simulation scenario, we assess the performance of the proposed method against two comparator approaches that treat the response variable as point-referenced at the centroid of the each block $B_i$, but differ in how they model the latent spatial effects.
\subsubsection{Comparator model A: centroids} 

    Our first comparator model
    assumes that $R(\cdot)$ is a spatially continuous process and treats the response from each block, $B_i$, as a point-referenced observation at the centroid of $B_i$. The sampling model is of the form $Y_i\Big|\mu_i\overset{\text{iid}}{\sim}\mathcal{F}$, with   mean $\mu_i$ and linear predictor 
\begin{align}\label{eq:comparatorA}
    g\big(\mu_i\big) = \beta_0+\beta_1 z_i^*+R(x_i).
\end{align}
    In Equation\eqref{eq:comparatorA}, $z_i^*$ is the sample mean of the 25 values $z_{ij}$ inside $B_i$. Also, $x_i$ is the centroid of $B_i$ and $R(\cdot)$ is a Mat\'ern field. Finally, the parameters $\beta_0$, $\beta_1$, $\rho_{R}$ and $\sigma_{R}$ are given the same priors as in Section \ref{subsec:studydomain_modelspec}.
    
\subsubsection{Comparator model B: Markov random field}

    Our second comparator model also uses $z_i^*$ sample means as covariate values input to the model, as in Equation \eqref{eq:comparatorA},
    but replaces the spatially continuous Mat\'ern field $R(\cdot)$ with a spatially discrete Markov random field (MRF) on the block centroids. The MRF approach is widely used for problems in which response data are available on a complete partition of the study region and prediction is only required for that same partition. The linear predictor for the MRF model is  
\begin{align}\label{eq:comparatorB}
    g\big(\mu_i\big) = \beta_0+\beta_1{z_i^*}+\dfrac{1}{\sqrt{\tau}}\Big( \sqrt{1-\phi}V_i+\sqrt{\phi}U_i \Big),
\end{align}
where $U_i$ follows the Besag model \citep{besag1974spatial},  scaled to have generalized variance equal to one, and  $V_i \overset{\text{iid}}{\sim}\mathcal{N}(0,1)$. 
This 
MRF is a reparametrisation of the
Besag-York-Molli\'e (BYM) model \citep{besag1991bayesian}, in
which $\tau$ is the common precision for $U_i$ and $V_i$, and $\phi$ is a mixing parameter
that represents the proportion of the marginal variance explained by the spatial effect $U_i$. We use the same diffuse priors for $\beta_0$ and $\beta_1$ as for the other two approaches. For $\tau$ and $\phi$ we use PC priors such that $\mathbb{P}(1/\sqrt{\tau}>1=0.01)$ and $\mathbb{P}(\phi>1=0.5)$.
    
\subsection{Performance scores}\label{subsec:scores}

We use the following scores to assess the performance of the three approaches.
\begin{enumerate}
    \item Dawid-Sebastiani (DS) score
    \begin{equation*}
       \dfrac{1}{m}\sum_{i=1}^n \Bigg[\dfrac{\big\{Y_i-\mathbb{E}[Y_i]\big\}^2}{{\rm Var}[Y_i]} + \log {\rm Var}[Y_i] \Bigg],
    \end{equation*}
    where the expectation and variance are computed with respect to ${\mathcal{F}_{pred}}$, the predictive distribution of $Y_i$.    
    \item Negative log score: $-\log \mathcal{F}_{pred}\Big(Y_i\Big)$
     \item Root mean square error (RMSE) of $ \hat{\mu}_i = \mathbb{E}_{{\mathcal{F}_{pred}}}[Y_i]: \;\;\;\;\sqrt{n^{-1}\sum_{i=1 }^n\Big\{\mu_i-\hat{\mu}_i\Big\}^2}$ 
    \item RMSE of $\hat{\mu}_{ij} = \mathbb{E}\Big[\hat{f}\Big(S(b_{ij})\Big)\Big|\mathbf{Y}\Big]\equiv\hat{f}\Big(S(b_{ij})\Big) : \;\;\;\;$
    \begin{equation*}
        \sqrt{\Big(m_{i}^{-1} \sum_{\forall j}\Big\{\hat{\mu}_{ij}- \mu_{ij}\Big\}^2\Big)}
    \end{equation*}
    where $m_{i}$ is the number of $b_{ij}$ in $B_i$. 
    \item Relative bias of parameter estimates: $\Big| (\hat{\beta}_i - \beta_i)/\beta_i \Big|\times 100\%$
    \item Coverage for $\mu_i$ and $\mu_{ij}$
\end{enumerate}

The first two scores are computed at the block level. The DS score is a strictly proper scoring rule \citep{gneiting2007strictly} that measures closeness between an observed quantity of interest and its predictive distribution ${\mathcal{F}_{pred}}$ \citep{dawid1999coherent}. It penalizes both biased predictions and overconfident predictions that underestimate predictive variance. The negative log score is also a proper score. It evaluates the predictive distribution at the observed value of $Y$. The third and fourth scores are computed with respect to the true $\mu(\cdot)$, either at the block-level $\mu_i$ or at the grid-level $\mu_{ij}$. The fifth score is the relative bias of the parameters $\beta_0$ and $\beta_1$.  Finally, for the coverage of $\mu_i$ and $\mu_{ij}$, we use
the actual coverage of nominal 95\% credible intervals. Predicting $\mu_{ij}$ is straightforward for the block aggregation approach, because covariates are observed at the resolution of $b_{ij}$, but not for the comparator models. In the case of the Gaussian sampling model, predictions of $\mu_{ij}$ for the comparator models are obtained from the posterior distribution of the 
quantity defined in Equation \eqref{eq:latentmodel} and using this as the predicted value of $\mu_{ij}$. We expect that this will yield predictions comparable to those from the block aggregation
approach because of the equivalence noted in Equation \eqref{eq:gaussian_model_equiv}. In the 
case of the Poisson sampling model, neither of the comparator models is designed for spatial disaggregation. To
obtain a crude predicted value of $\mu_{ij}$, we spread $\mu_i$ uniformly over
all 25 $b_{ij}$ within $B_i$, hence
$\hat{\mu}_{ij} =\hat{\mu}_i/25$.

\section{Simulation results}\label{sec:simresults}

\subsection{Gaussian sampling model}\label{subsec:results_gaussian}

\begin{figure}[htbp]
  \centering

  \begin{minipage}[b]{0.49\textwidth}
    \centering
    \includegraphics[width=\textwidth]{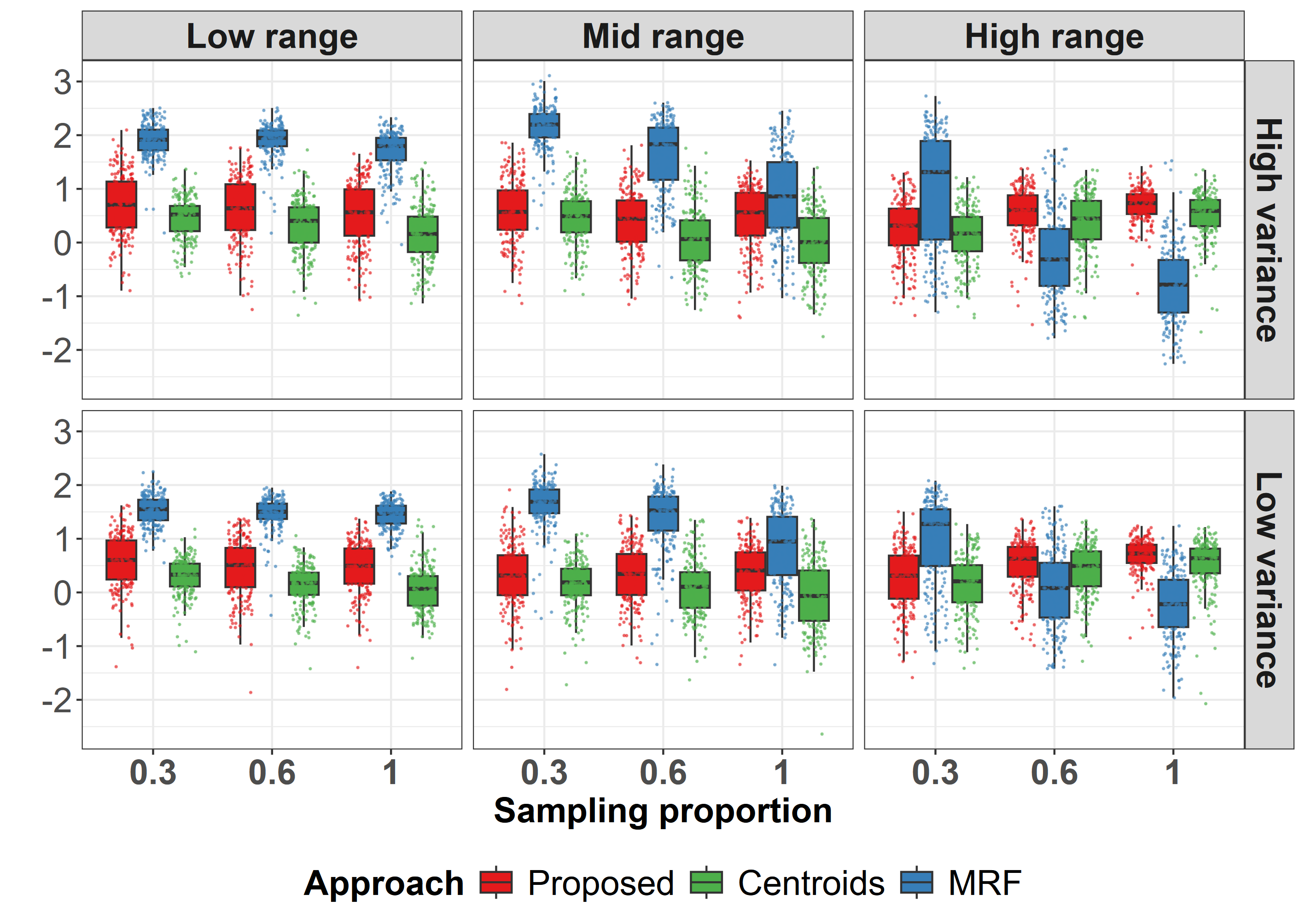}
    \vspace{0.5ex}
    \textbf{(a)}\; Dawid Sebastiani score
  \end{minipage}
  \hfill
  \begin{minipage}[b]{0.49\textwidth}
    \centering
    \includegraphics[width=\textwidth]{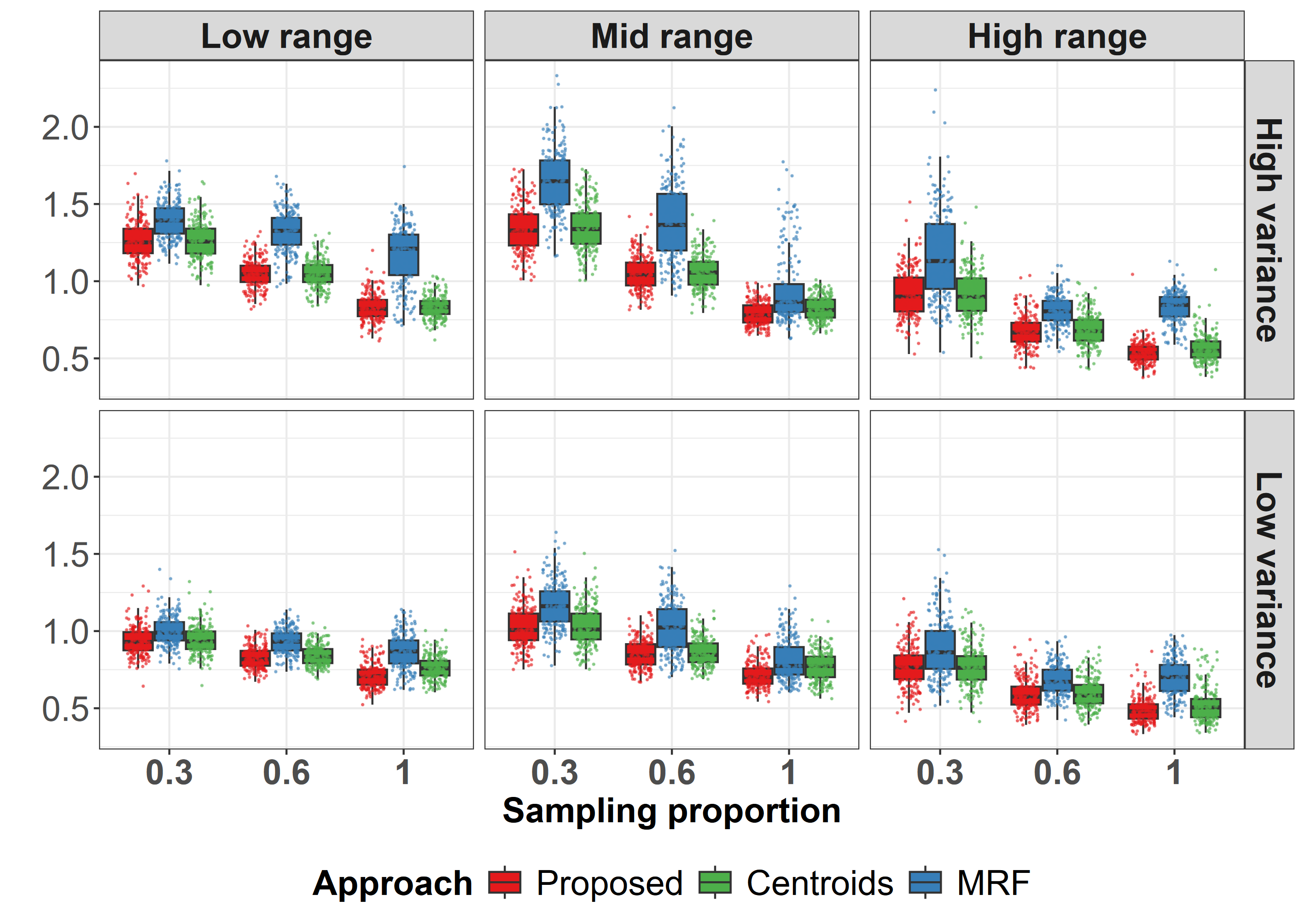}
    \vspace{0.5ex}
    \textbf{(b)}\; RMSE of $\hat{\mu}(B_i)$
  \end{minipage}

  \caption{Comparison of performance scores for the Gaussian case.}
  \label{fig:sim_res_gaussian}
\end{figure}

Figures \ref{fig:sim_res_gaussian}a and \ref{fig:sim_res_gaussian}b compare the DS scores and RMSE of $\hat{\mu}_i$ across all simulation scenarios and all three approaches. The block aggregation and centroid approaches perform similarly. The performance of the MRF approach is noticeably worse in the low-range and mid-range spatial correlation scenarios, for which the range parameter values, 0.05 and 0.1 units, respectively, are less than or equal to  
the linear dimension of each block. The block aggregation and centroid approaches, both of which treat $R(\cdot)$ as a spatially continuous process, are better able to capture the underlying spatial structure. In the high correlation range scenario (range parameter 0.4), the MRF approach has
a lower DS score than the other two approaches when the sampling proportion is either 60\% or 100\%.

The results for the negative log score show similar comparative insights (Supplementary material, Figure 6a). Additionally, there are no substantial differences among the three approaches with respect to the RMSE of $\hat{\mu}_{ij}$  (Supplementary material, Figure 6b) or the relative bias in the estimates of $\beta_0$ and $\beta_1$ (Supplementary material, Figures 7a and 7b). These results reflect our
observation in Section \ref{sec:model}, Equation \eqref{eq:gaussian_model_equiv}, that in the case of the Gaussian sampling model the block aggregation approach is equivalent to using the block-level average of the $\mathbf{z}_{ij}$ in a
block-level analysis.  Finally, the block aggregation and 
centroid models give coverages for each $\mu_i$ and $\mu_{ij}$ that are close to the nominal value of 95\%, whereas the MRF approach often gave low coverage (Supplementary material, Figures 8a and 8b).

\subsection{Poisson sampling model}\label{subsec:results_poisson} 

\begin{figure}[ht]
    \centering

    \begin{minipage}[b]{0.48\linewidth}
        \centering
        \includegraphics[width=\linewidth]{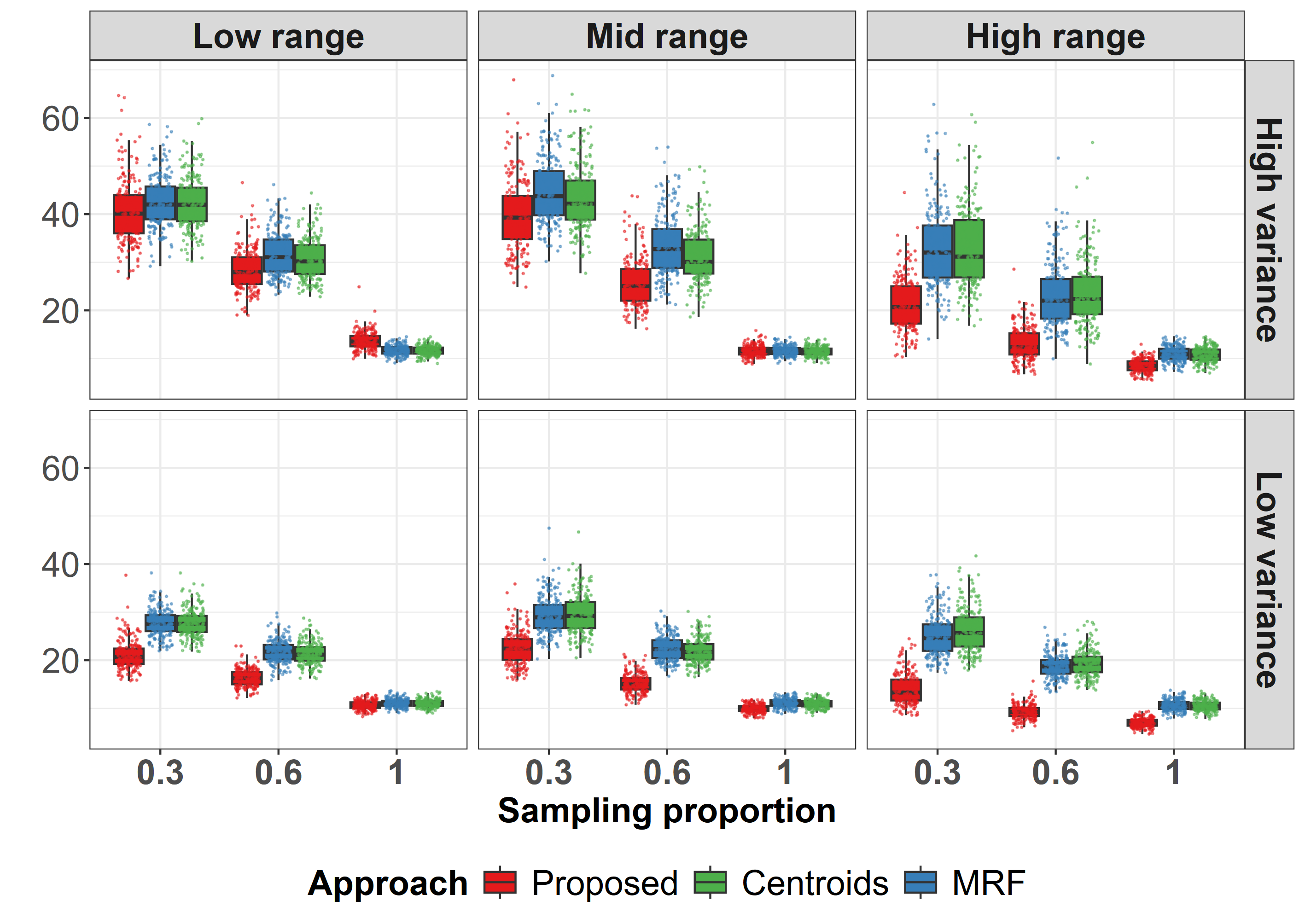}
        \vspace{0.5ex}
        \textbf{(a)} RMSE of $\hat{\mu}(B_i)$
    \end{minipage}
    \hfill
    \begin{minipage}[b]{0.48\linewidth}
        \centering
        \includegraphics[width=\linewidth]{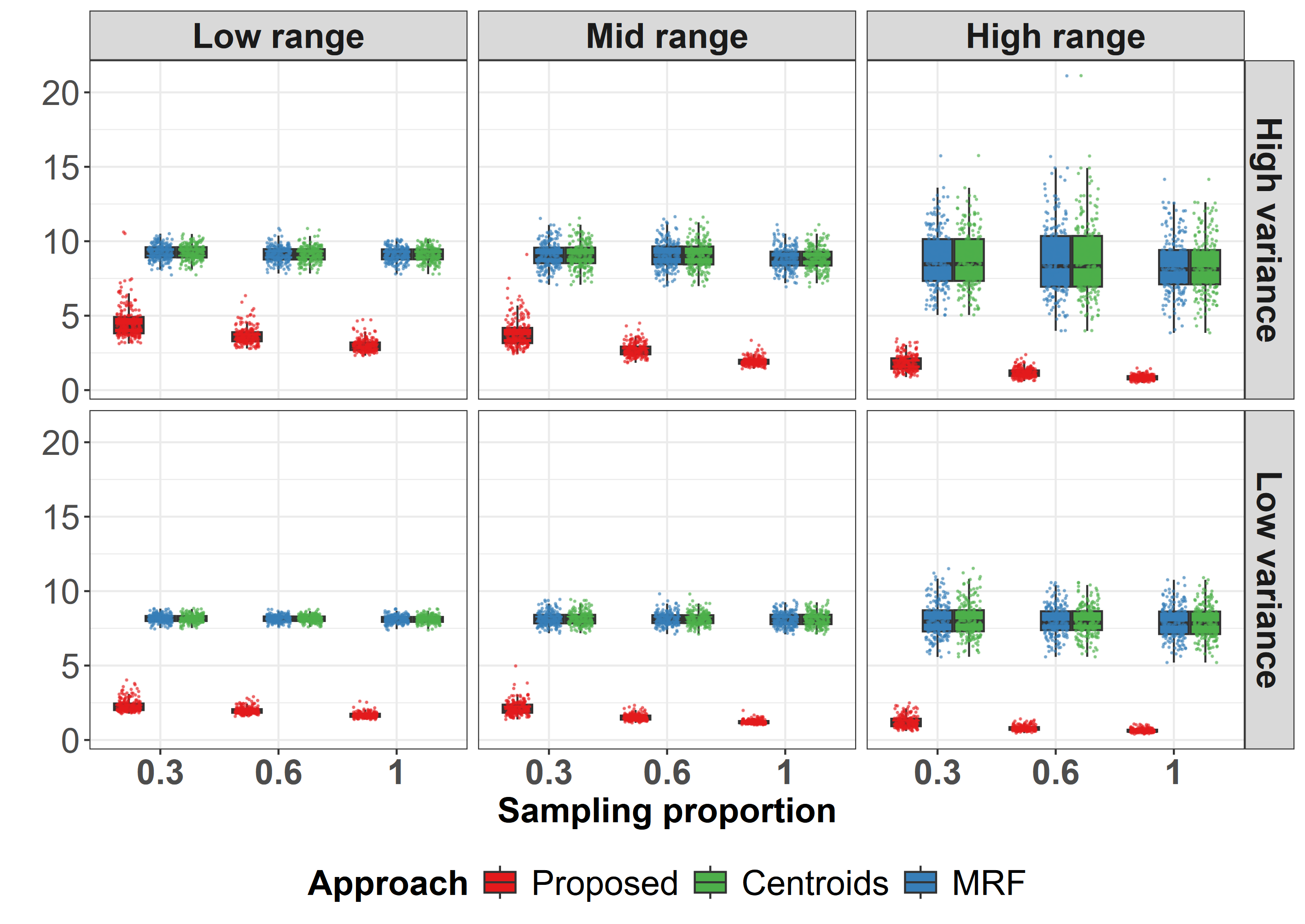}
        \vspace{0.5ex}
        \textbf{(b)} RMSE of $\hat{\mu}(b_{ij})$
    \end{minipage}

    \caption{Comparison of performance scores for the Poisson case}
    \label{fig:simres_Poisson_scores}
\end{figure}

Figures \ref{fig:simres_Poisson_scores}a show that the block aggregation approach
outperforms both comparator models with respect to the RMSEs for
$\hat{\mu}_i$ or $\hat{\mu}_{ij}$ The particularly  poor performance of both comparators
with respect to RMSEs for $\mu_{ij}$
is as expected, since they are not designed for spatial disaggregation. The coverage for $\mu_i$ is consistently close to the nominal 95\% level for all blocks under the proposed block approach, whereas both comparator models give very low coverage for some blocks, although their performance improves as the sampling proportion increases (Supplementary Material, Figure 11a). The block aggregation approach 
gives coverages of each $\hat{\mu}_{ij}$  close to the nominal value of 95\% (Supplementary Material, Figure 11b), and gives approximately unbiased estimates of the fixed effects $\beta _0$ and $\beta_1$ (Supplementary Material, Figures 10a and 10b). In contrast, the two comparator models suffer from aggregation bias \citep{greenland1990}. 

DS scores (Supplementary Material, Figure 9a) are very similar across all approaches when the range parameter is 0.4. However for range parameter values 0.05 and 0.1, the block aggregation method is outperformed by the comparator models, particularly so for the larger value of
the marginal variance of the spatial field. We observed the opposite pattern
in the case of the Gaussian sampling model. We observe similar comparative results with respect to the negative log score (Supplementary Material, Figure 9b). 

\subsection{Summary}

The overall benefits from the block
aggregation method are evident when the goal is to predict
both the $\mu_i$ and the $\mu_{ij}$. 
All three approaches generally give accurate results for block-level
predictions, except when the spatial correlation range is smaller than the linear dimensions of the blocks, in which case the MRF approach is
markedly inferior. 

\section{Applications}\label{sec:dataapplication}

This section presents a case study for each of the two sampling models considered in Section \ref{sec:model}. We also compare the proposed block aggregation method with the two comparator models introduced in Section \ref{sec:simstudy}.

\subsection{Virus concentrations in community wastewater} \label{subsec:wastewater}

\begin{figure}[htbp]
    \centering
    \includegraphics[width=.9\linewidth, trim={0 1.75cm 0 1.9cm}, clip]{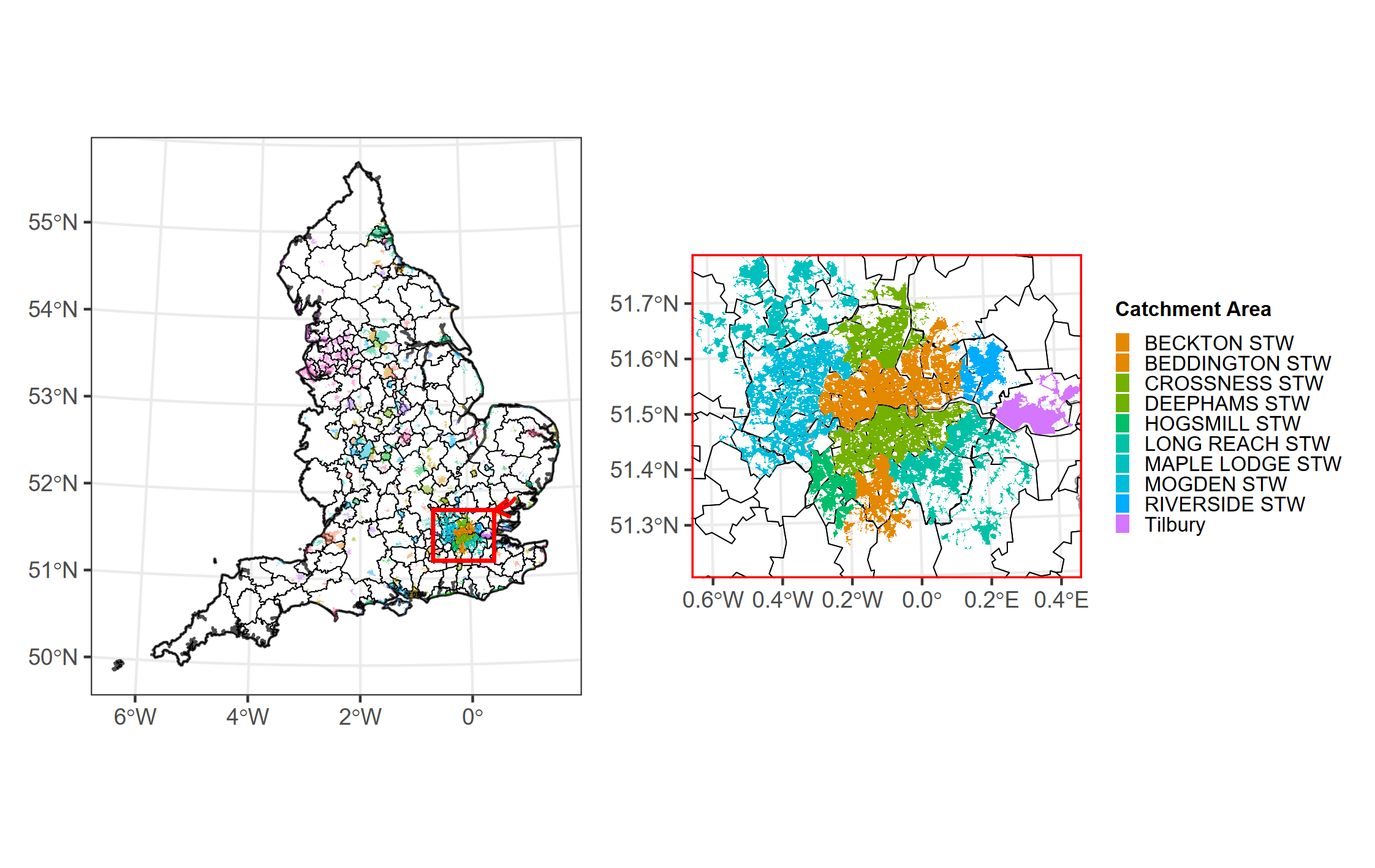}
    \caption{Catchment areas in England, and a zoomed in Greater London area. Shown also are the LTLA boundaries.}
    \label{fig:app_gaussian_data_england}
\end{figure}

For the Gaussian sampling model, we consider an environmental application relating to wastewater-based epidemiology. The response data consist of weekly average
values of log-transformed SARS-CoV-2 virus concentrations (number of gene copies of the pathogen per liter of wastewater) from a network of sewage treatment works (STW) in England \citep{ukhsa_wastewater_2022}. The inferential goal is to predict average concentrations for each of England's Lower Tier Local Authorities (LTLAs); their partition of England is not nested
within the sewage works catchments on which the response data are observed. We consider the six-day period $1^{\text{st}}$ to $6^{\text{th}}$ of June 2021. 
Figure \ref{fig:app_gaussian_data_england} shows the catchment areas for England, which are highly irregular in
shape and cover only part of the country.  Figure \ref{fig:app_gaussian_data_england} also shows the LTLA boundaries. We use population density at $1\;\text{km}\times 1\;\text{km}$ resolution as a single covariate, $z_{ij}$. 
We denote by $Y_i$ the observed log-transformed virus concentration for each of the 283 catchment areas, $B_i$.
The mean surface area of a catchment is approximately 36 $\text{km}^2$, substantially larger than the covariate resolution of 1 km $\times$ 1 km square cells, $b_{ij}$. We assume that $Y_i\big|\mu_i\overset{\text{iid}}{\sim}\mathcal{N}\Big(\mu_i,\sigma^2_e\Big)$ and fit the following model:
\begin{align*}
    \mu_i= m_{i}^{-1}\sum_{j=1}^{m_i} \mu_{ij} =  m_{i}^{-1} \sum_{j=1}^{m_i} \Big[\beta_0 + \beta_1 z_{ij} + R(b_{ij}) \Big],
\end{align*}
where $R(\cdot)$ is a Mat\'ern field with mean-squared differentiability parameter equal to 1. 
For model fitting, we use the SPDE approach
\citep{lindgren2011explicit} whose
mesh has inner and outer maximum edge-lengths of 10 km and 60 km, respectively
(Supplementary Material, Figure 12). For the two comparator models, we use the average of the population density over each $B_i$ as covariate. We specify diffuse priors for $\beta_0$ and $\beta_1$: $\beta_0 \sim \mathcal{N}(0,\infty)$ and $\beta_1 \sim \mathcal{N}(0,1000)$. We use PC priors for the Mat\'ern range and marginal variance parameters, $\rho_{R}$ and $\sigma^2_{R}$, such that $\mathbb{P}(\sigma_{R} > 1.49) = 0.5$ and $\mathbb{P}(\rho_{R} < 144 \;\text{km}) = 0.5$. For the parameters $\tau$ and $\phi$ of the MRF model (\ref{eq:comparatorB}),
we specify PC priors such that $\mathbb{P}(1/\sqrt{\tau}>1=0.01)$ and $\mathbb{P}(\phi>1=0.5)$.
Finally, for the precision parameter $1/\sigma^2_e$, we specify a diffuse Gamma prior with shape 1 and rate 0.00005.

The graph for the MRF catchment-level model, using the conventional definition of neighbours as catchments 
that share a common boundary segment, contains disconnected components and singletons. This presents a problem for fitting the MRF model (Equation \eqref{eq:comparatorB}), which we circumvent by using the approach in \cite{freni2018note}. This scales each connected component with more than one node so that the parameter $\tau$ in Equation \eqref{eq:comparatorB} can be interpreted as the typical precision \citep{sorbye2014scaling}.
Additionally, the priors for singletons are 
specified as independent Normal distributions with mean 0 and precision $\tau$. Finally,
we impose a sum-to-zero constraint on the set of random effects within each connected component.  

Table \ref{tab:app_gaussian_fixedeffs} shows the estimated fixed effects $\beta_0$ and $\beta_1$ for the three approaches. The posterior estimates for the proposed method and the centroids approach are very similar, whereas the estimates for the MRF model have slightly smaller posterior standard deviation. Population density and wastewater virus concentration of SARS-CoV-2 are positively associated, as expected. This is also consistent with the results in \cite{li2023spatio}, who fitted a spatio-temporal geostatistical model to the same data source. The block 
aggregation  and  centroids approaches give very similar estimates of the parameters of the Mat\'ern field (Supplementary Material, Table 1).
 The estimated standard deviation of the spatial component of the MRF model is smaller than the estimate of $\sigma_{R}$ from the other approaches.

\begin{table}[htbp]
\centering
\begin{tabular}{|l | l r r r r|}
  \hline
  \textbf{Approach} & \textbf{Parameter} & \textbf{Mean} & \textbf{SD} & \textbf{P$2.5^{\text{th}}$} & \textbf{P$97.5^{\text{th}}$} \\ 
  \hline
  \multirow{2}{*}{Centroids} & $\beta_0$ & 3.878 & 1.114 & 1.696 & 6.061 \\ 
                             & $\beta_1$ & 0.377 & 0.153 & 0.077 & 0.677 \\ \hline
  \multirow{2}{*}{MRF}       & $\beta_0$ & 3.424 & 0.975 & 1.512 & 5.336 \\ 
                             & $\beta_1$ & 0.449 & 0.133 & 0.188 & 0.710 \\ \hline
  \multirow{2}{*}{Block aggregation}  & $\beta_0$ & 3.850 & 1.114 & 1.667 & 6.034 \\ 
                             & $\beta_1$ & 0.381 & 0.153 & 0.081 & 0.682 \\ 
  \hline
\end{tabular}
\caption{Posterior estimates of fixed effects $\beta_0$ and $\beta_1$ for three approaches: centroids, MRF and block aggregation.}
\label{tab:app_gaussian_fixedeffs}
\end{table}

A model comparison using cross-validated  DS score, negative log score, and RMSE showed only a slight comparative advantage for the block aggregation approach over the centroids and MRF approaches (Supplementary Material, Table 2). Figures 13 and 14 in the Supplementary Material compare the predicted values of wastewater virus concentrations $\mu_i$ at catchment areas around the Greater London area. The predicted values are very similar for the three approaches although the posterior standard deviation for the MRF approach is smaller; see Table \ref{tab:app_gaussian_fixedeffs}.

\subsubsection{Predicting $\mu(\cdot)$ at LTLA level}

 Prediction of wastewater viral concentration at LTLA-level
 is straightforward using the block aggregation and centroids approaches, for which both the covariate data and the spatial effect $R(\cdot)$ are defined on
 continuous space; we first predict the values of $\mu_{ij}$ at the resolution of the population density raster, then aggregate to LTLA-level using spatial averaging.

 For the MRF model, the spatial effect depends on the specific catchment-level graph, which does not translate directly to any LTLA-level graph. To compute LTLA-level predictions using the MRF approach, we compute LTLA-level predictions of the virus concentration using the approach in \cite{mills2024utility}, as follows.
 We first calculate the area of overlap between each LSOA and each catchment area and use these proportions
 to estimate the population in the $j^{\text{th}}$ LSOA serviced by at least one sewage treatment works, say $\hat{p}_j$.
 We then estimate the population in the $k^{\text{th}}$ LTLA that is serviced by at least one STW company, say $\hat{p}_k$, by summing $\hat{p}_j$ over all LSOAs inside the $k^{\text{th}}$ LTLA. The estimated wastewater virus concentration for the $k^{\text{th}}$ LTLA, denoted by $\tilde{Y}_k$ is then given by:
\begin{align*}
    \tilde{Y}_k = \sum_{i \in k^{\text{th}}\; \text{LTLA}} \text{w}_{ik}Y_i,
\end{align*}
where $\text{w}_{ik}$ is the relative contribution of the $i^{\text{th}}$ catchment area to the total serviced population in the $k^{\text{th}}$ LTLA. We then fit the MRF model using the $\tilde{Y}_k$ as the response variable in Equation \eqref{eq:comparatorB}, in conjunction with the
neighbourhood graph based on contiguity of LTLA boundaries.

\begin{figure}[htbp]
    \centering

    \begin{minipage}[b]{0.45\textwidth}
        \centering
        \includegraphics[width=\linewidth]{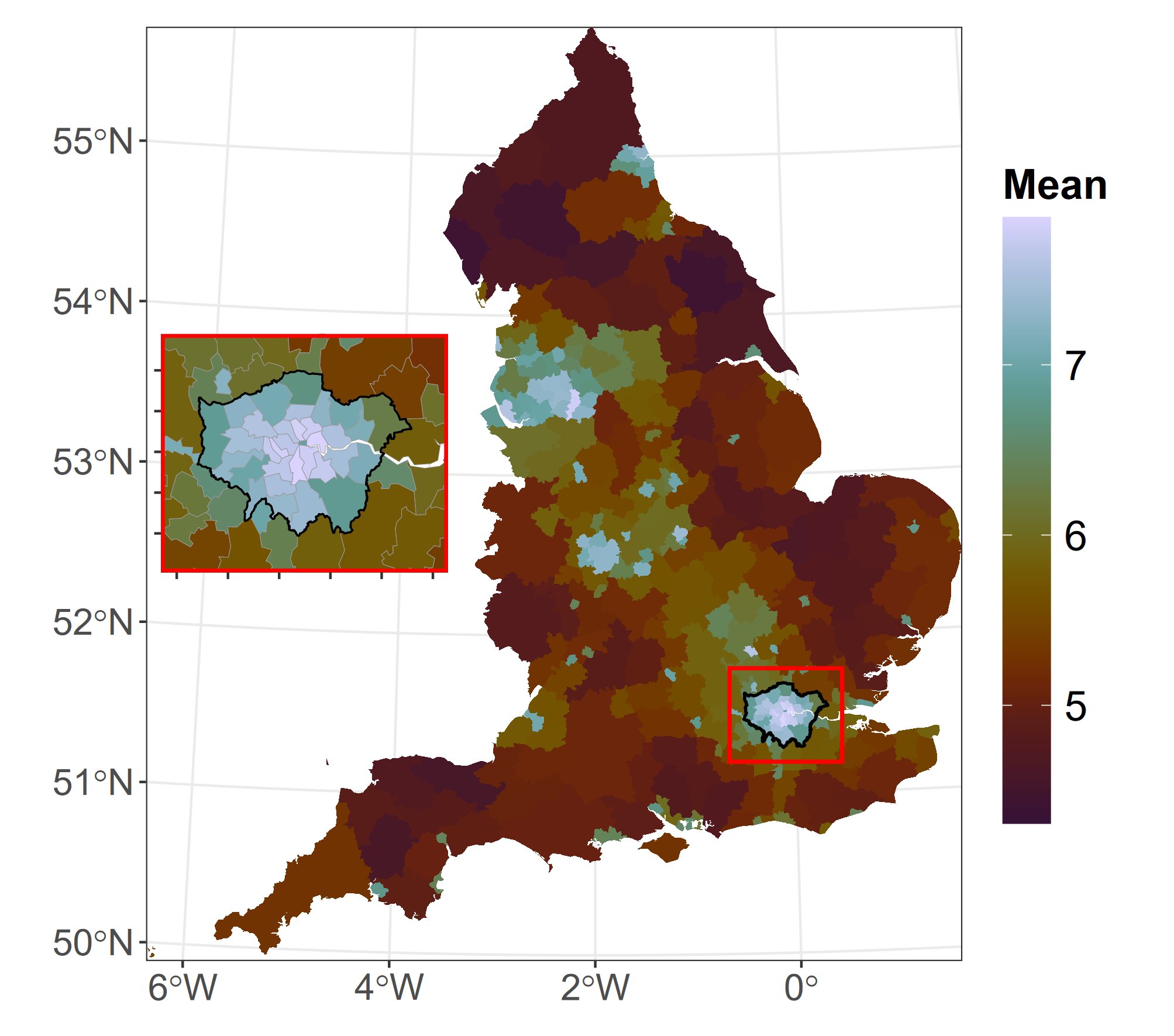}
        \vspace{0.6ex}
        \textbf{(a)}\; Centroid approach
    \end{minipage}
    \hspace{2mm}
    \begin{minipage}[b]{0.45\textwidth}
        \centering
        \includegraphics[width=\linewidth]{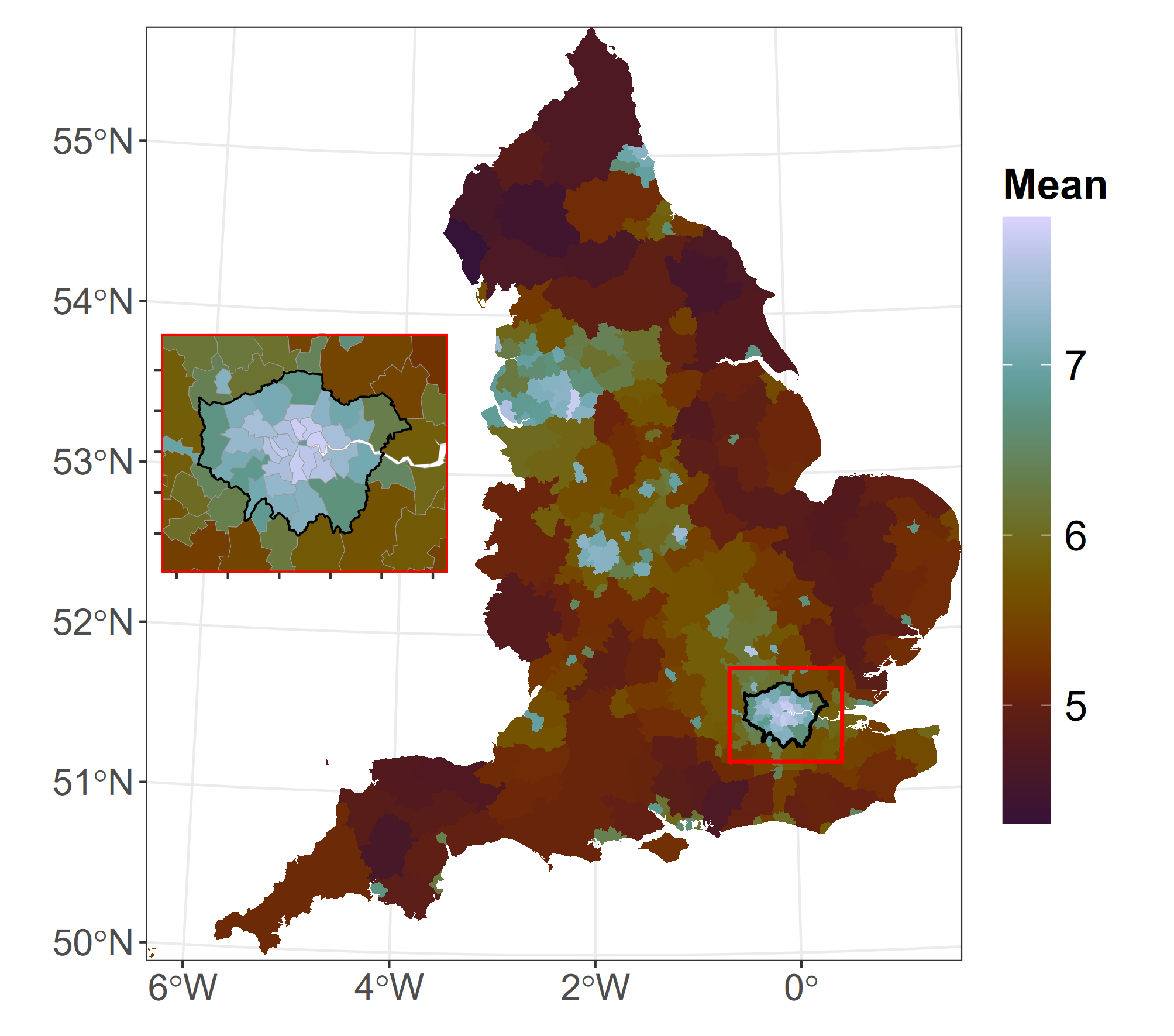}
        \vspace{0.6ex}
        \textbf{(b)}\; Block aggregation approach
    \end{minipage}

    \vspace{2ex}

    \begin{minipage}[b]{0.45\textwidth}
        \centering
        \includegraphics[width=\linewidth]{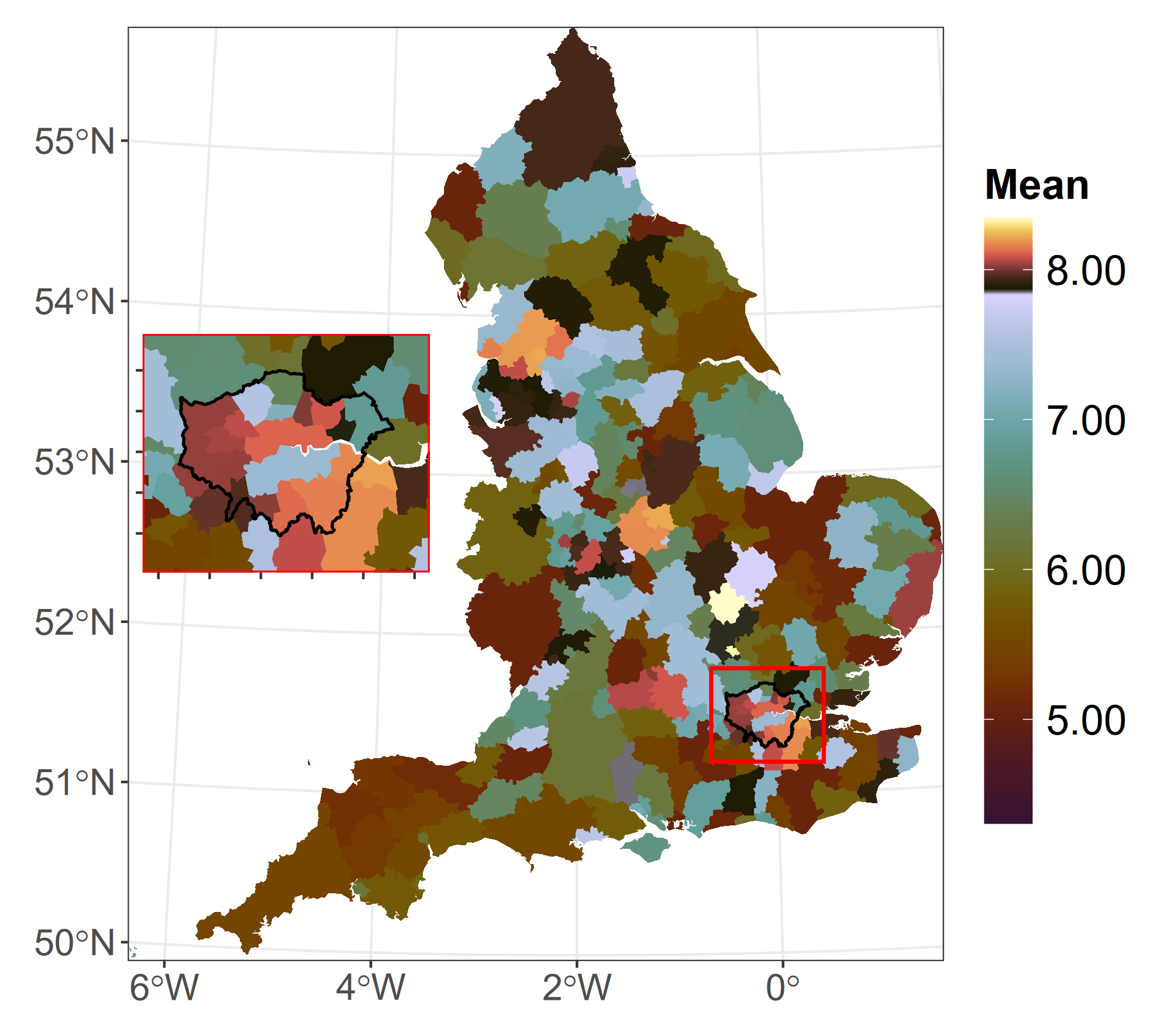}
        \vspace{0.6ex}
        \textbf{(c)}\; MRF approach
    \end{minipage}
    \hspace{2mm}
    \begin{minipage}[b]{0.45\textwidth}
        \centering
        \includegraphics[width=\linewidth]{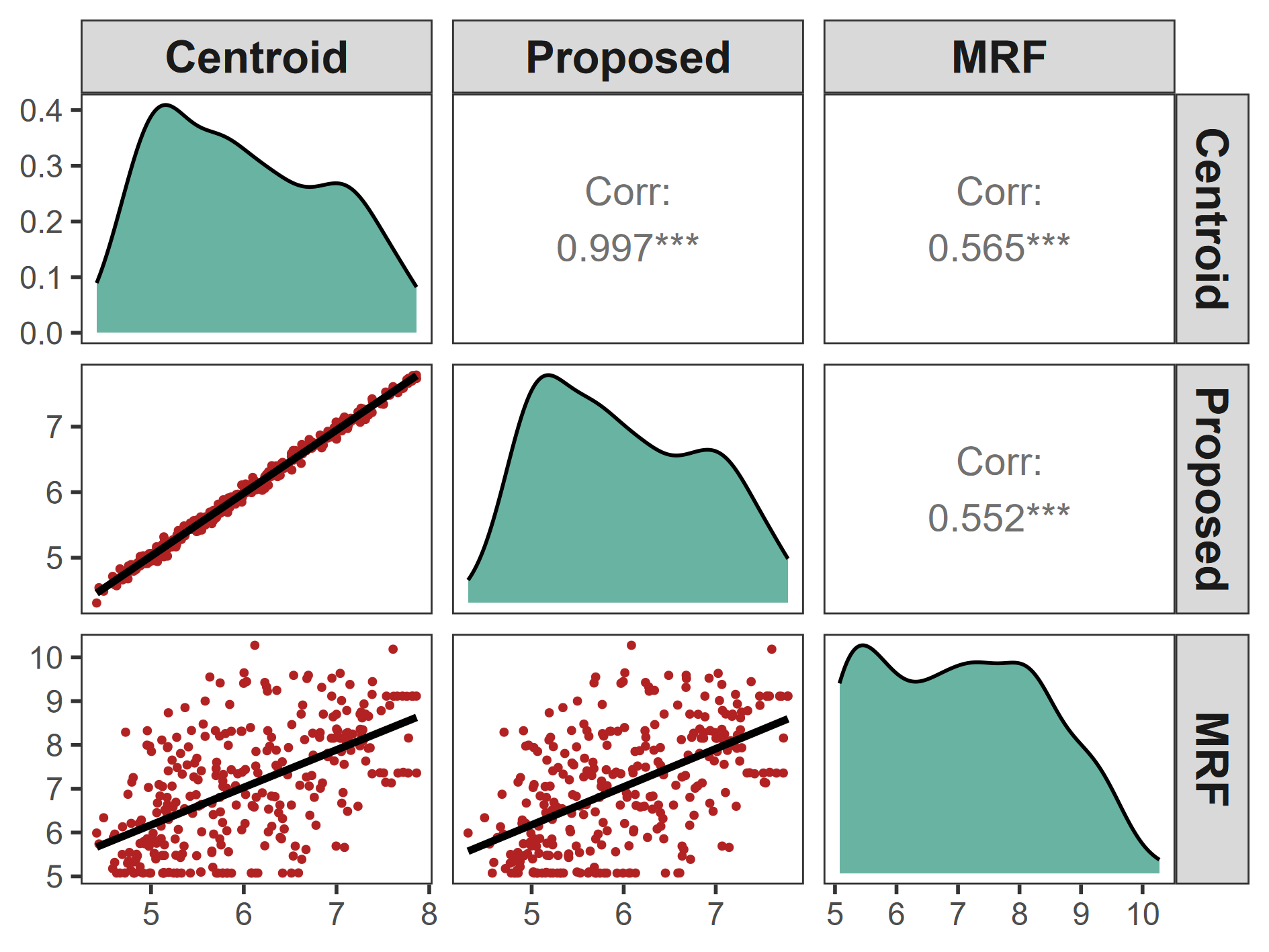}
        \vspace{0.6ex}
        \textbf{(d)}\; Agreement among the approaches
    \end{minipage}

    \caption{Comparison of predicted LTLA-level wastewater virus concentrations using 
    (a) the centroid approach, 
    (b) the block aggregation approach, 
    (c) the MRF approach, and 
    (d) comparison of predicted means across the three approaches.}
    \label{fig:app_gaussian_ltla_preds_combined}
\end{figure}

Figures \ref{fig:app_gaussian_ltla_preds_combined}a -- \ref{fig:app_gaussian_ltla_preds_combined}c show the predicted LTLA-level wastewater virus concentrations for the three approaches. As expected, in view of the equivalence noted in Equation \eqref{eq:gaussian_model_equiv}, the LTLA-level predictions are very similar for the centroids and block aggregation approaches. The London area has high virus concentrations, reflecting its high population density. The predicted LTLA-level values obtained from the MRF approach are markedly different. Figure \ref{fig:app_gaussian_ltla_preds_combined}d uses pairwise scatterplots to
compare the predictions of LTLA-level virus concentrations obtained from the three approaches. The predictions from the centroids and block aggregation approaches are almost identical, whereas those from the MRF approach are substantially different, and more variable.

\subsection{Cardiovascular-related hospitalisations}

\begin{figure}[ht]
    \centering
    \includegraphics[width=.85\linewidth]{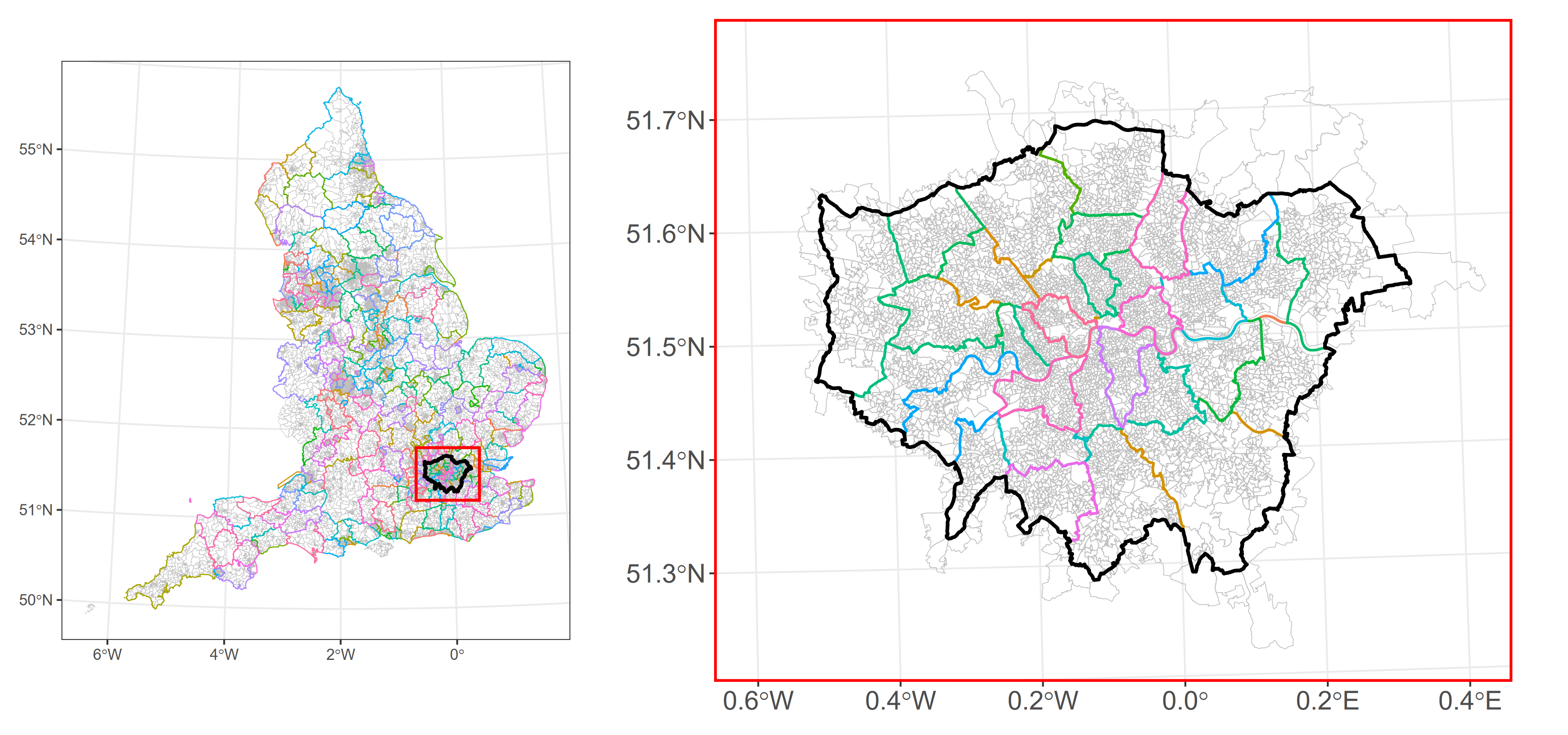}
    \caption{\textit{Left}: Plot of the LTLA boundaries (in color) and the LSOA boundaries (in gray); \textit{Right}: Zoomed-in map of Greater London area}
    \label{fig:app_Poisson_ltla_lsoa}
\end{figure}

\begin{figure}[ht]
    \centering

    \begin{minipage}[b]{0.48\linewidth}
        \centering
        \includegraphics[width=\linewidth]{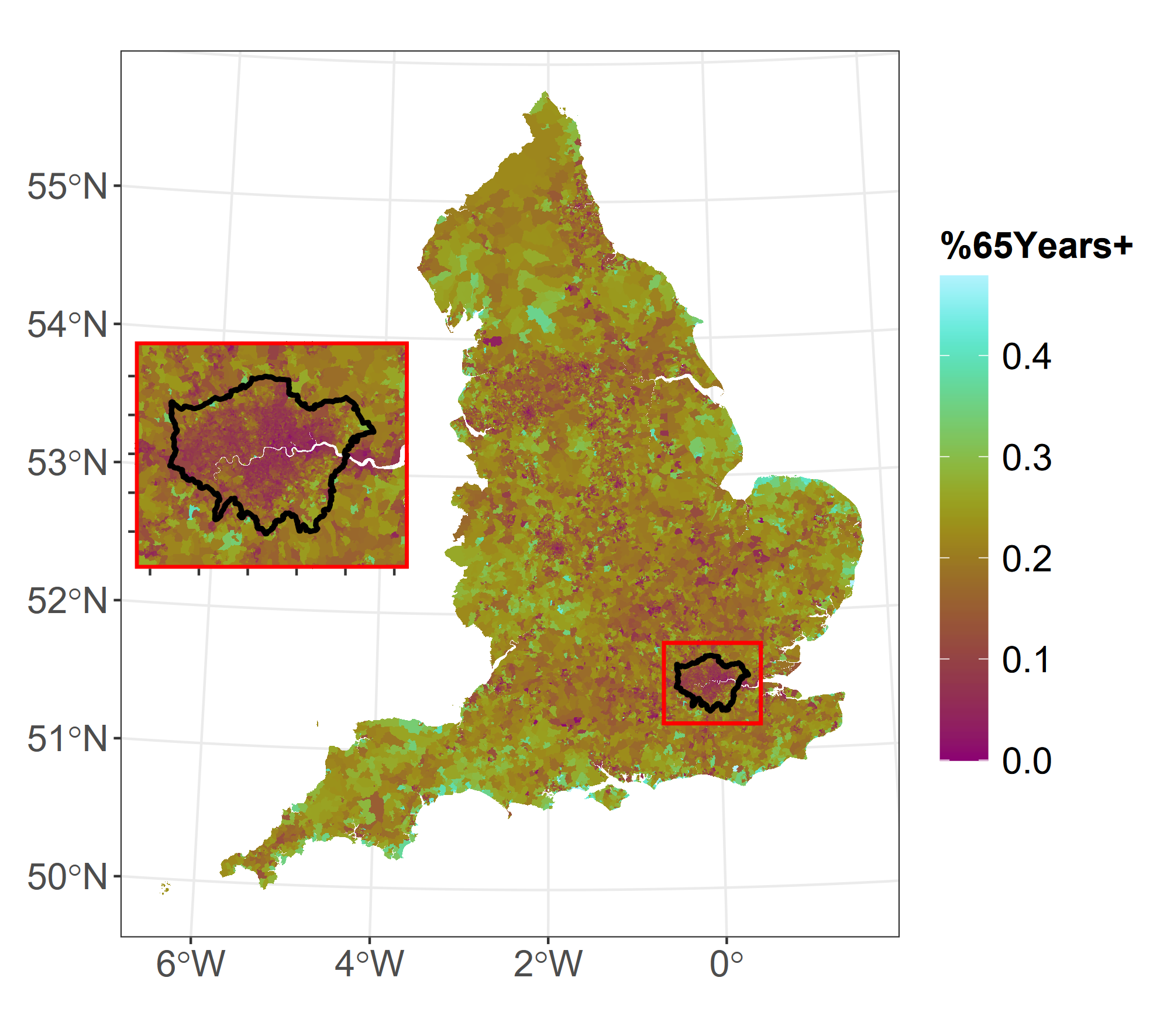}
        \vspace{0.6ex}
        \textbf{(a)}\; $\texttt{OLD}_{ij}$ at LSOA-level
    \end{minipage}\hfill
    \begin{minipage}[b]{0.48\linewidth}
        \centering
        \includegraphics[width=\linewidth]{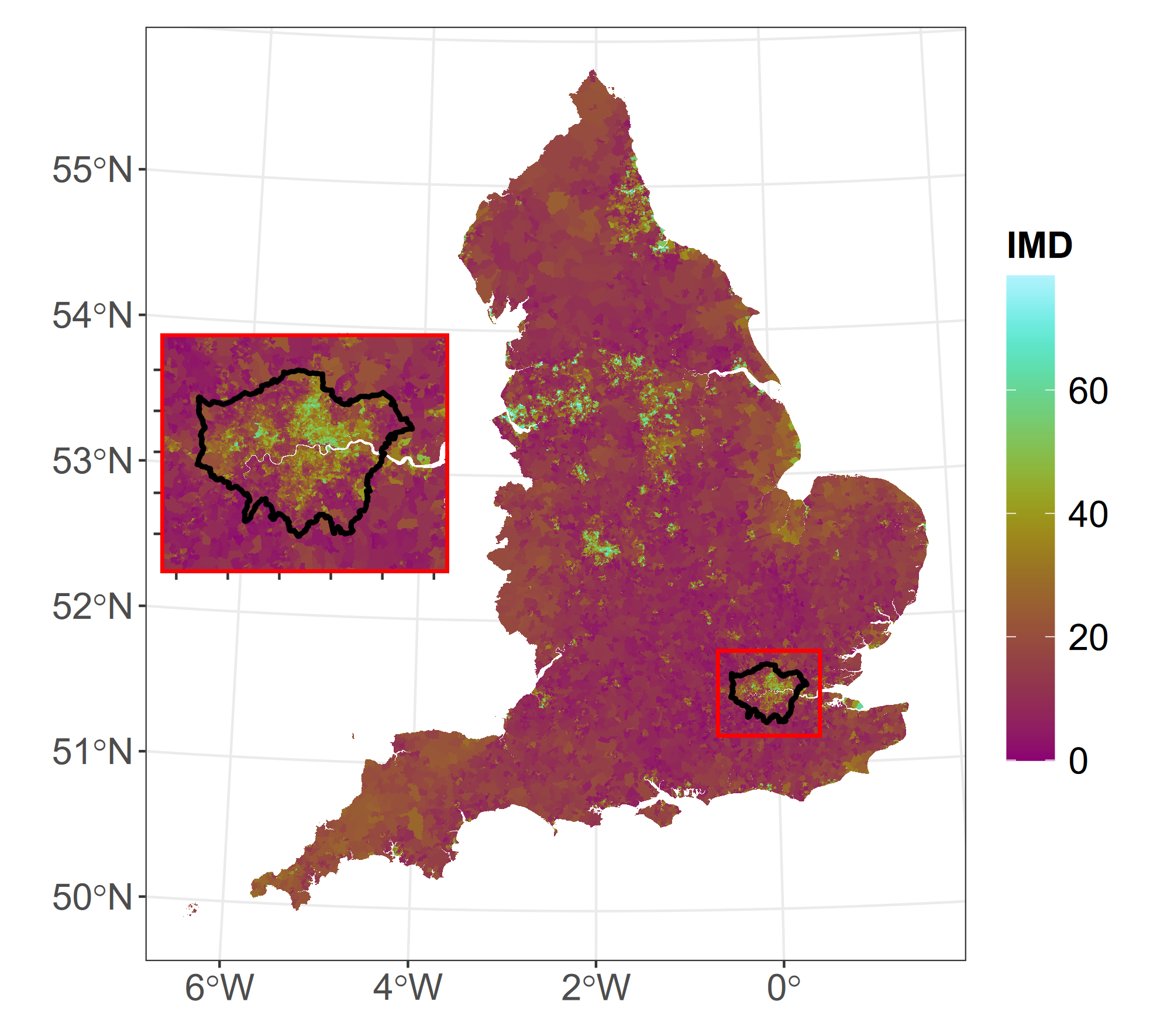}
        \vspace{0.6ex}
        \textbf{(b)}\; $\texttt{IMD}_{ij}$ at LSOA-level
    \end{minipage}

    \vspace{2ex}

    \begin{minipage}[b]{0.48\linewidth}
        \centering
        \includegraphics[width=\linewidth]{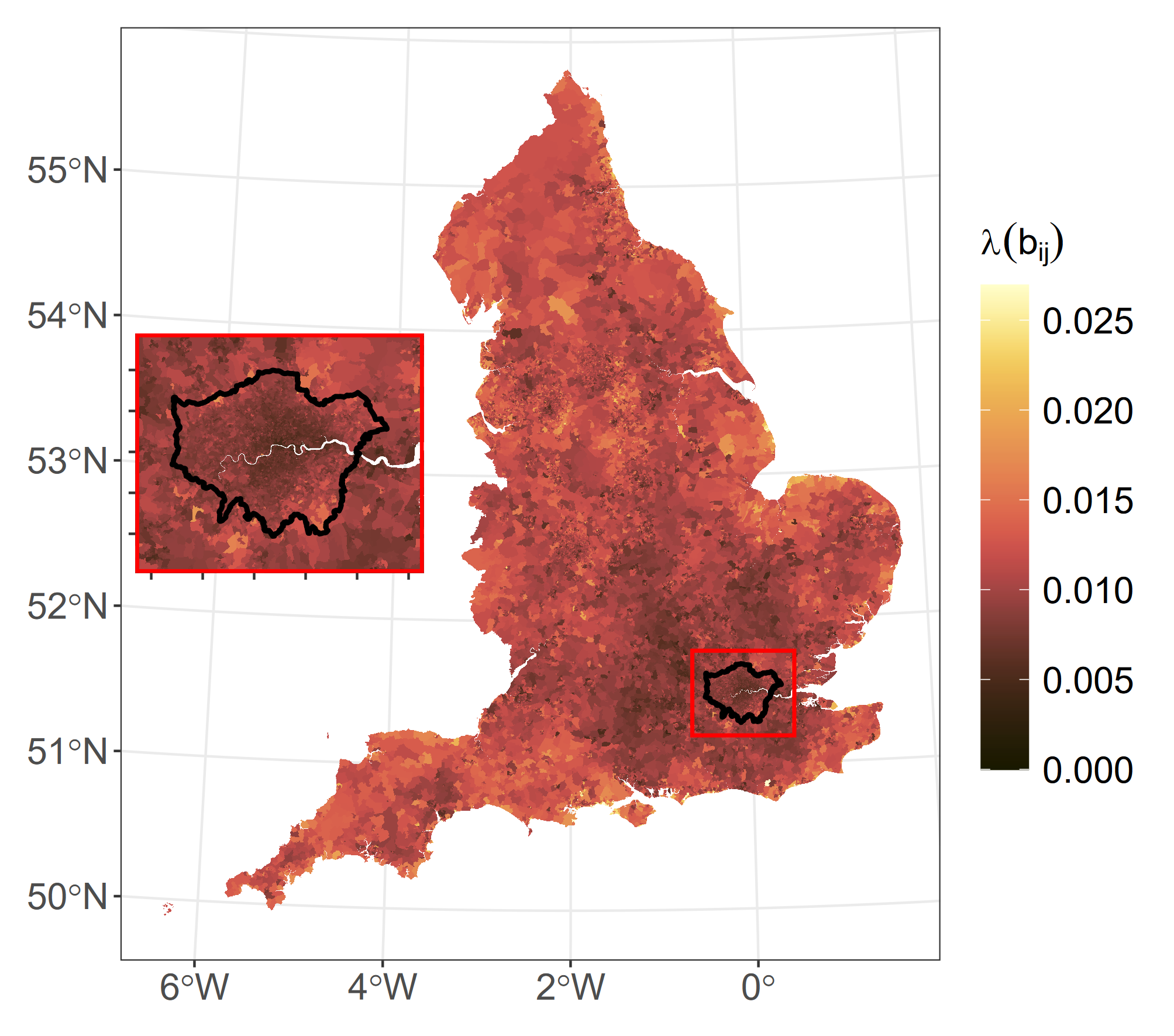}
        \vspace{0.6ex}
        \textbf{(c)}\; posterior mean of $\lambda_{ij}$
    \end{minipage}\hfill
    \begin{minipage}[b]{0.48\linewidth}
        \centering
        \includegraphics[width=\linewidth]{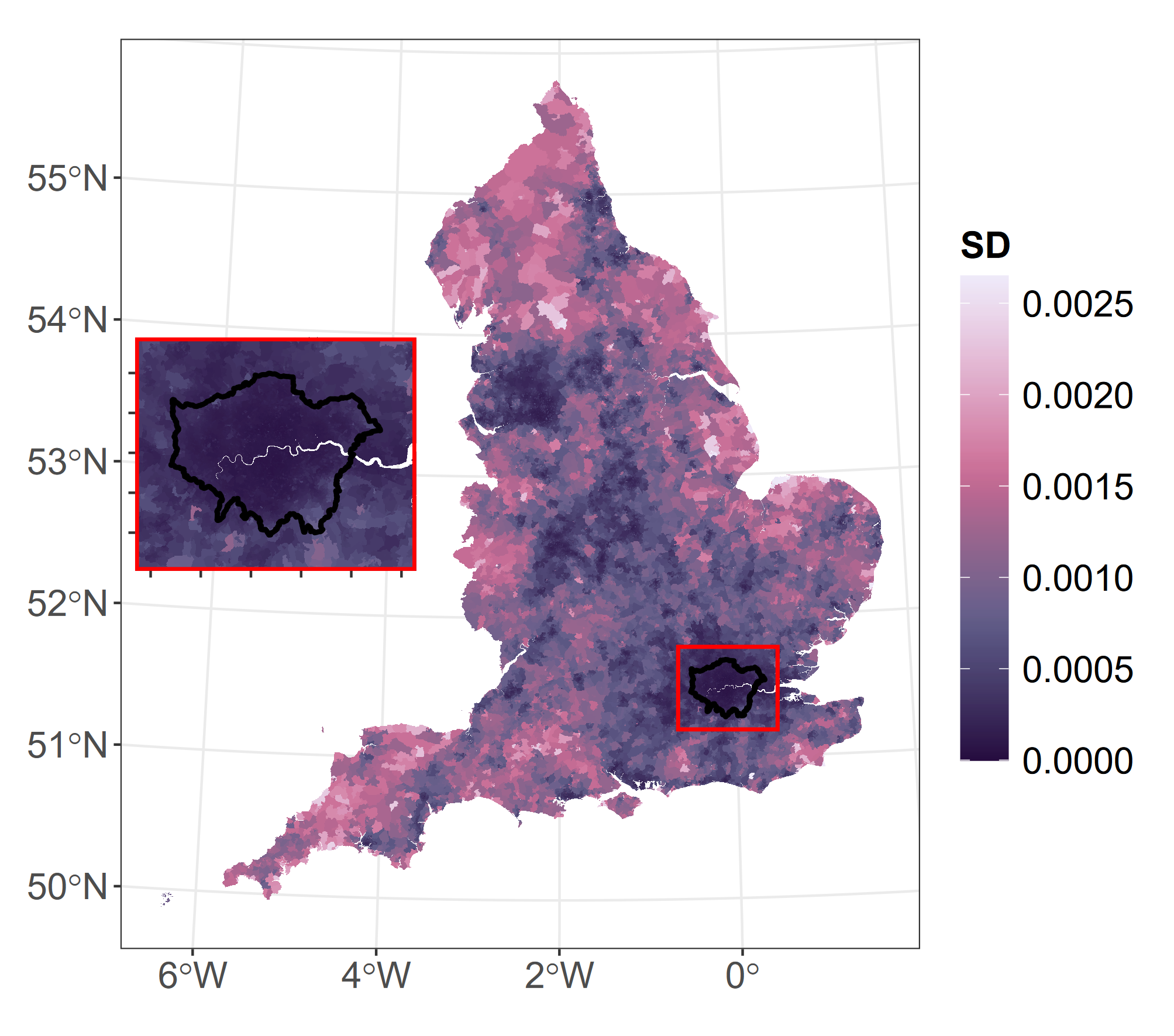}
        \vspace{0.6ex}
        \textbf{(d)}\; Posterior standard deviation of $\lambda_{ij}$
    \end{minipage}

    \caption{LSOA-level covariates and predicted  values of disease incidence, $\lambda_{ij}$.}
    \label{fig:app_poisson_combined}
\end{figure}

For our second application, the response is the number of cardiovascular-related hospitalisations for erach LTLA in England for the year 2011, and our goal is to predict hospitalisation numbers at LSOA-level. As LSOA-level covariates, we consider: (i) the  Index of Multiple Deprivation (IMD) for year 2011 (Figure \ref{fig:app_poisson_combined}a); (ii) the percentage of 65+ years old population (Figure \ref{fig:app_poisson_combined}b). Figure \ref{fig:app_Poisson_ltla_lsoa} shows the boundaries of the 322 LTLAs (in color) and 32,844 LSOAs (in grey) in England for 2011. It also shows a zoomed-in portion of Greater London. On average,  each LTLA 
includes around 100 LSOAs. In contrast to the wastewater application
described in Section \ref{subsec:wastewater},
 the LTLAs (blocks) partition the whole of England.

We denote by $Y_i$ the number of hospitalisations for the LTLA $B_i$, $\texttt{OLD}_{ij}$ as the
LSOA-level proportion of the population 65 years old or above, and $\texttt{IMD}_{ij}$ as the
LSOA-level IMD score. 
 
  We fit the following model:
\begin{equation}\label{eq:app_poisson_model}
    \begin{aligned}
    Y_i\Big|\mu_i &\overset{\text{iid}}{\sim} \text{Poisson}\big(\mu_i\big) \\ 
    & \mathbb{E}[Y_i] = \mu_i = \sum_{j=1}^{m_i}\mu_{ij}= \sum_{j=1}^{m_i}\big(\lambda_{ij}\times P_{ij}\big)\\
    &\lambda_{ij} = \exp\Big\{\beta_0 + \beta_1\texttt{OLD}_{ij} + \beta_2\texttt{IMD}_{ij} + R(b_{ij}) \Big\},
\end{aligned}
\end{equation}
where the offset $P_{ij}$  is the total population of the cell $b_{ij}$, $\lambda_{ij}$ is the incidence rate for $b_{ij}$ and $R(b_{ij})$ is a Mat\'ern field with mean-squared differentiability parameter
fixed to 1.

For the centroids aproach,
we use the population-weighted average of the covariates $\texttt{OLD}_{ij}$ and $\texttt{IMD}_{ij}$, giving the  LTLA-level predictor expression
{\begin{equation}\label{eq:app_poisson_model_centroids}
    \begin{aligned}
    & \mathbb{E}[Y_i] = \mu_i = \lambda_i\times \sum_{j=1}^{m_i} P(b_{ij})\\
    &\lambda_i = \exp\Bigg[\beta_0 + \beta_1\dfrac{\sum_{j=1}^{m_i} P_{ij}\times\texttt{OLD}_{ij}}{\sum_{j=1}^{m_i}P_{ij}} + \beta_2\dfrac{\sum_{j=1}^{m_i} P_{ij}\times\texttt{IMD}_{ij}}{\sum_{j=1}^{m_i} P_{ij}} + R(x_i) \Bigg],
\end{aligned}
\end{equation}}
where $x_i$ is the centroid of $B_i$.
 The MRF model uses the same covariates, but
 represents the spatial random effect as a spatially discrete graph with neighbourhoods 
 defined by contiguity of LTLA boundaries. For inference, we use the same  mesh and prior distributions as in Section \ref{subsec:wastewater}.

Table \ref{tab:app_Poisson_fixedeffs} shows the estimated fixed effects for the three modelling approaches. The posterior means are very similar, while the posterior SD is slightly larger for the MRF model.  More deprived LTLAs, and LTLAs with higher proportion of people who are 65 years or older have higher incidence of cardiovascular-related hospitalisations. The estimates of model hyperparameters are shown in Table 3 of the Supplementary Material. The estimated spatial range of $R(\cdot)$ is approximately 27.5 km and 18.5 km for the centroids and block aggregation approaches, respectively. The respective estimated marginal standard deviations $\sigma_{R}$ are 0.124 and 0.169. For
the MRF model, the
estimated marginal standard deviation is 0.109 
and the estimated mixing parameter $\phi$ is 0.80, indicating strong spatial dependence. 

The MRF model gives the best performance with
respect to leave-one-area-out cross-validation at LTLA-level (Supplementary Material, Table 4
and Figure 15).  This agrees with
the results from the simulation study, where we found that the three models give similar predictive performance at block-level predictions.

\begin{table}[htbp]
\centering
\begin{tabular}{|l|lrrrr|}
  \hline
  \textbf{Approach} & \textbf{Parameter} & \textbf{Mean} & \textbf{SD} &\textbf{P$2.5^{\text{th}}$} & \textbf{P$97.5^{\text{th}}$}  \\
  \hline
  \multirow{3}{*}{Centroids} 
    & $\beta_0$ & -5.313 & 0.042 & -5.396 & -5.231 \\
    & $\beta_1$ & 3.339  & 0.160 & 3.025  & 3.653 \\
    & $\beta_2$ & 0.009  & 0.001 & 0.007  & 0.010 \\
    \hline
  \multirow{3}{*}{MRF} 
    & $\beta_0$ & -5.327 & 0.045 & -5.416 & -5.238 \\
    & $\beta_1$ & 3.375  & 0.193 & 2.997  & 3.753 \\
    & $\beta_2$ & 0.008  & 0.001 & 0.007  & 0.010 \\
  \hline
  \multirow{3}{*}{Block aggregation} 
    & $\beta_0$ & -5.306 & 0.043 & -5.391 & -5.221 \\
    & $\beta_1$ & 3.193  & 0.157 & 2.885  & 3.500 \\
    & $\beta_2$ & 0.008  & 0.001 & 0.007  & 0.010 \\
    \hline
\end{tabular}
\caption{Posterior estimates of fixed effects $\beta_0$, $\beta_1$, and $\beta_2$ for  centroids, MRF, and block aggregation approaches.}
\label{tab:app_Poisson_fixedeffs}
\end{table}

The advantage of the block aggregation approach lies in its ability to perform spatial disaggregation of the health outcome to the resolution of the covariate information.  Figures \ref{fig:app_poisson_combined}c and  \ref{fig:app_poisson_combined}d show the predicted LSOA-level incidence $\hat{\lambda}_{ij}$ and predictive standard deviation. Figure \ref{fig:app_poisson_combined} shows that the predicted values $\hat{\lambda}_{ij}$ are strongly correlated with $\texttt{OLD}_{ij}$, and weakly correlated with $\texttt{IMD}_{ij}$ (see also Figure 17 in the Supplementary Material). The predicted LSOA-level disease counts $\mu_{ij}$ and the corresponding posterior uncertainties are shown in Figure 16 in the Supplementary Material.

\section{Discussion}\label{sec:discussion}

We have proposed a block aggregation approach for spatial prediction and spatial disaggregation 
when response data are
only available on a set of {\it blocks}, i.e. areal units that partition some or all of the area of interest, and covariate data are available for a set 
of {\it cells} at
a finer spatial resolution
that approximates to a set of points in continuous space.
Typically, cells correspond to pixels in
a raster image. If cells are irregular polygons,
we accommodate them by overlaying a pixel grid and either replicating 
covariate values for all
pixels within any one cell,
or distributing the cell value 
proportionately, according to context.

We adopt a spatially continuous model formulation whose parameters can be interpreted independently of the scale on which the response data are recorded. 
The model provides a principled framework for understanding covariate-response relationships
at finer spatial resolution than is possible with current models, 
and for predictions
where no response data are observed,
at whatever spatial resolution is required in a particular application. This flexibility contrasts with widely used Markov Random Field models whose interpretation is limited to the set of areal units on which the response data are observed.

We have shown, through simulations and applications,
that the performance of the block aggregation approach is comparable to that of existing approaches when
the inferential goal is
block-level prediction, but brings clear advantages
when cell-level prediction is required.
The two applications presented in
Section \ref{sec:dataapplication} demonstrate that
 the block aggregation model
 can be used to address significant,
 national-scale environmental and public health research questions at a finer spatial
 resolution than is possible with
current methods.

The block-level approach combines a latent, spatially continuous {\it process model} with a spatially discrete {\it sampling model} for the block-level response conditional on the underlying process. The sampling model is analogous to a generalized linear mixed  model, involving a distributional form for the response and  an inverse link function between the latent spatial process to the scale of the response and spatial integration to specify the block-level mean response. We have implemented the approach
for linear Gaussian and log-linear Poisson sampling models. The approach extends to other sampling models, with the proviso that our current implementation requires the
distributional form of the sampling model to be closed under aggregation. We are currently exploring ways of relaxing this requirement through the use of distributional approximations. The extension to a spatio-temporal  model with latent process $R(x,t)$ is conceptually straightforward but computationally more demanding as, even when $R(x,t)$ incorporates Markov dependence in time, the response data have non-Markov structure. One way to control the computational demand is to fit a spatio-temporal model to sequences of data over moving fixed time-intervals, $(t,t+k)$ say.

Our long-term goal is to incorporate
our approach into near-real-time health surveillance systems, for which purpose we require an implementation that is computationally fast and 
can be run reliably with minimal supervision. For this reason, we have used a linearised INLA
method that does not need tuning to each new application, as would be the case for an MCMC algorithm. The linearisation requires iteration of the INLA algorithm to optimise the linearisation point; in our experience, around 3 iterations
are required, based on the convergence criteria stated in Section \ref{sec:inference}.

The proposed approach reflects methodological innovations made possible not only by the availability of data at multiple spatial resolutions, but also by advances in efficient computational methods for fitting complex models. This approach showcases that it is possible to accommodate data of varying spatial resolution, without relying on pre-processing steps, such as computing spatial averages as input to the model, which can inevitably discard information through aggregation. The proposed model provides a principled framework for understanding covariate-response relationships, performing spatial disaggregation, and generating predictions for arbitrary block configurations by adopting a \textit{continuous} perspective, i.e., treating areal observations as being governed by an underlying spatial continuous process.

\section*{Funding}

The work has been funded by the Medical Research Council (MRC), award UKRI078: "Incorporating wastewater-based epidemiology into a real-time, multiplex public health surveillance system". MB also acknowledges partial support from the MRC Centre for Environment and Health, funded by the UK Medical Research Council, Grant number: MR/L01341X/1. MB, EW,  and SV acknowledge infrastructure support for the Department of Epidemiology and Biostatistics provided by the NIHR Imperial Biomedical Research Centre (BRC).

\section*{Authors contribution}

Conceptualization: SJV, PJD, GL, MB\\
Methodology: SJV, PJD, FL, HR, GL, EW, MB\\
Software: FL, HR\\
Validation: PJD, MB\\
Formal Analysis: SJV, PJD, GL, MB\\
Investigation: SJV, PJD, GL, EW, MB\\
Resources: SJV, PJD, MB\\
Writing – Original Draft: SJV, PJD, MB\\
Writing – Review \& Editing: PJD, GL, MW, MB\\
Visualization: SJV\\
Supervision: PJD, MB\\
Project Administration: MB\\
Funding Acquisition: PJD, GL, MW, MB

\section*{Acknowledgments}

We acknowledge the contributions from the following individuals who are also named under the UKRI078 grant: Christopher Williams and Alisha Davies from Public Health Wales, and Professor Davey Jones from Bangor University.

\bibliography{references} 

\clearpage
\pagestyle{plain}

\setcounter{section}{0}
\setcounter{figure}{0}
\setcounter{table}{0}

\section*{Supplementary Material}

\section{Study domain for simulation study}

\begin{figure}[htbp]
        \centering
        \includegraphics[width=1\linewidth]{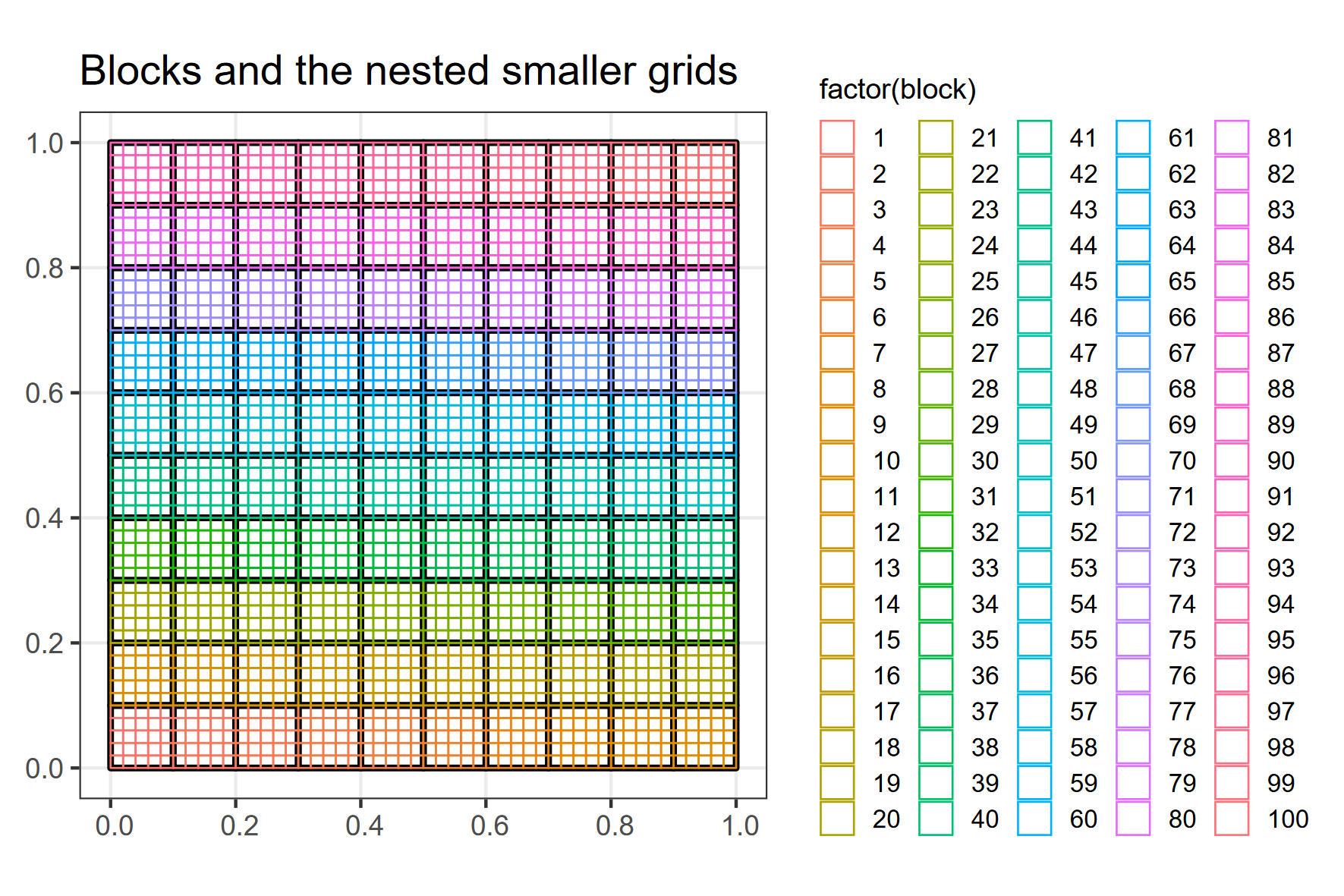}
        \caption{Blocks $B_i, i=1,\ldots,100$; and nested grids $b_{ij}$ for each $B_i$}
        \label{fig:areas_grids}
    \end{figure}

\newpage

\section{Simulated data example}

\begin{figure}[ht]
    \centering

    \begin{minipage}[b]{0.24\linewidth}
        \centering
        \includegraphics[width=\linewidth]{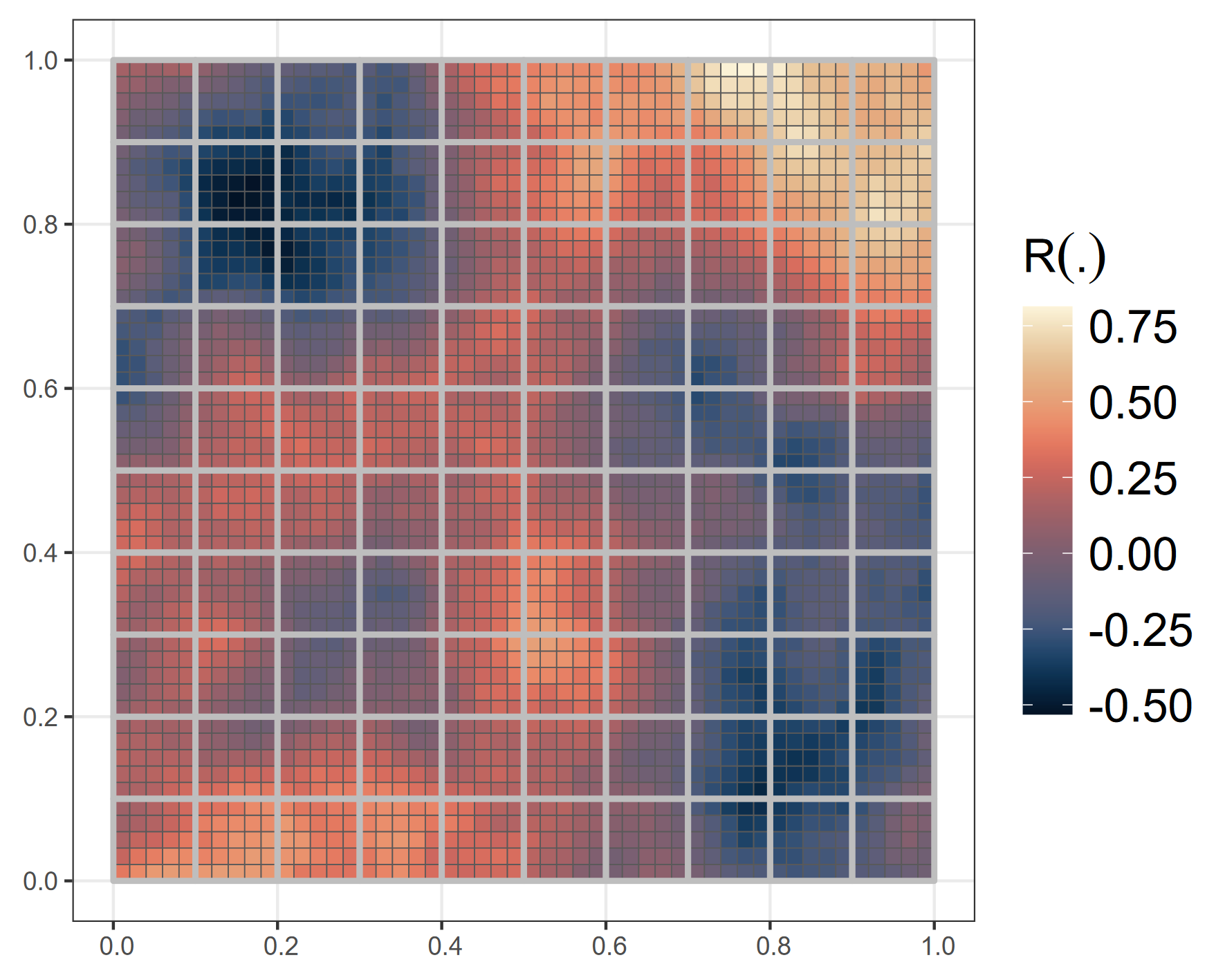}
        \vspace{0.4ex}
        \textbf{(a)}\; $R(b_{ij})$
        \label{fig:illust_poisson_data_a}
    \end{minipage}
    \hfill
    \begin{minipage}[b]{0.24\linewidth}
        \centering
        \includegraphics[width=\linewidth]{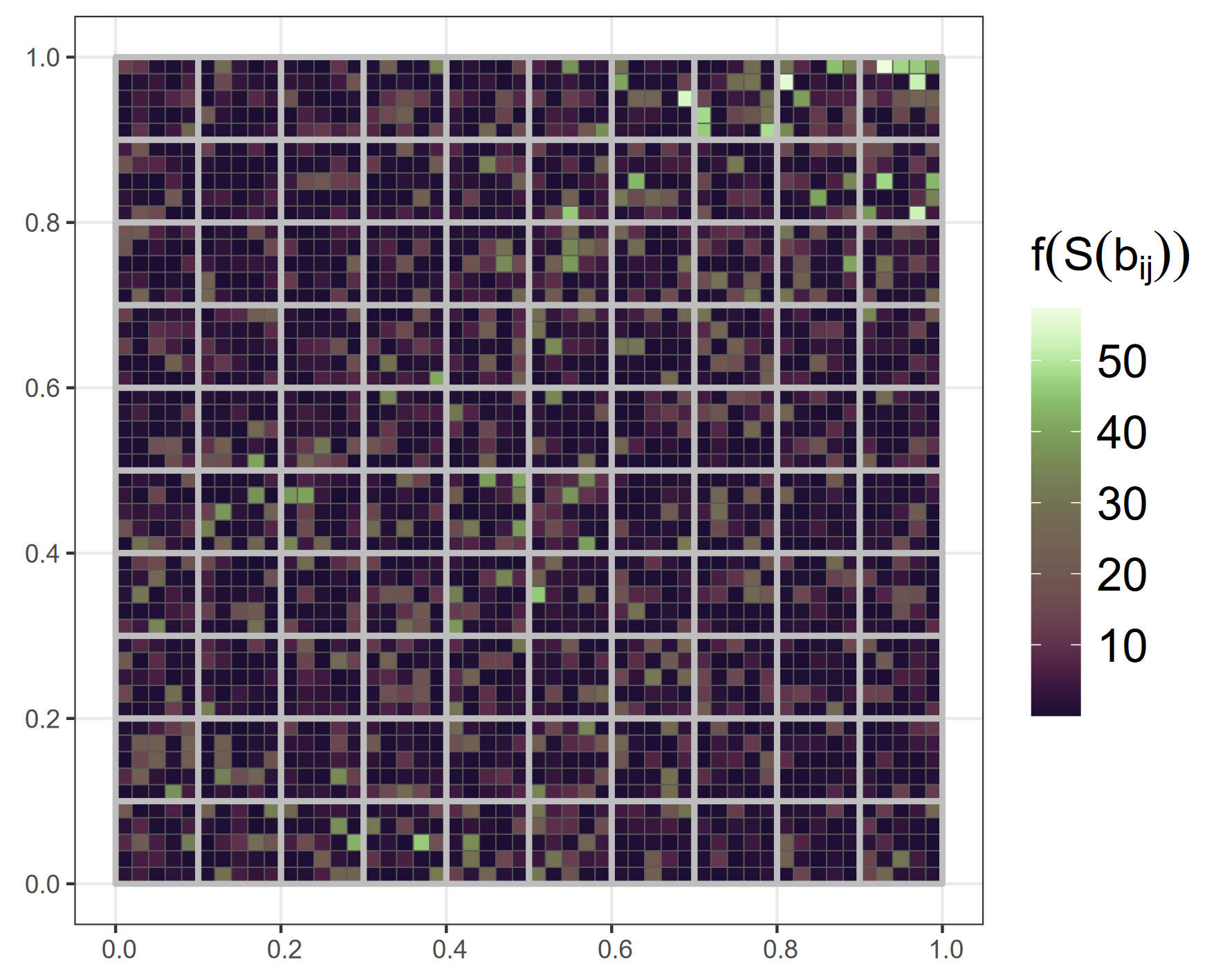}
        \vspace{0.4ex}
        \textbf{(b)}\; $f(S(b_{ij}))$
        \label{fig:illust_poisson_data_b}
    \end{minipage}
    \hfill
    \begin{minipage}[b]{0.24\linewidth}
        \centering
        \includegraphics[width=\linewidth]{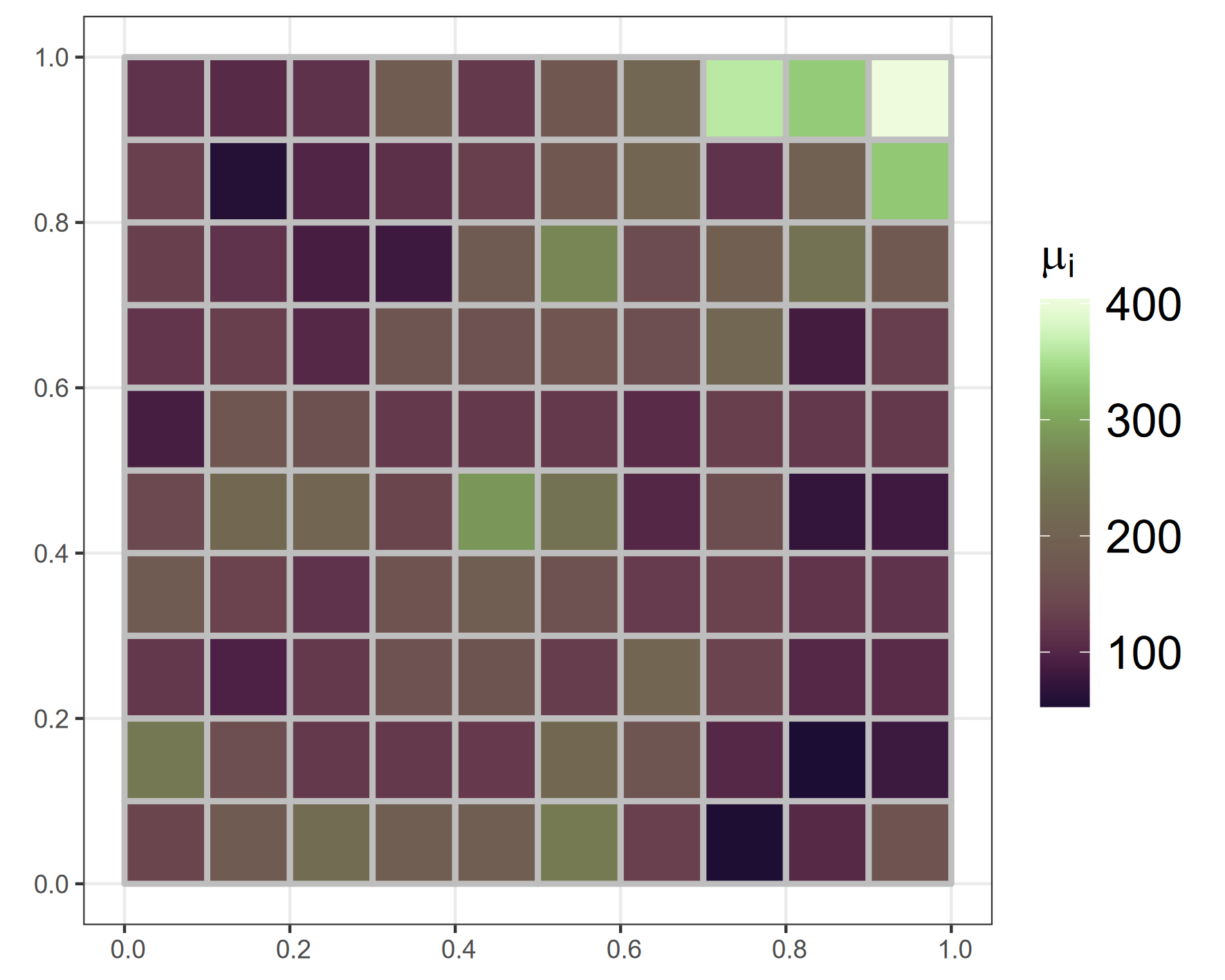}
        \vspace{0.4ex}
        \textbf{(c)}\; $\mu_i$
        \label{fig:illust_poisson_data_c}
    \end{minipage}
    \hfill
    \begin{minipage}[b]{0.24\linewidth}
        \centering
        \includegraphics[width=\linewidth]{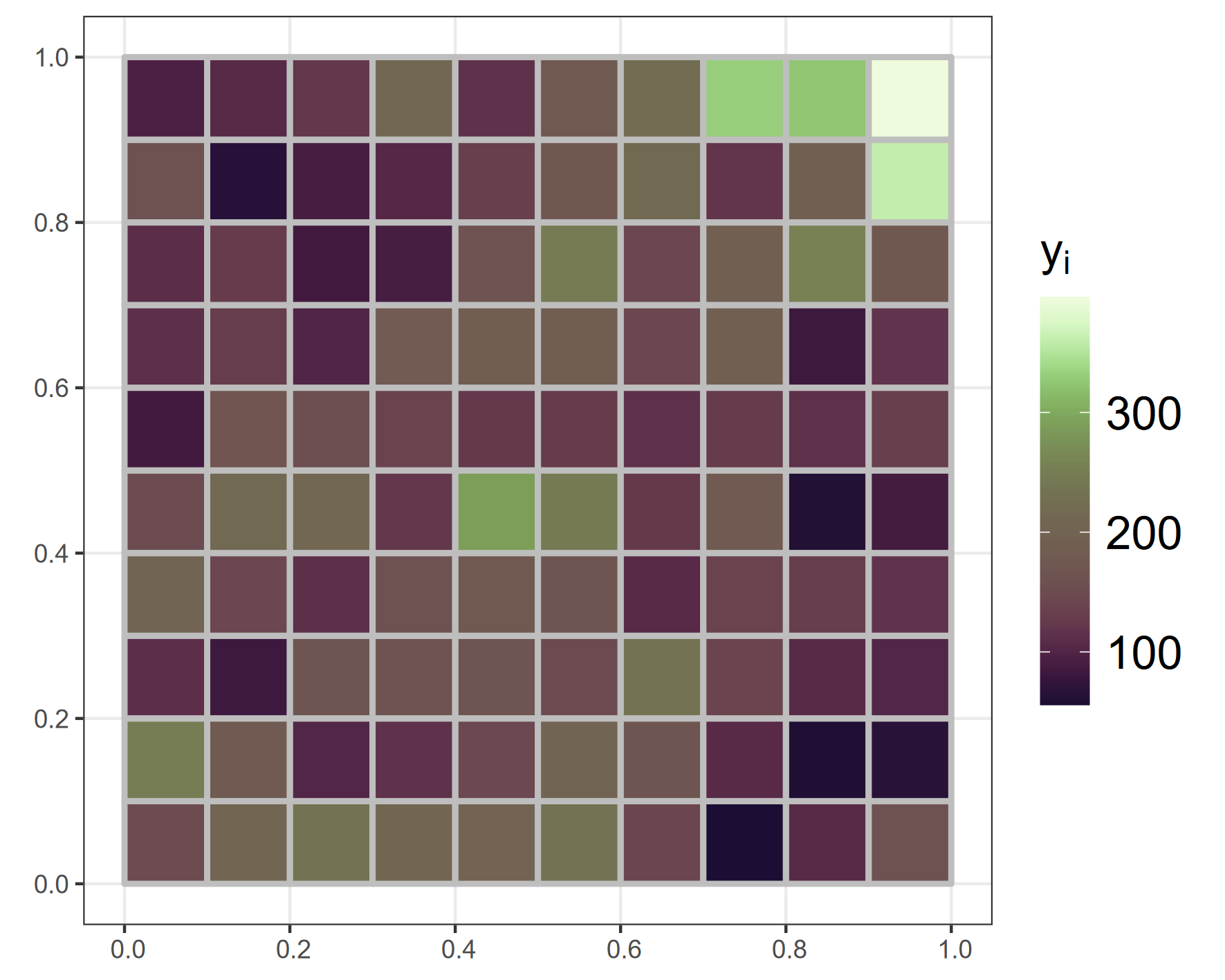}
        \vspace{0.4ex}
        \textbf{(d)}\; $y_i$
        \label{fig:illust_poisson_data_d}
    \end{minipage}

    \caption{Simulated values for a Poisson block aggregation model outcome.}
    \label{fig:illust_poisson_data}
\end{figure}

This section presents a simulated data example for a Poisson sampling model. Figure \ref{fig:illust_poisson_data}a shows a simulated Mat\'ern field, with a range parameter of 0.4 units and a marginal standard deviation of 0.15. Figure \ref{fig:illust_poisson_data}b shows the simulated $f\Big( S(b_{ij})\Big)$, while Figure \ref{fig:illust_poisson_data}c shows the aggregated values $\mu_i$. Figure \ref{fig:illust_poisson_data}d shows a simulated set of data ${Y_i}$ given the $\mu_i$ in Figure \ref{fig:illust_poisson_data}c. 

\begin{figure}[htbp]
    \centering
    \includegraphics[width=.9\linewidth]{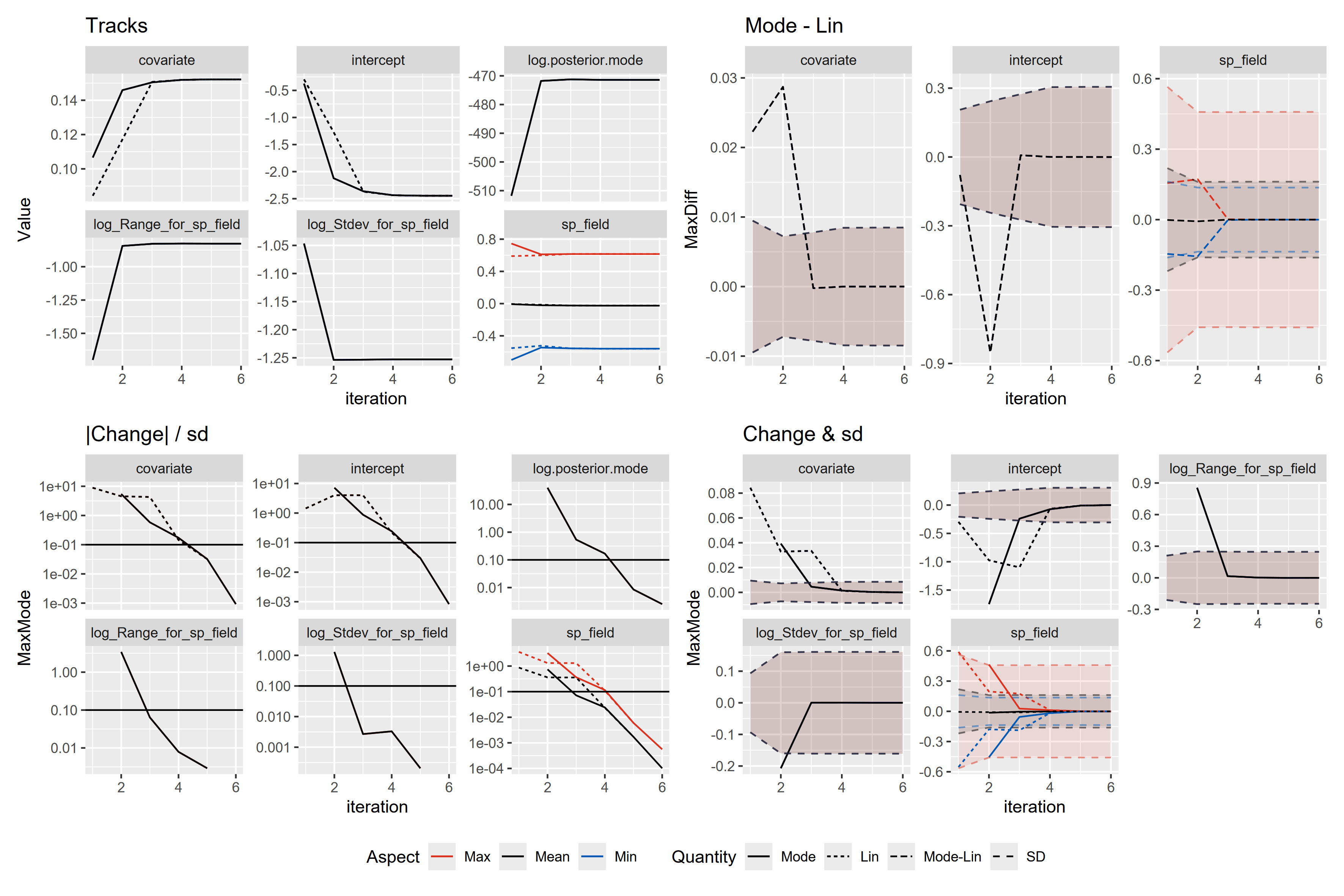}
    \caption{Convergence plot from the iterative linearised INLA for the simulated data in Figure \ref{fig:illust_poisson_data}}
    \label{fig:illust_poisson_convergence plot}
\end{figure}

Figure \ref{fig:illust_poisson_convergence plot} shows some convergence plots for the iterative linearised INLA method. The top left panel shows a comparison of the linearisation points $\bm{u}_0^{(k)}$ (dashed line) and the mode $\hat{\bm{u}}_0^{(k)}$ (solid line) from the linearised model. Note that the values of $\beta_0$ and $\beta_1$ 
coincide after the third iteration. For the Mat\'ern field $R(\cdot)$, the plot shows the values of the maximum and minimum weights at the mesh nodes. The linearisation point and the conditional posterior mode for the max and min mesh node weight have almost coincided after the second iteration. For the Mat\'ern covariance parameters $\rho_{R}$ and $\sigma_{R}$, the linearisation point and mode almost
coincide after the first iteration. The top right panel shows the difference between $\bm{u}_0^{(k)}$ and $\hat{\bm{u}}_0^{(k)}$, which is another way of examining the correspondence of the values shown in the top left panel. These plots indicate that the line search method has arrived at $\alpha\approx 1$, since there is a close correspondence between the linearisation point and the conditional posterior mode after a few iterations. The bottom left panel shows another convergence criterion, the change in the values of both the linearisation point and the conditional posterior mode relative to the posterior standard deviation across iterations. Here, the set threshold of 0.10, which is shown as the horizontal solid line, is clearly satisfied. Finally, the bottom right panel shows similar insights from the bottom left panel, but this time using the absolute change in both $\bm{u}_0^{(k)}$ and $\hat{\bm{u}}_0^{(k)}$ across iterations.

\begin{figure}[t]
    \centering

    \begin{minipage}[b]{0.24\linewidth}
        \centering
        \includegraphics[width=\linewidth]{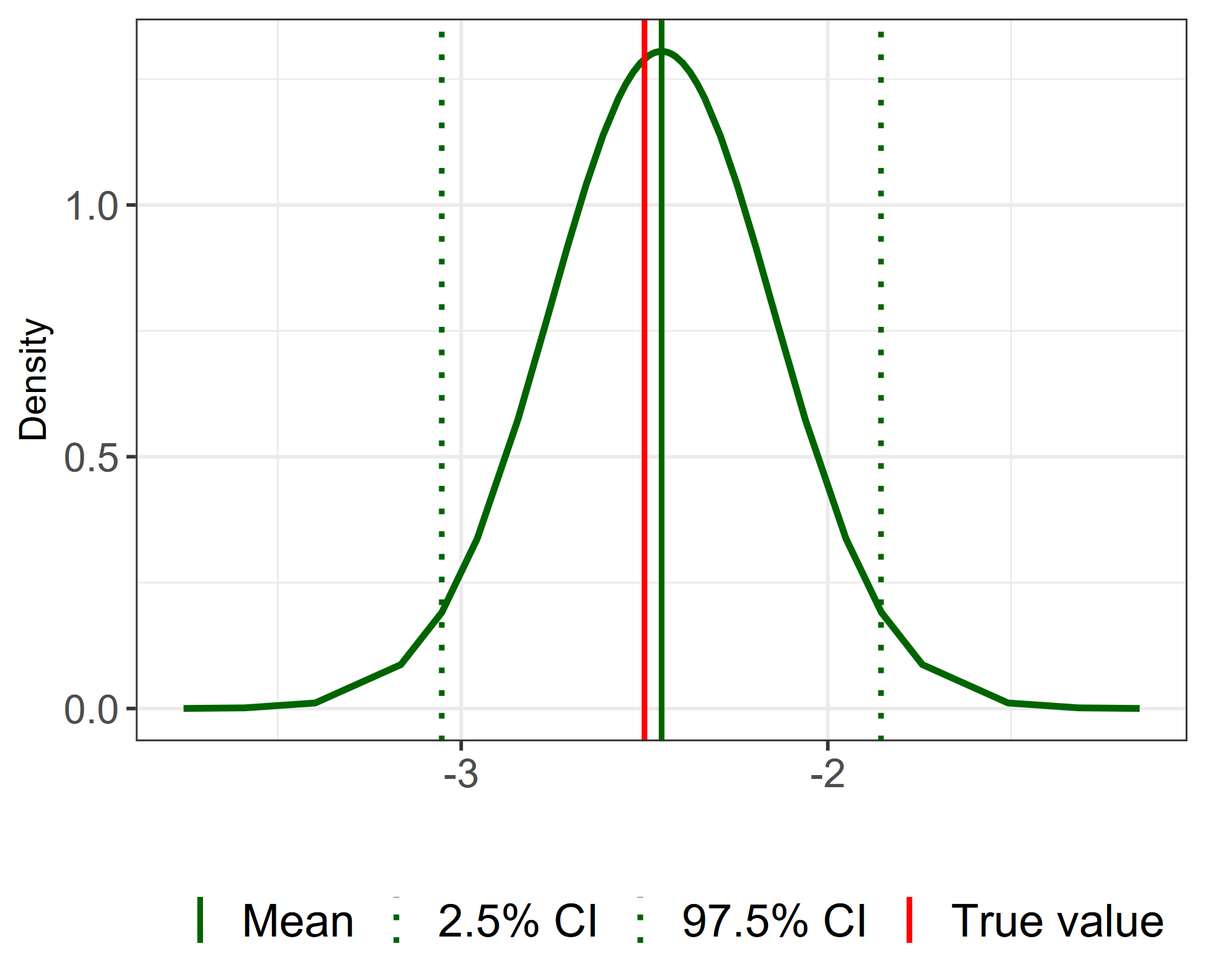}
        \vspace{0.5ex}
        \textbf{(a)}\; Estimated $\beta_0$
        \label{fig:illust_poisson_results_a}
    \end{minipage}
    \hfill
    \begin{minipage}[b]{0.24\linewidth}
        \centering
        \includegraphics[width=\linewidth]{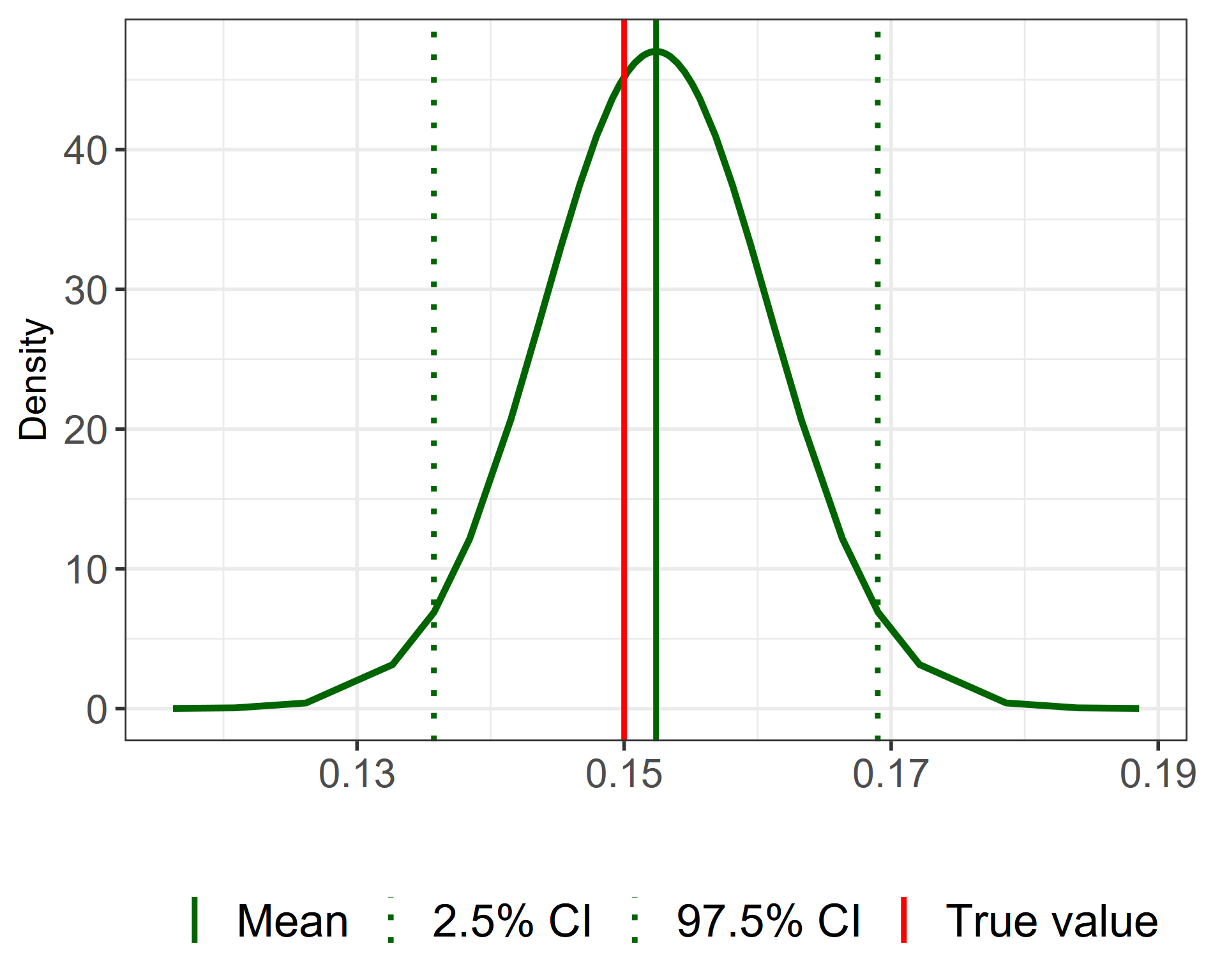}
        \vspace{0.5ex}
        \textbf{(b)}\; Estimated $\beta_1$
        \label{fig:illust_poisson_results_b}
    \end{minipage}
    \hfill
    \begin{minipage}[b]{0.24\linewidth}
        \centering
        \includegraphics[width=\linewidth]{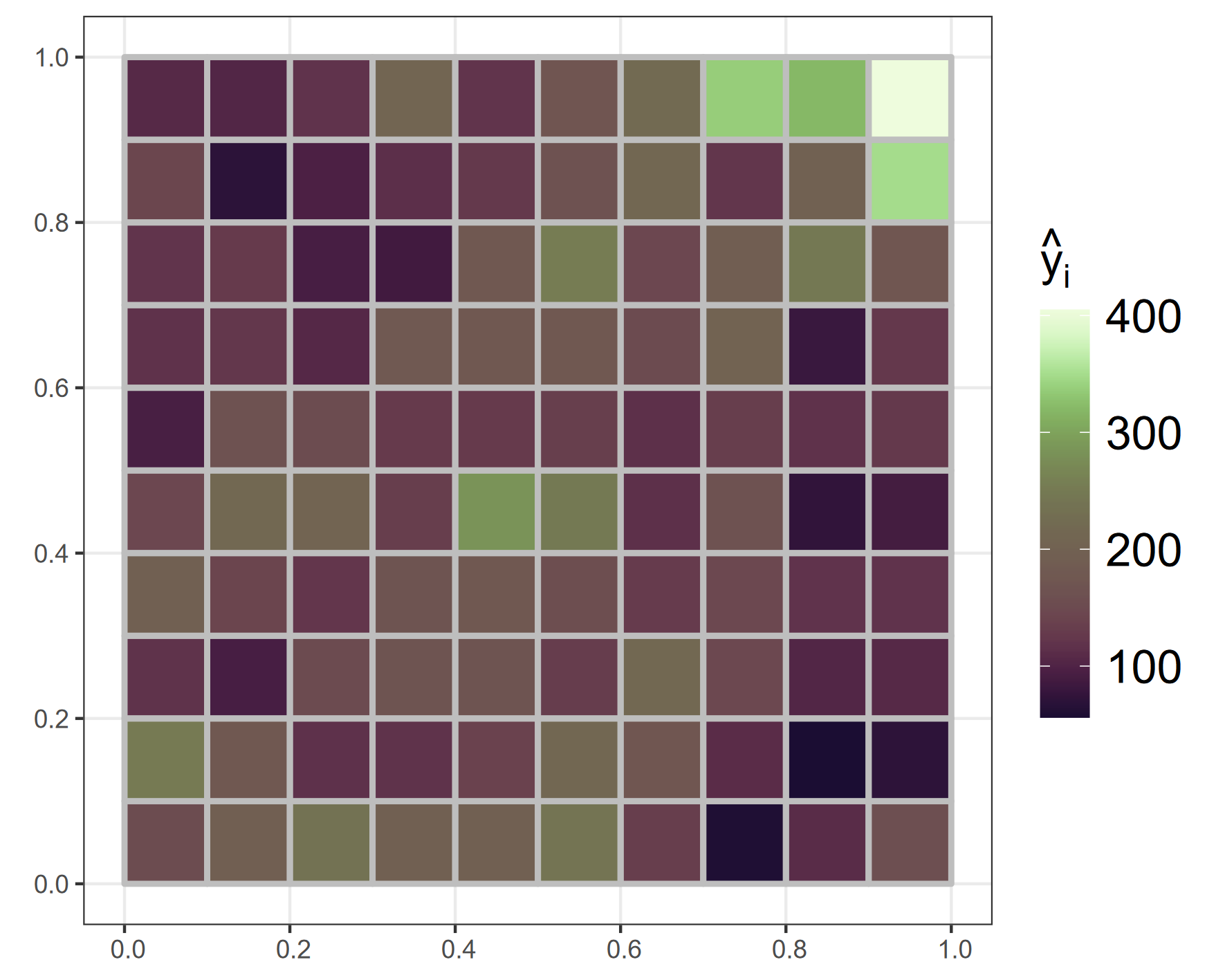}
        \vspace{0.5ex}
        \textbf{(c)}\; Predicted $Y_i$
        \label{fig:illust_poisson_results_c}
    \end{minipage}
    \hfill
    \begin{minipage}[b]{0.24\linewidth}
        \centering
        \includegraphics[width=\linewidth]{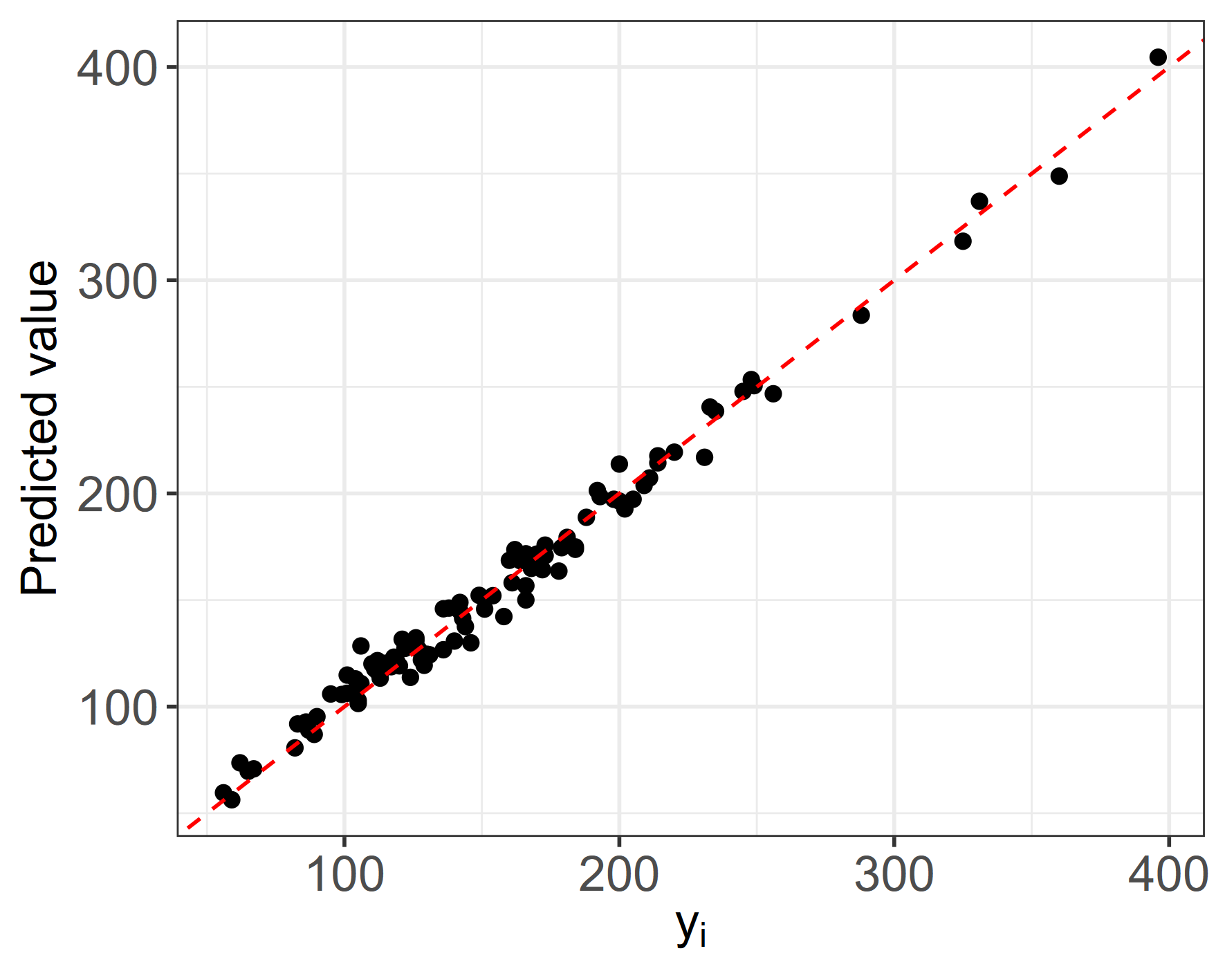}
        \vspace{0.5ex}
        \textbf{(d)}\;$Y_i$ vs $\hat{Y}_i$
        \label{fig:illust_poisson_results_d}
    \end{minipage}

    \caption{Illustration of Poisson model results from the simulated data in Figure~\ref{fig:illust_poisson_data}.}
    \label{fig:illust_poisson_results}
\end{figure}

\begin{figure}[t]
    \centering

    \begin{minipage}[b]{0.28\linewidth}
        \centering
        \includegraphics[width=\linewidth]{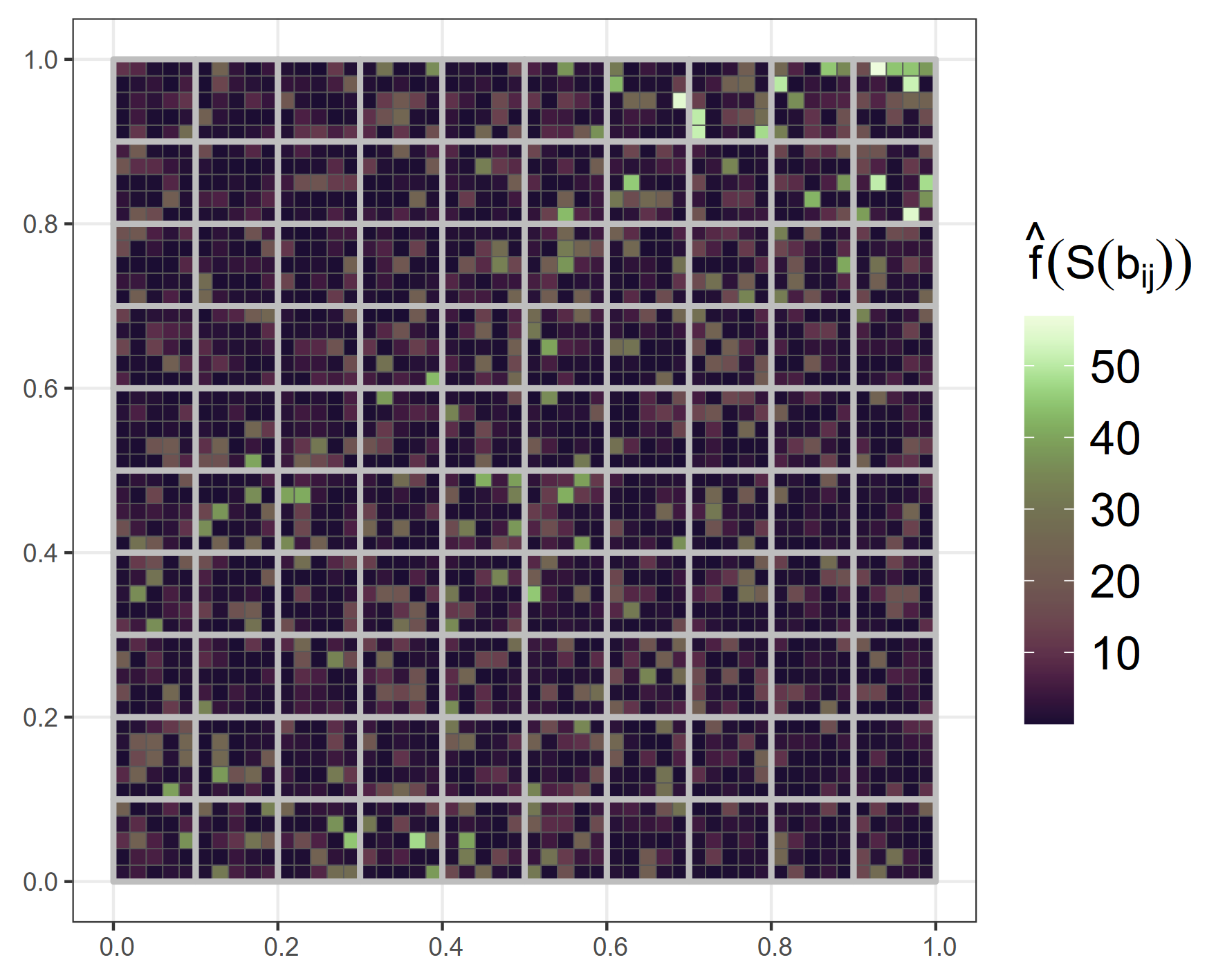}
        \vspace{0.5ex}
        \textbf{(a)}\; Estimated $f(S(b_{ij}))$
        \label{fig:illust_poisson_results_2_a}
    \end{minipage}
    \hfill
    \begin{minipage}[b]{0.28\linewidth}
        \centering
        \includegraphics[width=\linewidth]{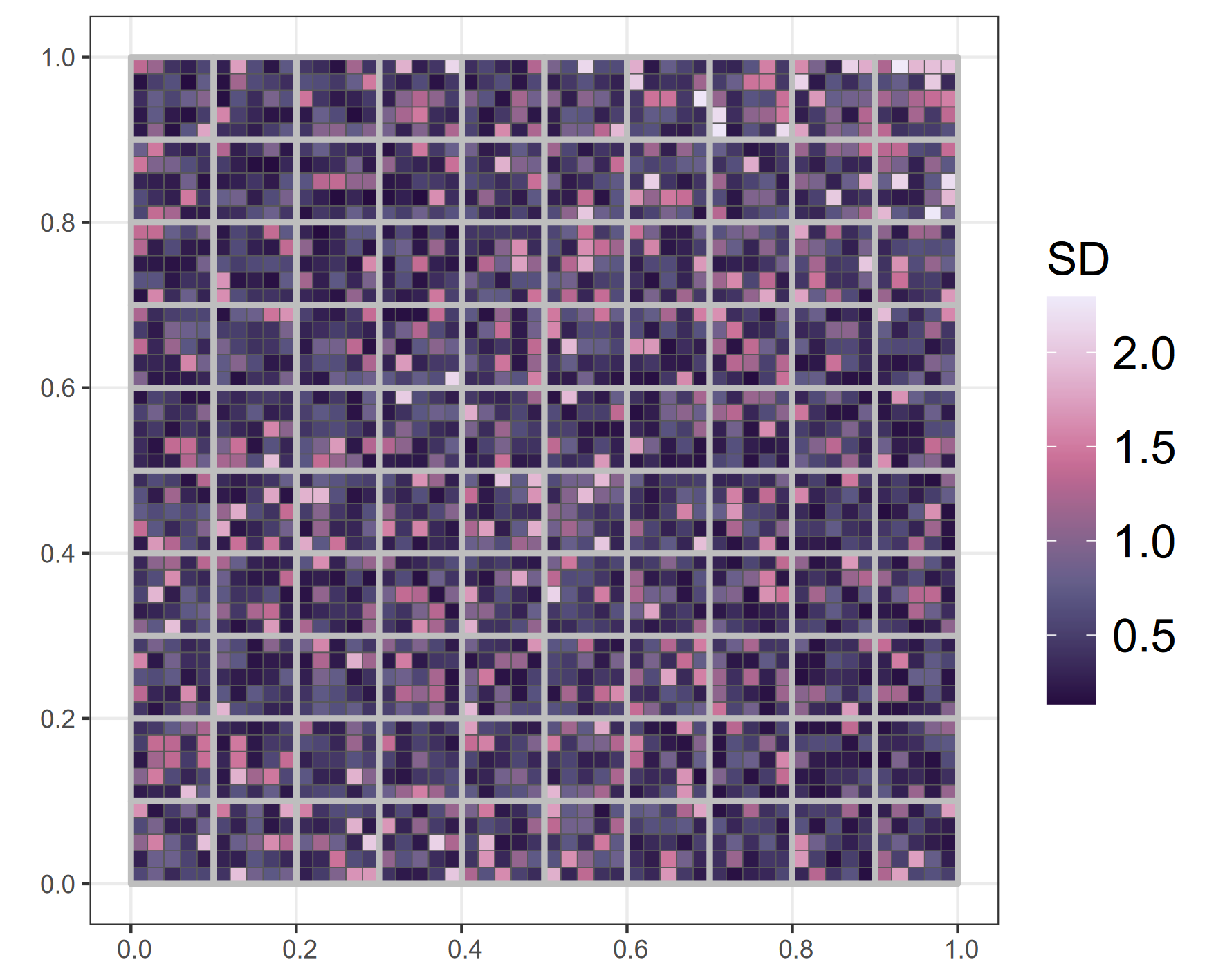}
        \vspace{0.5ex}
        \textbf{(b)}\; Posterior SD of $\hat{f}()$
        \label{fig:illust_poisson_results_2_b}
    \end{minipage}
    \hfill
    \begin{minipage}[b]{0.28\linewidth}
        \centering
        \includegraphics[width=\linewidth]{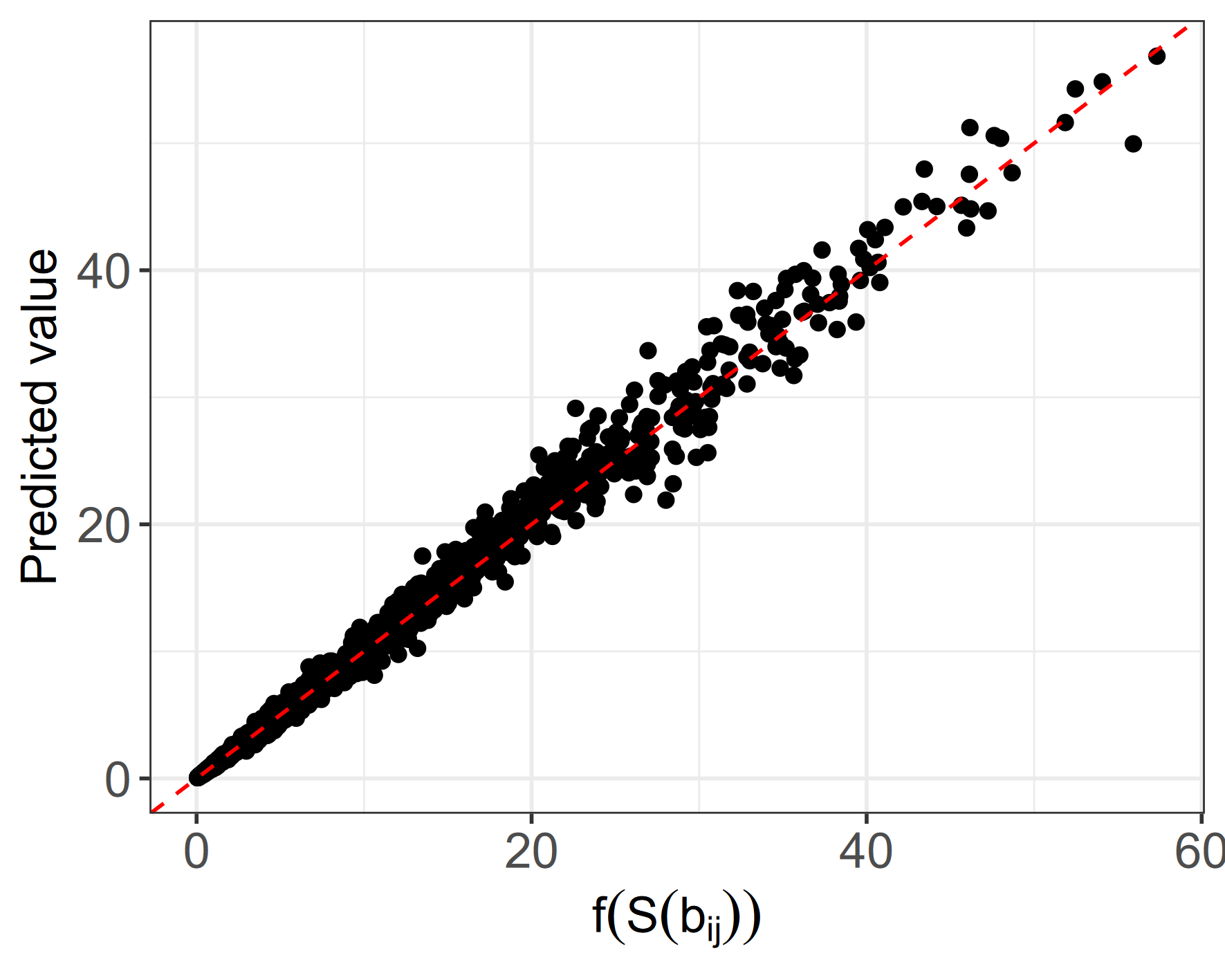}
        \vspace{0.5ex}
        \textbf{(c)}\; $f(S(b_{ij}))$ vs $\hat{f}()$
        \label{fig:illust_poisson_results_2_c}
    \end{minipage}

    \caption{Illustration of Poisson model results from the simulated data in Figure~\ref{fig:illust_poisson_data}.}
    \label{fig:illust_poisson_results_2}
\end{figure}

Figures \ref{fig:illust_poisson_results}a and \ref{fig:illust_poisson_results}b show the estimated marginal posteriors of $\beta_0$ and $\beta_1$. The posterior means are very close to the true values. Moreover, Figure \ref{fig:illust_poisson_results}c shows the predicted values of $Y_i$, while the scatterplot in Figure \ref{fig:illust_poisson_results}d shows a a close agreement between the predicted and observed values of the response, $Y_i$.  Figure \ref{fig:illust_poisson_results_2} shows results for the prediction of the latent values $f\Big(S(b_{ij})\Big)$. The predicted values are shown in Figure \ref{fig:illust_poisson_results_2}a, while the corresponding posterior SDs are in Figure \ref{fig:illust_poisson_results_2}b. Figure \ref{fig:illust_poisson_results_2}c shows a scatterplot between the predicted and true values of $f\Big(S(b_{ij})\Big)$. The aforementioned results show that the proposed model is able to recover correctly the latent process, and is able to perform spatial disaggregation. 

\clearpage

\section{Simulation results}

\subsection{Gaussian case}

\begin{figure}[htbp]
    \centering

    \begin{minipage}[b]{0.42\textwidth}
        \centering
        \includegraphics[width=\textwidth]{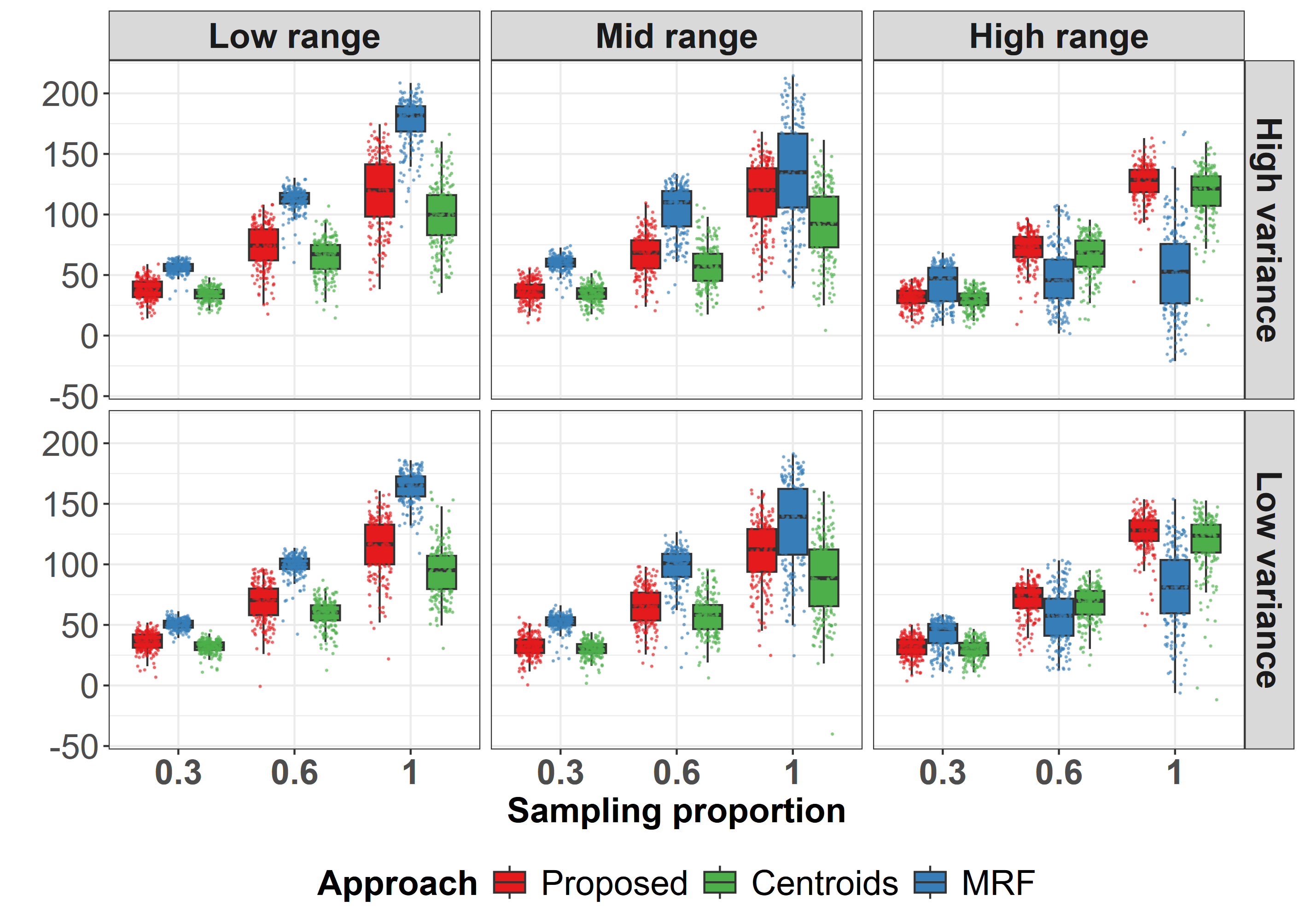}
        \vspace{0.6ex}
        \textbf{(a)}\; Negative log score
    \end{minipage}
    \hfill
    \begin{minipage}[b]{0.42\textwidth}
        \centering
        \includegraphics[width=\textwidth]{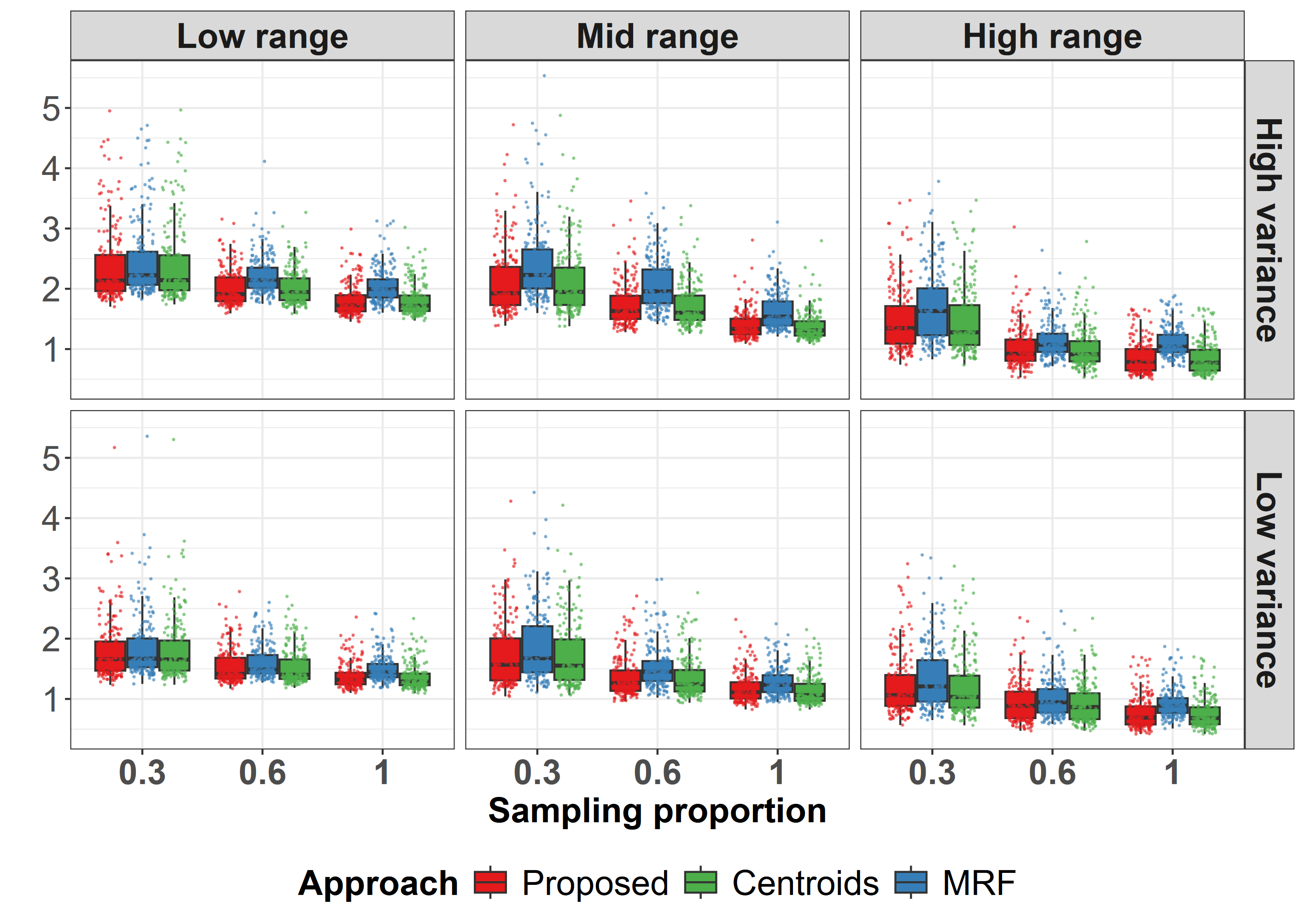}
        \vspace{0.6ex}
        \textbf{(b)}\; RMSE of $\hat{\mu}_{ij}$
    \end{minipage}

    \caption{Plot of negative log score and RMSE of $\hat{\mu}_{ij}$}
\end{figure}

\begin{figure}[htbp]
    \centering

    \begin{minipage}[b]{0.42\linewidth}
        \centering
        \includegraphics[width=\linewidth]{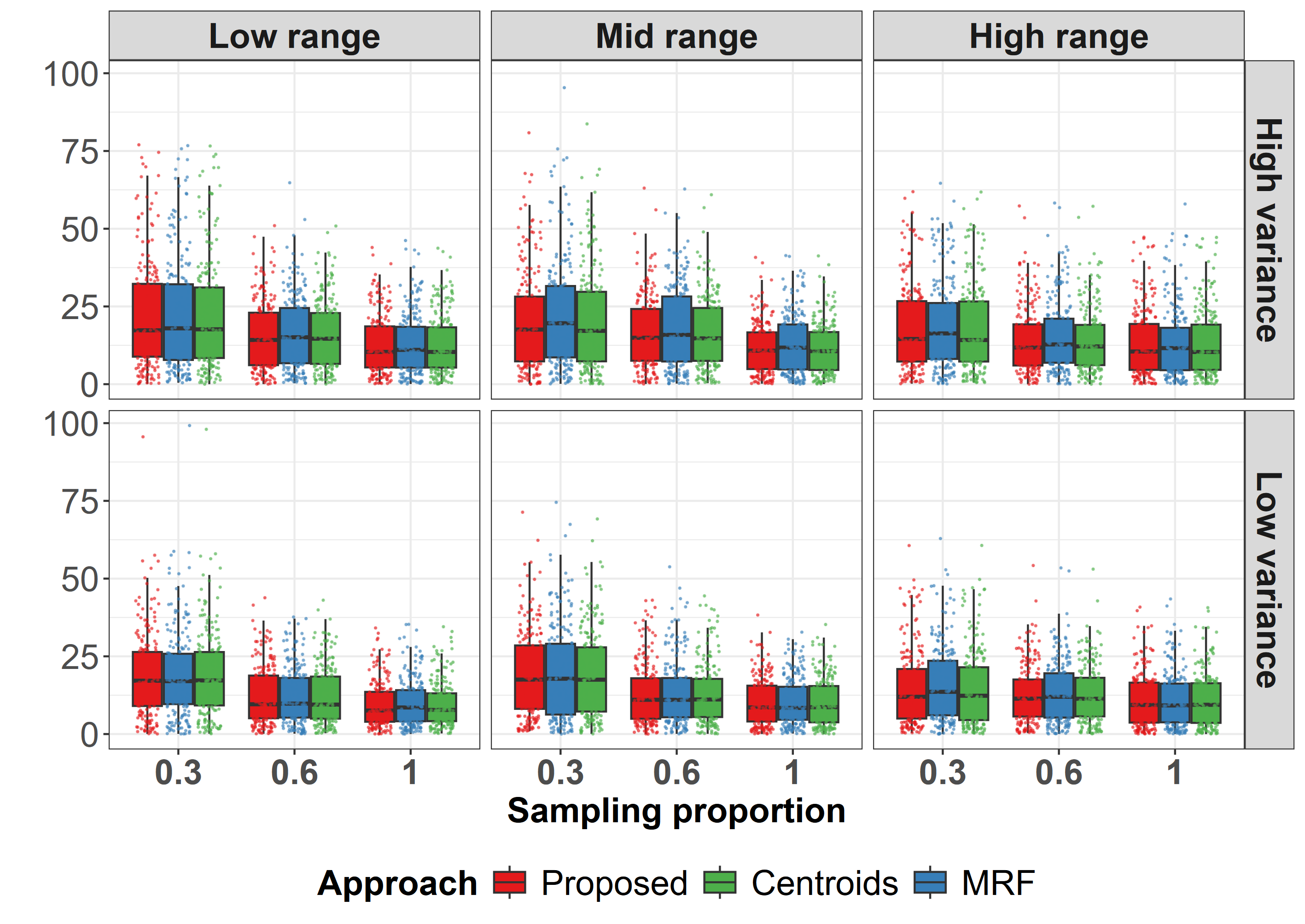}
        \vspace{0.5ex}
        \textbf{(a)}\; $\beta_0$
        \label{fig:simres_Gaussian_scores_relbias_a}
    \end{minipage}
    \hfill
    \begin{minipage}[b]{0.42\linewidth}
        \centering
        \includegraphics[width=\linewidth]{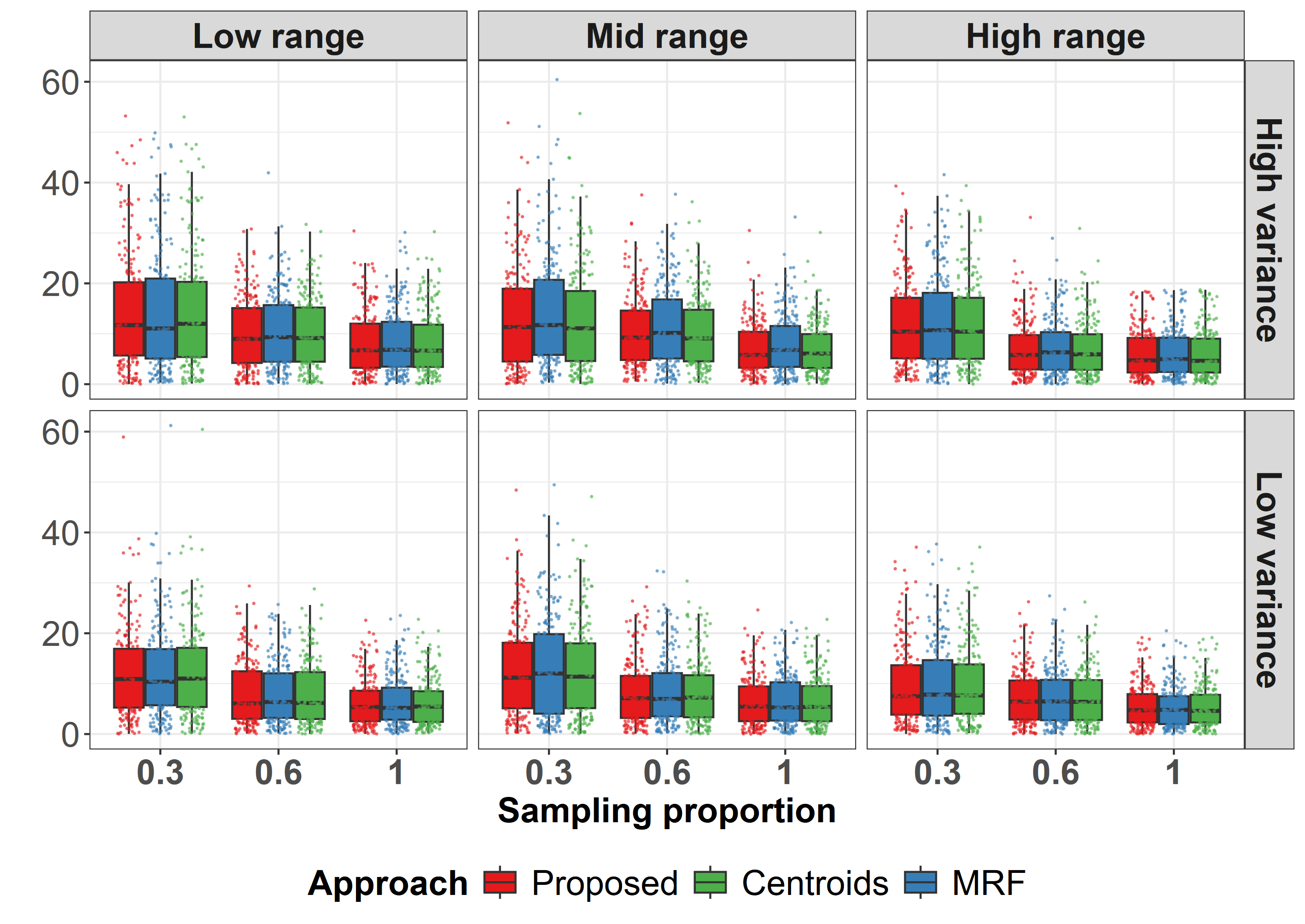}
        \vspace{0.5ex}
        \textbf{(b)}\; $\beta_1$
        \label{fig:simres_Gaussian_scores_relbias_b}
    \end{minipage}

    \caption{Plot of relative bias (in \%) for the fixed effects $\beta_0$ and $\beta_1$.}
    \label{fig:simres_Gaussian_scores_relbias}
\end{figure}

\begin{figure}[htbp]
    \centering

    \begin{minipage}[t]{0.42\linewidth}
        \centering
        \includegraphics[width=\linewidth]{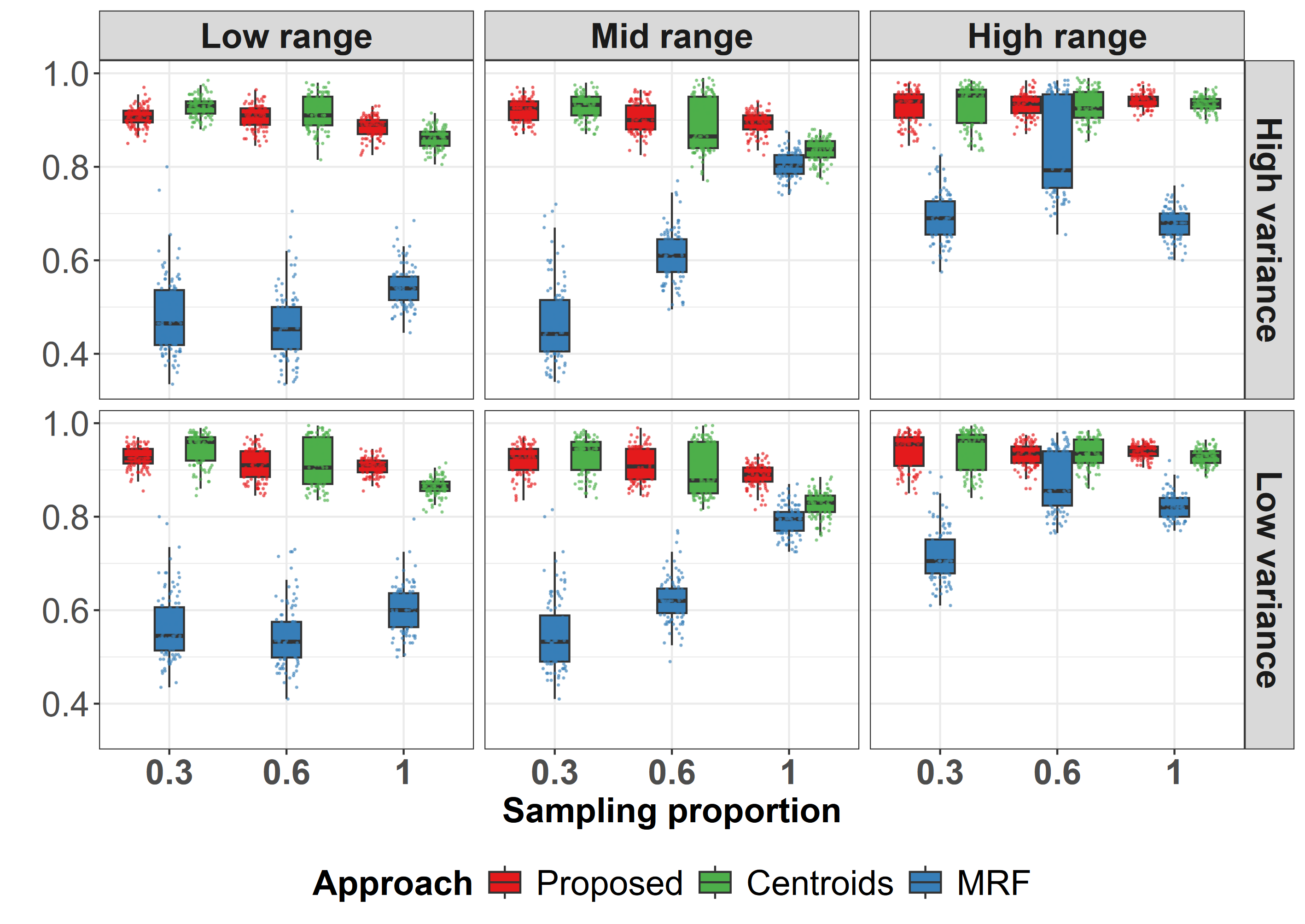}
        \vspace{0.6ex}
        \textbf{(a)}\; Coverage for $\mu_i$
        \label{fig:coverage_mu_large}
    \end{minipage}
    \hfill
    \begin{minipage}[t]{0.42\linewidth}
        \centering
        \includegraphics[width=\linewidth]{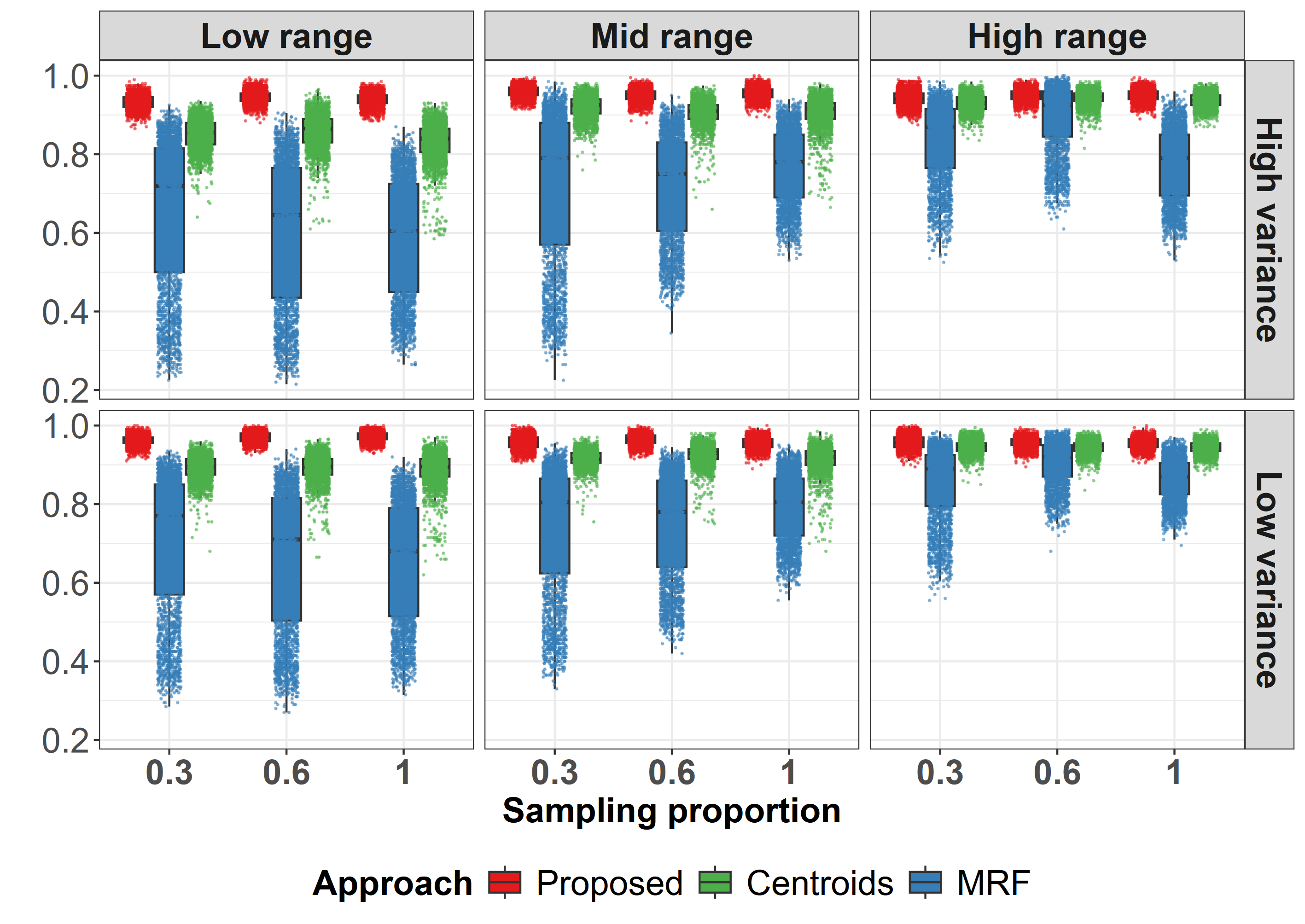}
        \vspace{0.6ex}
        \textbf{(b)}\; Coverage for $\mu_{ij}$
        \label{fig:coverage_mu_small}
    \end{minipage}

    \caption{Plots of the coverage for $\mu_i$ and $\mu_{ij}$. 
    Each point in the boxplots corresponds to  a block, $B_i$ (left-hand panel)
    or a cell, $b_{ij}$ (right-hand panel)}
    \label{fig:combined_coverage_mu}
\end{figure}

\clearpage

\subsection{Poisson case}

\begin{figure}[htbp]
    \centering

    \begin{minipage}[b]{0.41\linewidth}
        \centering
        \includegraphics[width=\linewidth]{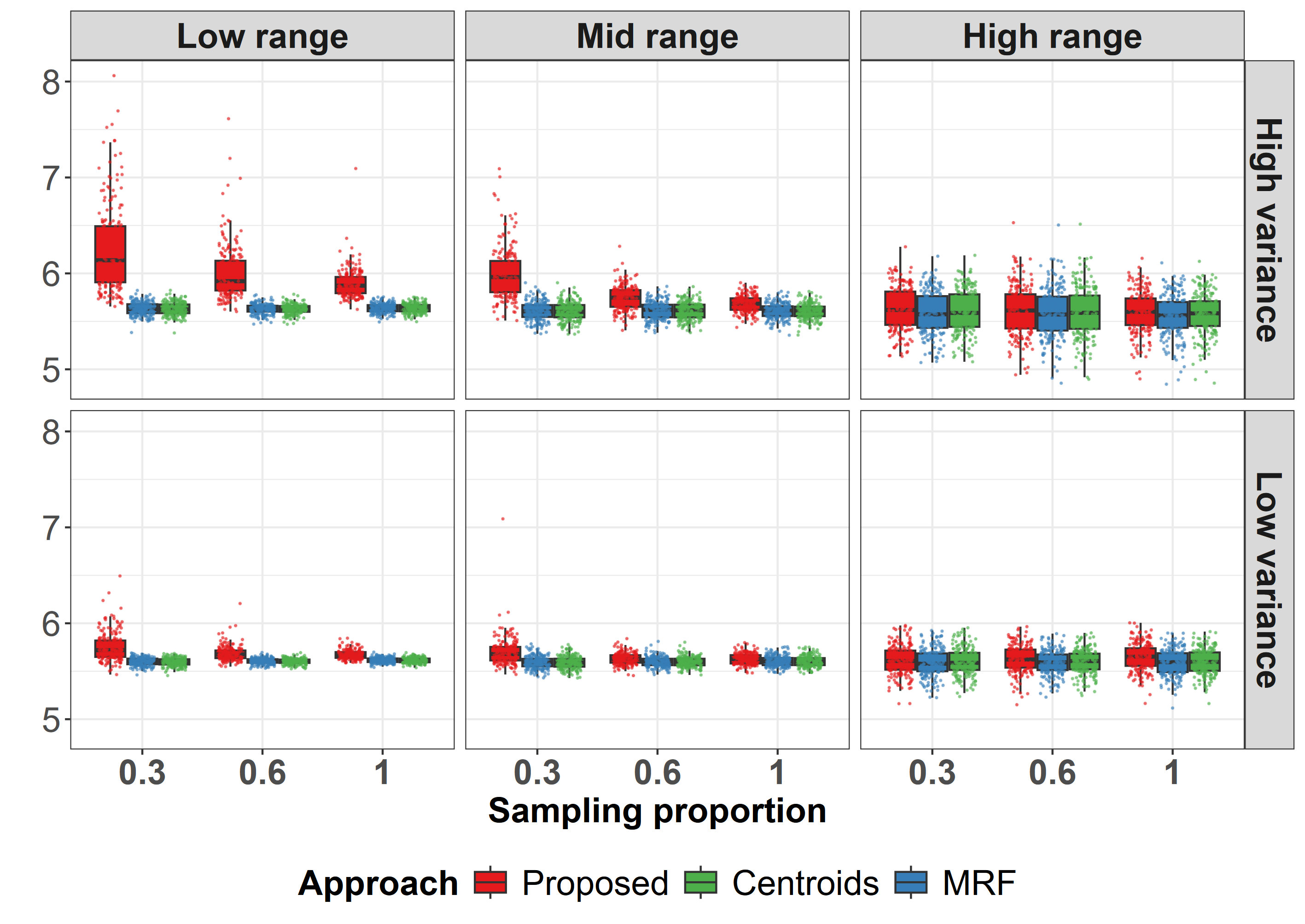}
        \vspace{0.6ex}
        \textbf{(a)}\; Dawid Sebastiani score
    \end{minipage}
    \hfill
    \begin{minipage}[b]{0.41\linewidth}
        \centering
        \includegraphics[width=\linewidth]{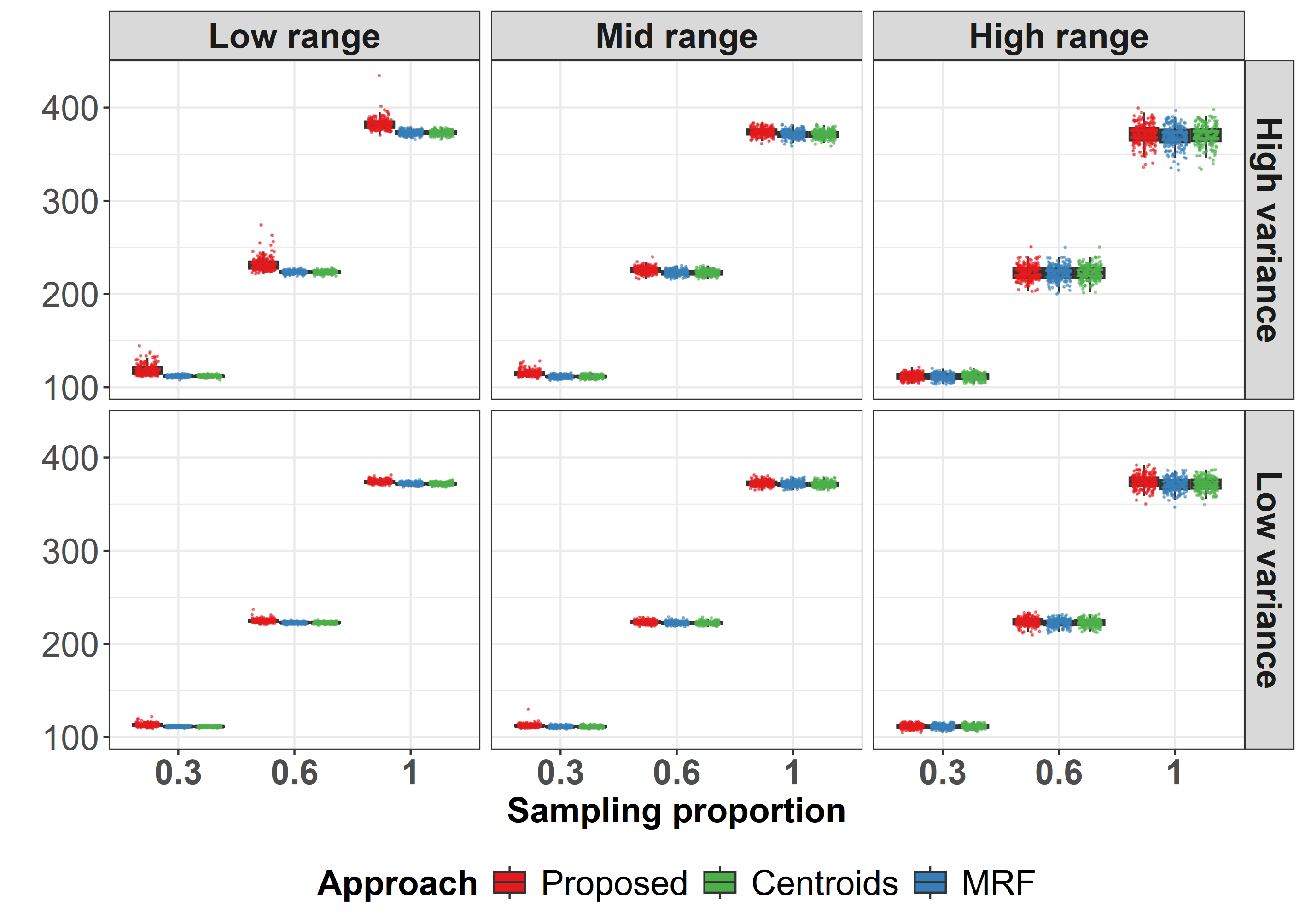}
        \vspace{0.6ex}
        \textbf{(b)}\; Negative log score
    \end{minipage}

    \caption{Comparison of scores for the Poisson case.}
\end{figure}

\begin{figure}[htbp]
    \centering

    \begin{minipage}[b]{0.41\linewidth}
        \centering
        \includegraphics[width=\linewidth]{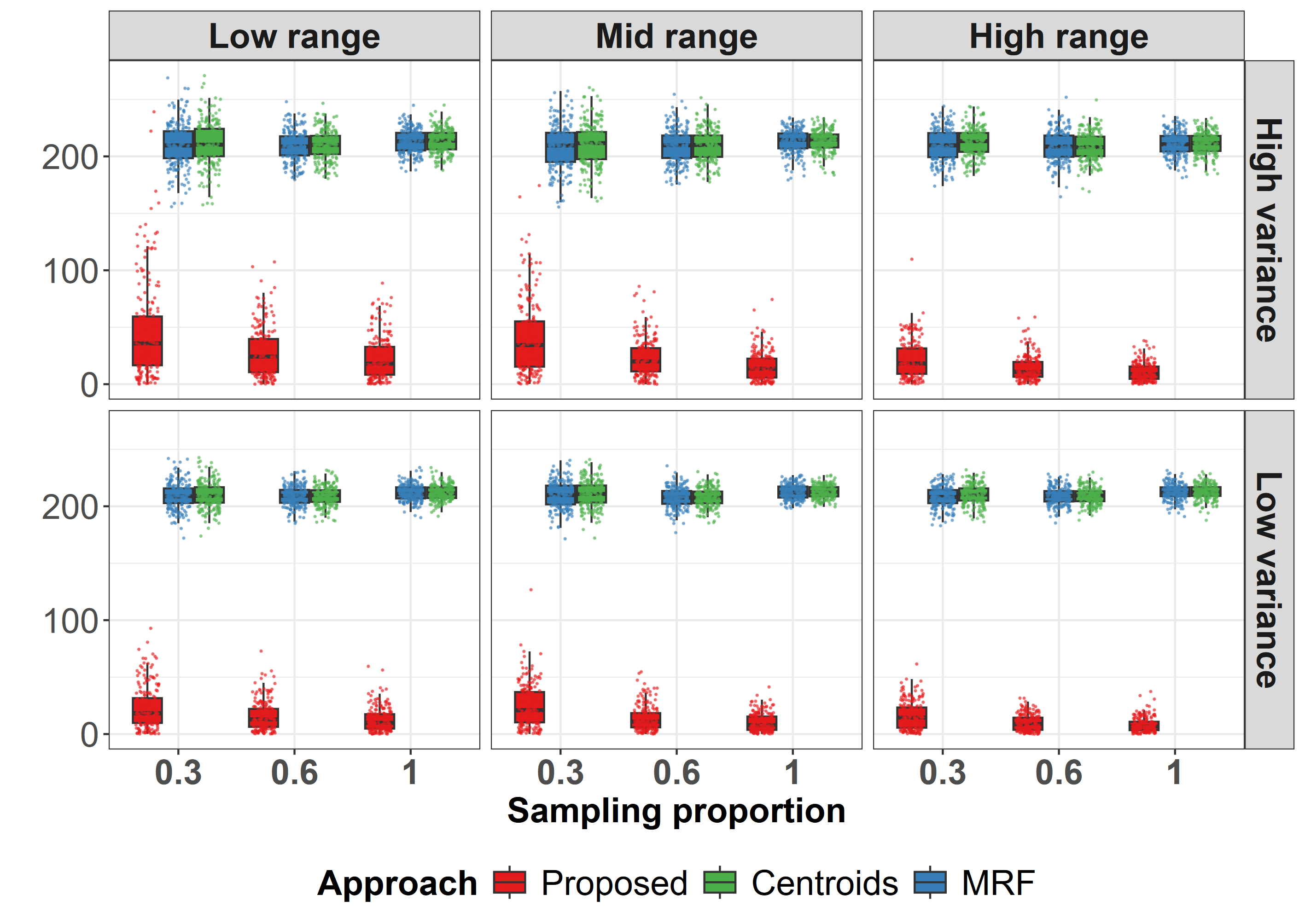}
        \vspace{0.6ex}
        \textbf{(a)}\; $\beta_0$
        \label{fig:simres_Poisson_scores_relbias_a}
    \end{minipage}
    \hfill
    \begin{minipage}[b]{0.41\linewidth}
        \centering
        \includegraphics[width=\linewidth]{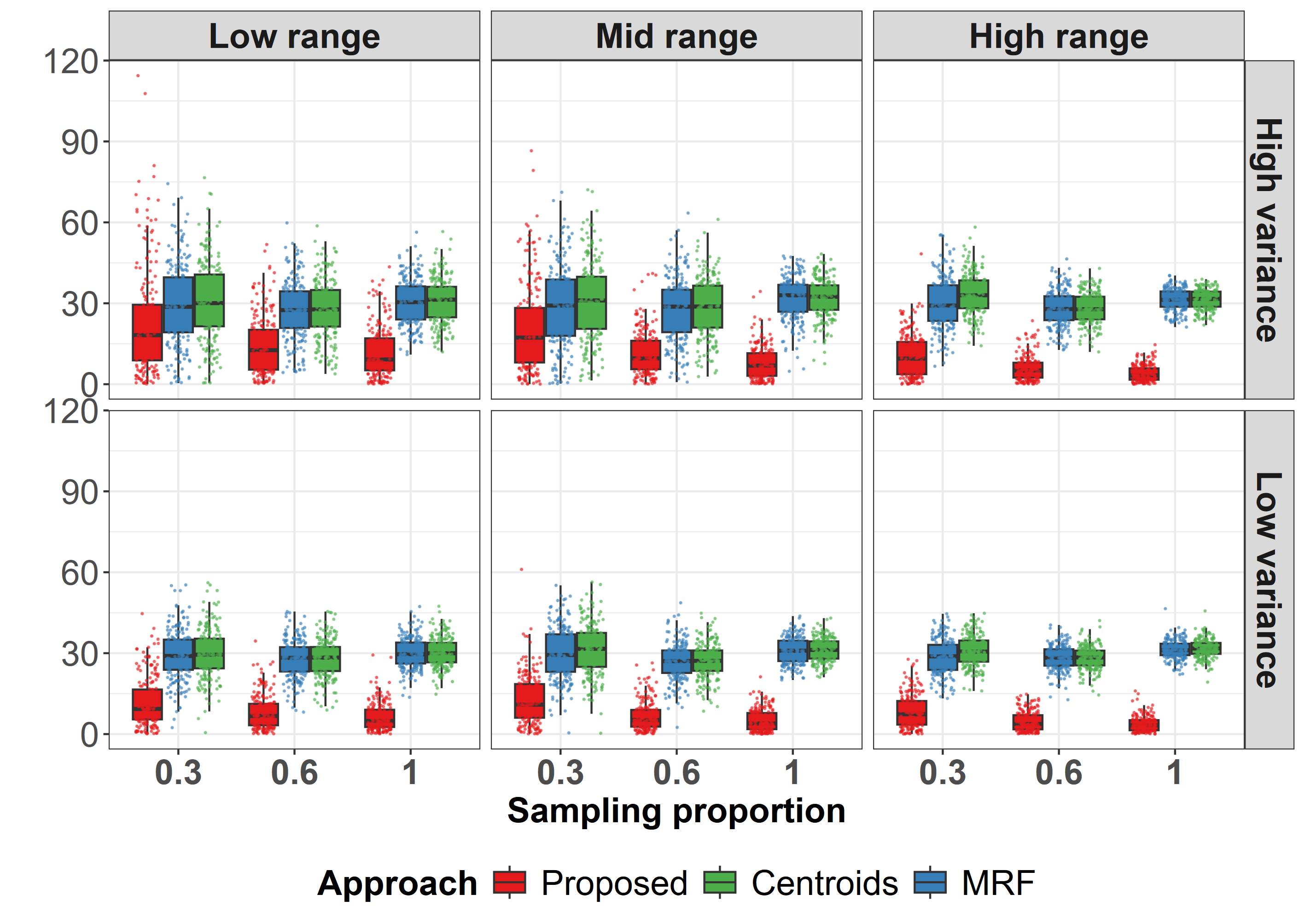}
        \vspace{0.6ex}
        \textbf{(b)}\; $\beta_1$
        \label{fig:simres_Poisson_scores_relbias_b}
    \end{minipage}

    \caption{Plot of relative bias (in \%) for the fixed effects $\beta_0$ and $\beta_1$.}
    \label{fig:simres_Poisson_scores_relbias}
\end{figure}

\begin{figure}[htbp]
    \centering

    \begin{minipage}[b]{0.41\linewidth}
        \centering
        \includegraphics[width=\linewidth]{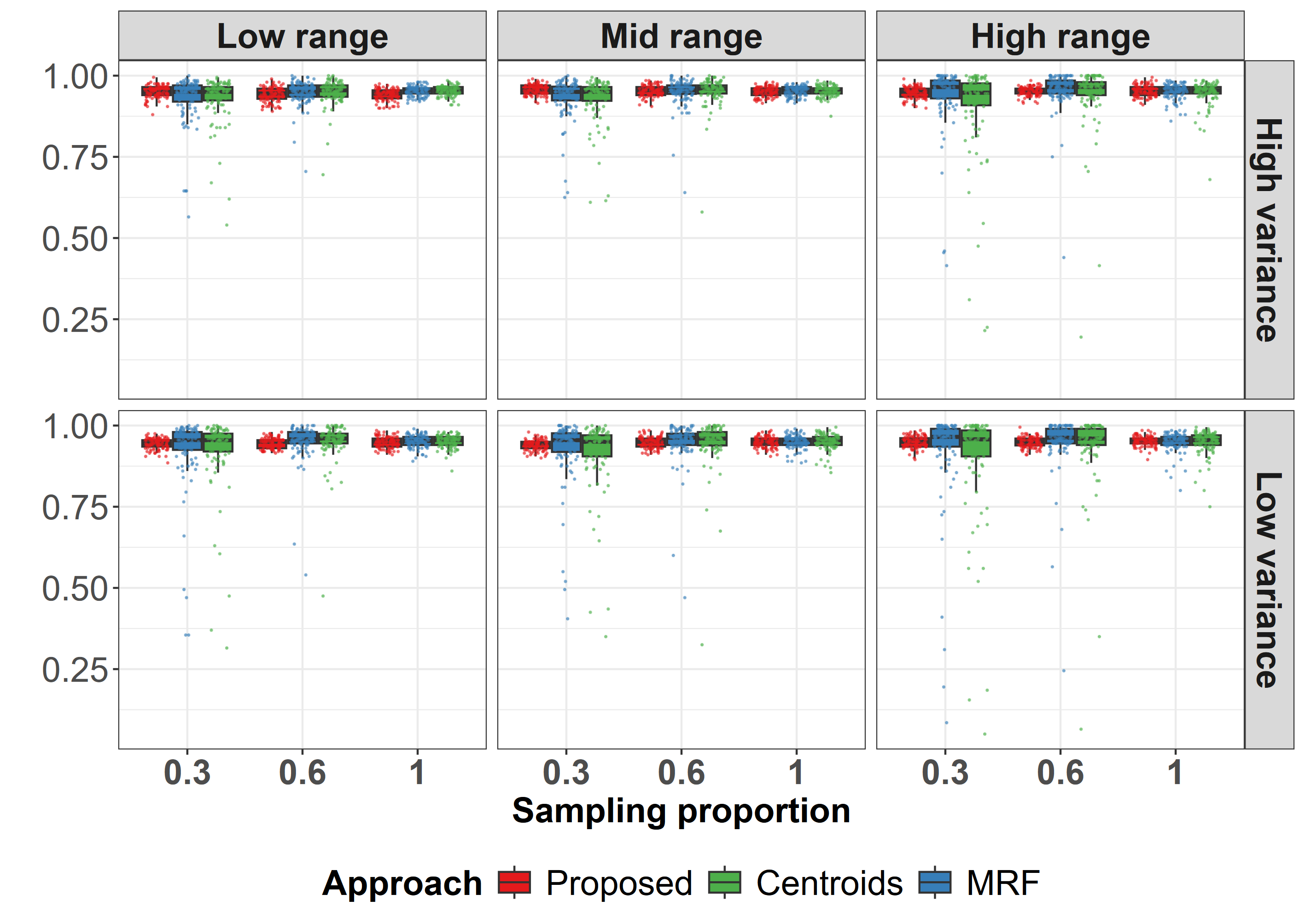}
        \vspace{0.6ex}
        \textbf{(a)}\; Coverage for $\mu_i$
        \label{fig:mu_coverage}
    \end{minipage}
    \hfill
    \begin{minipage}[b]{0.41\linewidth}
        \centering
        \includegraphics[width=\linewidth]{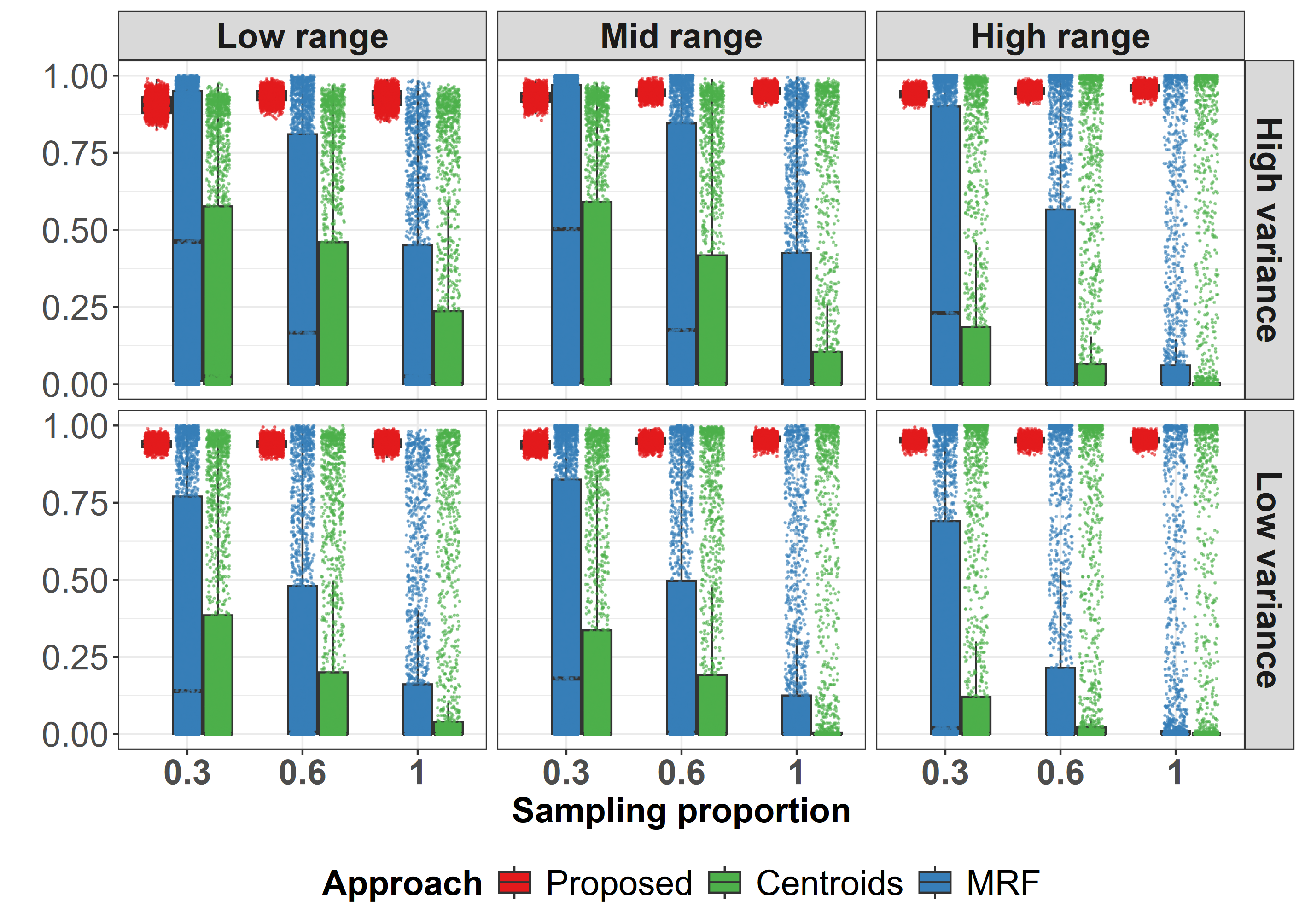}
        \vspace{0.6ex}
        \textbf{(b)}\; Coverage for $\mu_{ij}$
        \label{fig:mu_coverage_small}
    \end{minipage}

    \caption{Plots of the coverage for $\mu_i$ and $\mu_{ij}$.  
    Each point in the boxplots corresponds to  a block, $B_i$ (left-hand panel)
    or a cell, $b_{ij}$ (right-hand panel)}
\end{figure}

\clearpage

\section{Application}

\subsection{Virus concentrations in community wastewater}

\begin{figure}[htbp]
    \centering
    \includegraphics[width=0.65\linewidth]{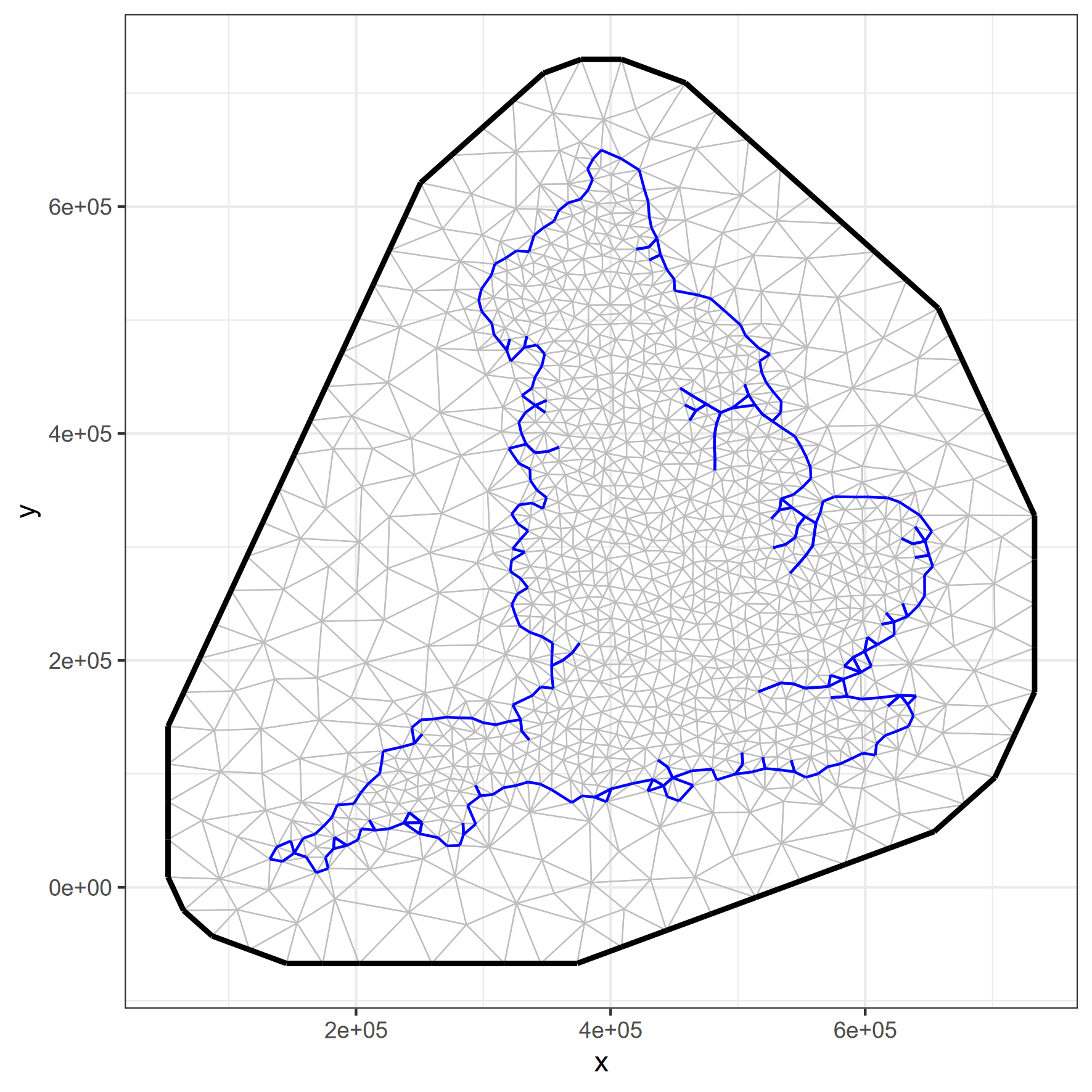}
    \caption{England mesh for model fitting}
\end{figure}

\begin{table}[htbp]
\centering
\begin{tabular}{|l | l r r r r|}
  \hline
  \textbf{Approach} & \textbf{Parameter} & \textbf{Mean} & \textbf{SD} & \textbf{P$2.5^{\text{th}}$} & \textbf{P$97.5^{\text{th}}$} \\
  \hline
  
  \multirow{3}{*}{Centroids} 
    & $1/\sigma^2_e$ & 0.555 & 0.066 & 0.434 & 0.694 \\
    & $\rho_R$ (km)   & 80.597 & 44.918 & 25.816 & 196.473 \\
    & $\sigma_R$ & 0.594 & 0.163 & 0.325 & 0.959 \\
  \hline
  
  \multirow{3}{*}{MRF}
    & $1/\sigma^2_e$ & 0.477 & 0.051 & 0.378 & 0.578 \\
    & $\tau$         & 8076.945 & 169331.008 & 2.420 & 29825.468 \\
    & $\phi$         & 0.444 & 0.275 & 0.035 & 0.944 \\
  \hline
  
  \multirow{3}{*}{Proposed}
    & $1/\sigma^2_e$ & 0.551 & 0.064 & 0.434 & 0.685 \\
    & $\rho_R$ (km)   & 82.046 & 45.649 & 26.427 & 199.874 \\
    & $\sigma_R$ & 0.591 & 0.164 & 0.320 & 0.959 \\
  \hline

\end{tabular}
\caption{Posterior estimates of Mat\'ern covariance parameters for the centroids,  MRF, and block aggregation approaches}
\end{table}

\begin{table}[htbp]
\centering
\begin{tabular}{|l|rrr|}
  \hline
  \textbf{Approach} & \textbf{MDS} & \textbf{ TNLS}& \textbf{RMSE} \\ 
  \hline
 Centroids & 1.804 & 1.821 & 1.487 \\ 
 MRF & 1.769 & 1.803 & 1.463 \\ 
  Block aggregation & 1.745 & 1.792 & 1.443 \\ 
   \hline
\end{tabular}
\caption{Scores from doing leave-one-area-out cross validation for the Gaussian application: Mean DS (MDS) score, total negative log score (TNLS), and RMSE of the predicted $Y_i$}
\label{tab:app_gaussian_scores}
\end{table}

\begin{figure}[H]
    \centering
    \includegraphics[width=\linewidth, trim={0 3.25cm 0 3.25cm}, clip]{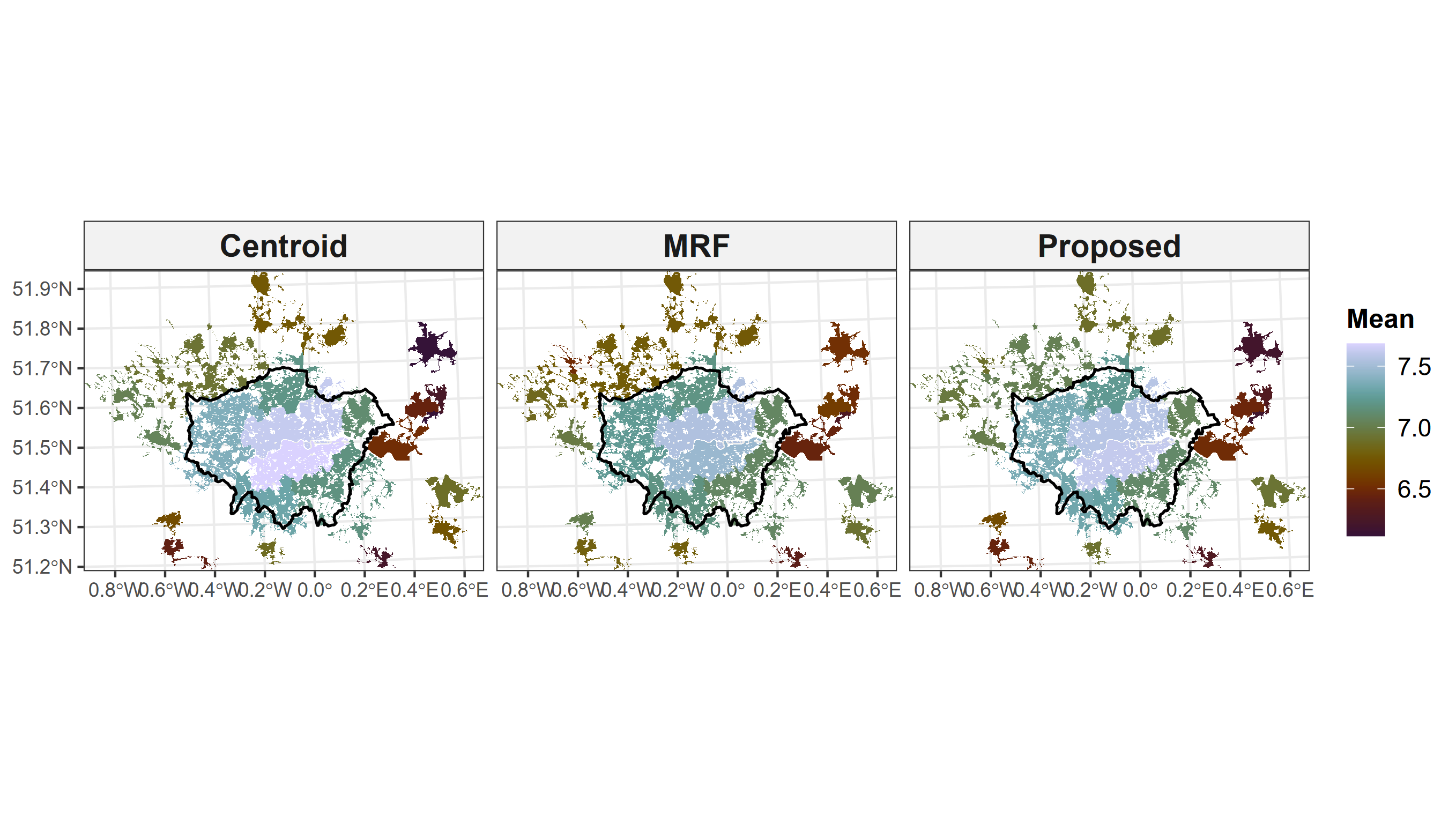}
    \caption{Comparison of predicted (posterior mean) wastewater virus concentration at the catchment areas around the Greater London area. The predicted values of $Y_i$ are very similar for the three approaches. However, the posterior SD for the MRF are smaller. }
    \label{fig:app_gaussian_preds_catchment}
\end{figure}

\begin{figure}[H]
    \centering
    \includegraphics[width=\linewidth, trim={0 3.25cm 0 3.25cm}, clip]{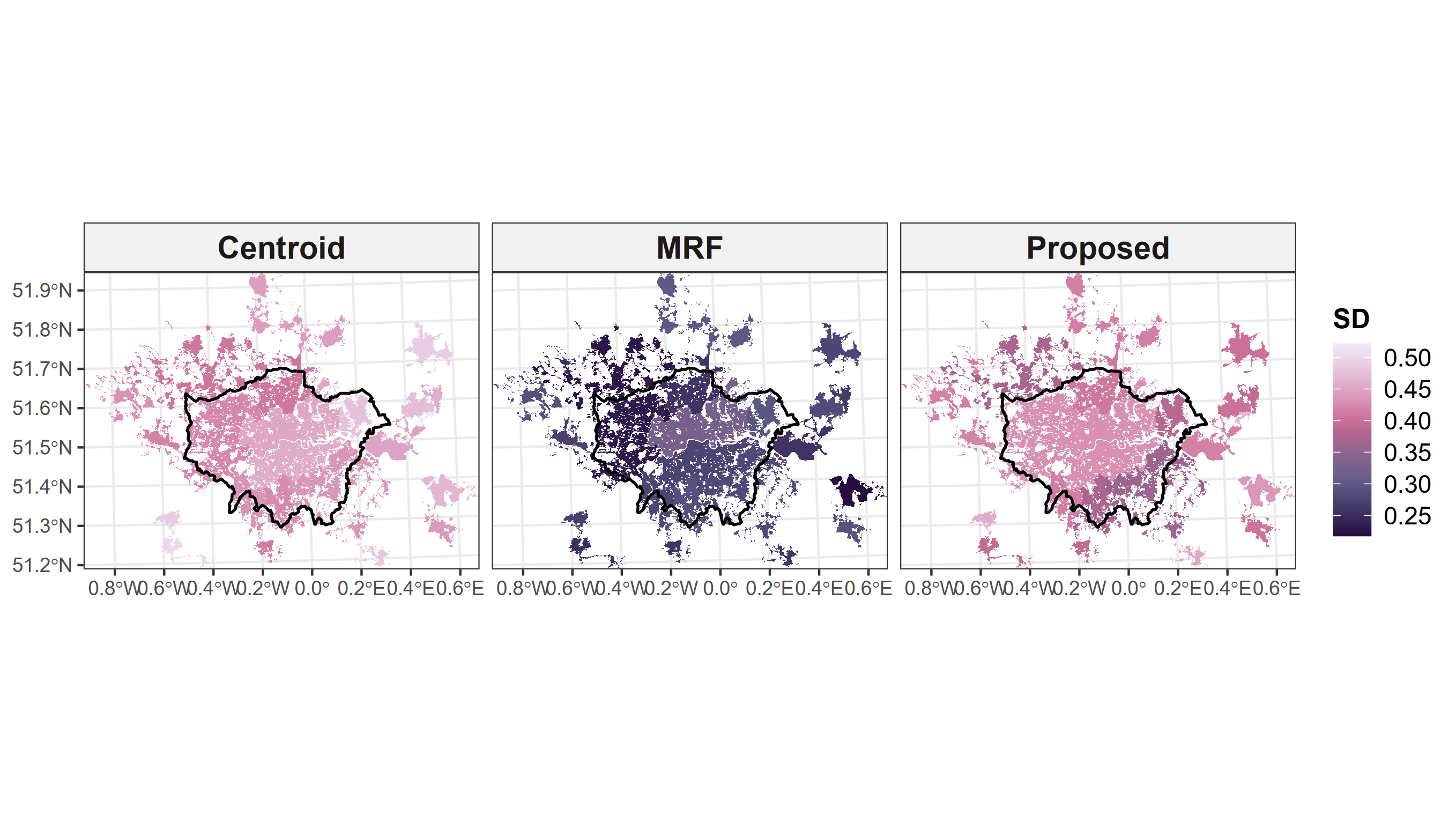}
    \caption{Comparison of posterior standard deviation of wastewater virus concentration at the catchment areas around the Greater London area. The posterior SDs for the MRF approach are smaller.}
    \label{fig:app_gaussian_preds_catchment_sd}
\end{figure}

\subsection{Cardiovascular-related hospitalisations}

\begin{table}[htbp]
\centering
\begin{tabular}{|l|lrrrr|}
  \hline
 \textbf{Approach} & \textbf{Parameter} & \textbf{Mean} & \textbf{SD} & \textbf{P$2.5^{\text{th}}$} & \textbf{P$97.5^{\text{th}}$} \\
  \hline

  \multirow{2}{*}{Centroids}
    & $\rho_{R}$ (km)    & 27.441 & 3.513 & 21.115 & 34.923 \\
    & $\sigma_{R}$ & 0.124     & 0.006    & 0.113     & 0.137     \\
  \hline

  \multirow{2}{*}{MRF}
    & $\tau$ & 86.060 & 11.051 & 65.801 & 109.221 \\
    & $\phi$ & 0.805  & 0.088  & 0.600  & 0.939   \\
  \hline

  \multirow{2}{*}{Block aggregation}
    & $\rho_{R}$ (km)   & 18.574 & 2.616 & 13.828 & 24.103 \\
    & $\sigma_{R}$ & 0.169     & 0.011    & 0.148     & 0.192     \\
  \hline
\end{tabular}
\caption{Posterior estimates of Mat\'ern covariance
parameters for the centroids, MRF, and block aggregation approaches. }
\end{table}

\begin{table}[htbp]
\centering
\begin{tabular}{|l|rrr|}
  \hline
 \textbf{Approach} & \textbf{MDS} & \textbf{TNLS}& \textbf{RMSE} \\ 
 \hline
  Centroids & 12.182 & 7.042 & 202.901 \\ 
  MRF & 10.570 & 6.212 & 158.120 \\
Block aggregation & 11.952 & 6.943 & 190.959 \\  
   \hline
\end{tabular}
\caption{Scores from doing leave-one-area-out cross validation for the Poisson application: Mean DS (MDS) score, total negative log score (TNLS), and RMSE of the predicted $Y_i$}
\label{tab:app_poisson_scores}
\end{table}

\begin{figure}[H]
    \centering
    \includegraphics[width=0.8\linewidth]{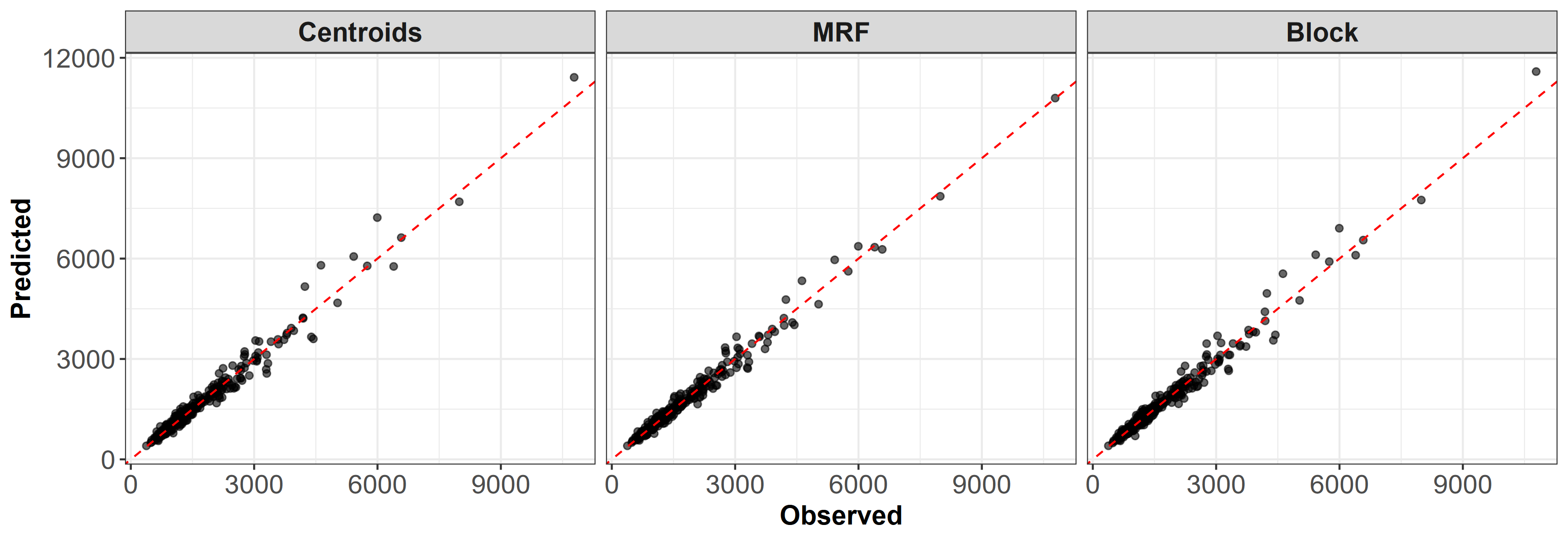}
    \caption{Scatterplot of the observed cases versus the predicted cases from the LOOCV for the centroids, MRF, and block aggregation approaches}
    \label{fig:app_Poisson_scatterplot}
\end{figure}

\begin{figure}[H]
    \centering

    \begin{minipage}[b]{0.48\linewidth}
        \centering
        \includegraphics[width=\linewidth]{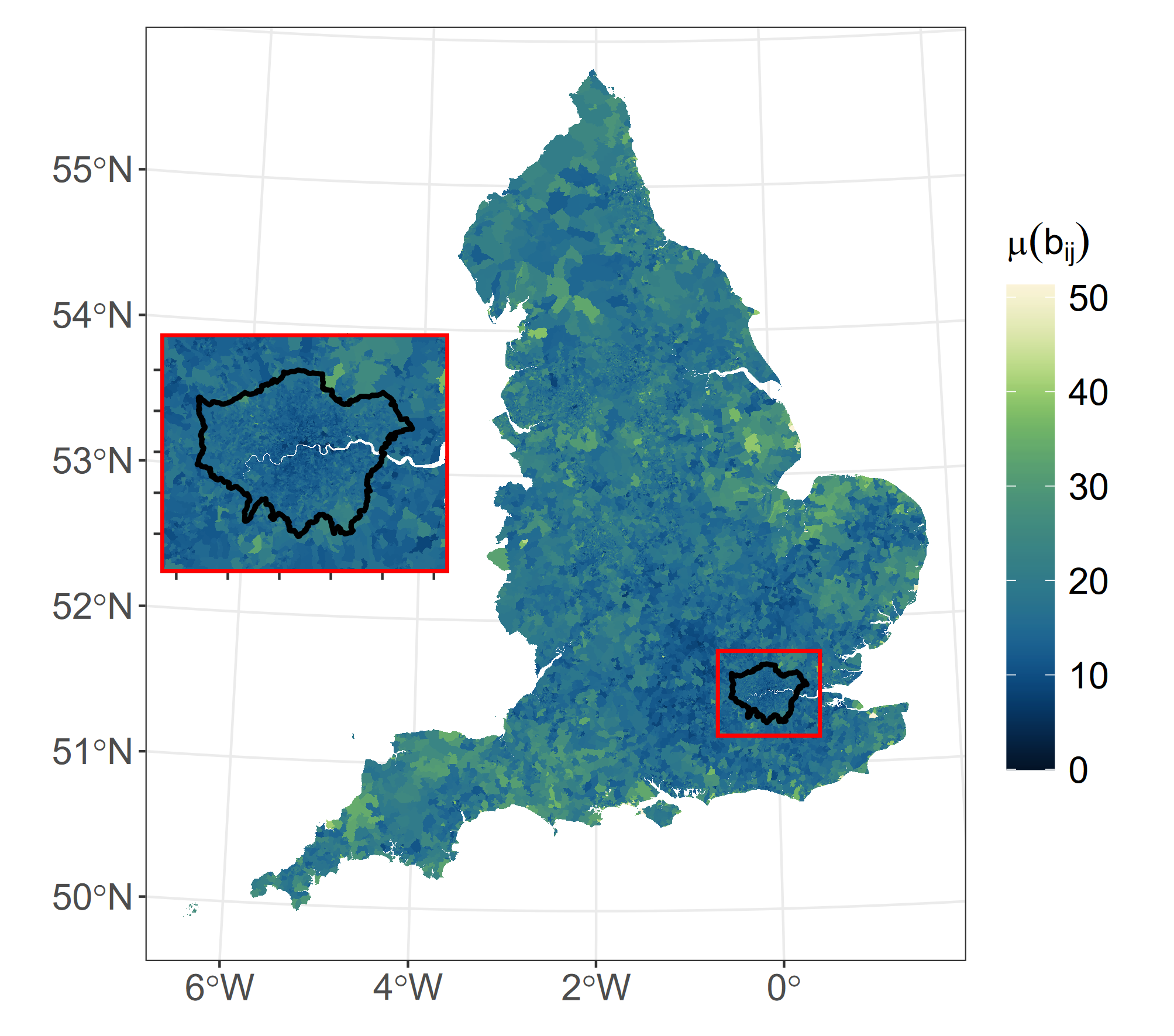}
        \vspace{0.6ex}
        \textbf{(a)}\; Estimated (posterior mean) $\mu_{ij}$
        \label{fig:app_poisson_disagg_mu_mean}
    \end{minipage}
    \hfill
    \begin{minipage}[b]{0.48\linewidth}
        \centering
        \includegraphics[width=\linewidth]{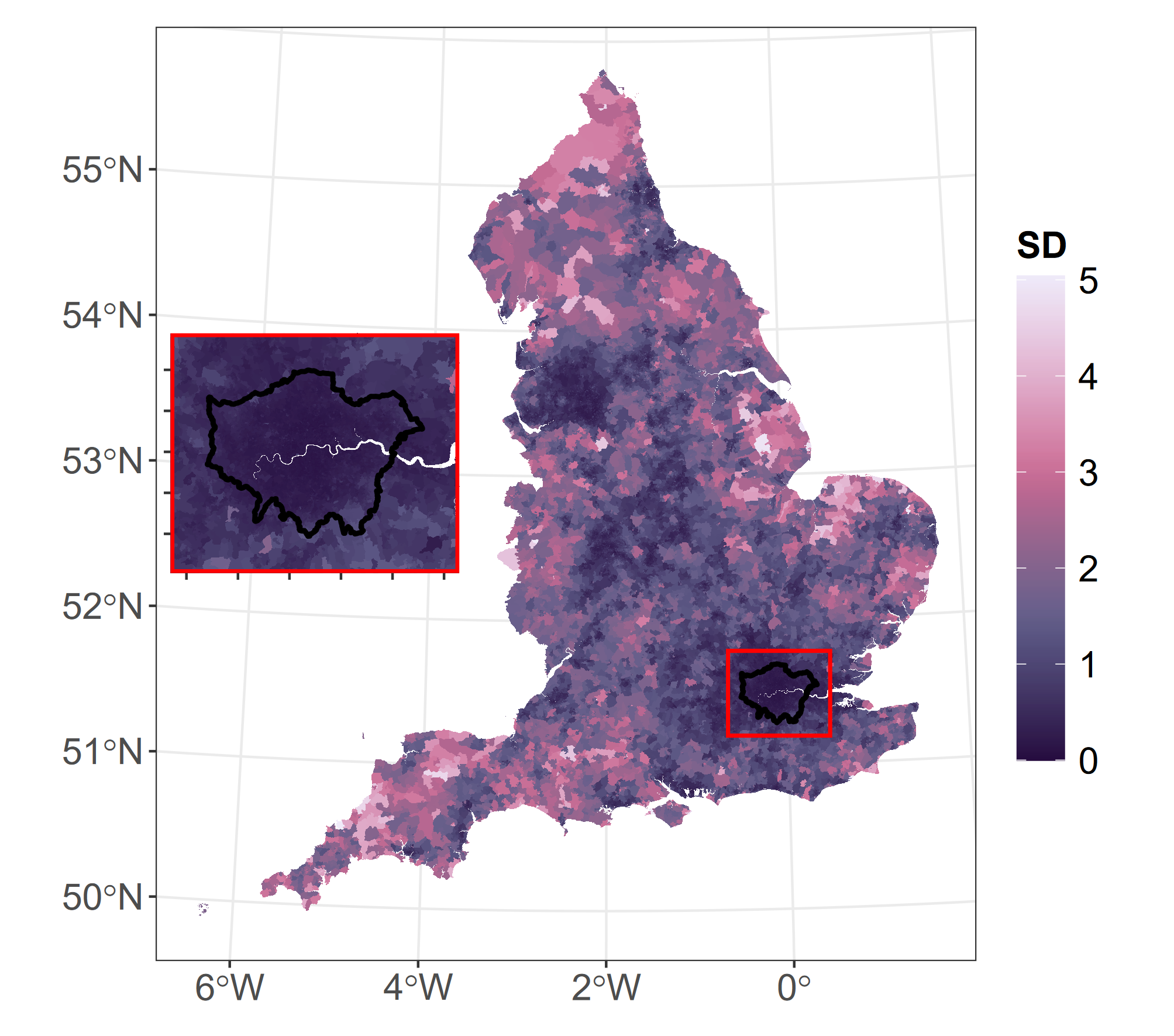}
        \vspace{0.6ex}
        \textbf{(b)}\; Posterior SD of $\mu_{ij}$
        \label{fig:app_poisson_disagg_mu_sd}
    \end{minipage}

    \caption{Predicted disaggregated values of LSOA-level disease count $\mu_{ij}$.}
    \label{fig:app_poisson_disagg_mu}
\end{figure}

\begin{figure}[H]
    \centering

    \begin{minipage}[b]{0.4\linewidth}
        \centering
        \includegraphics[width=\linewidth]{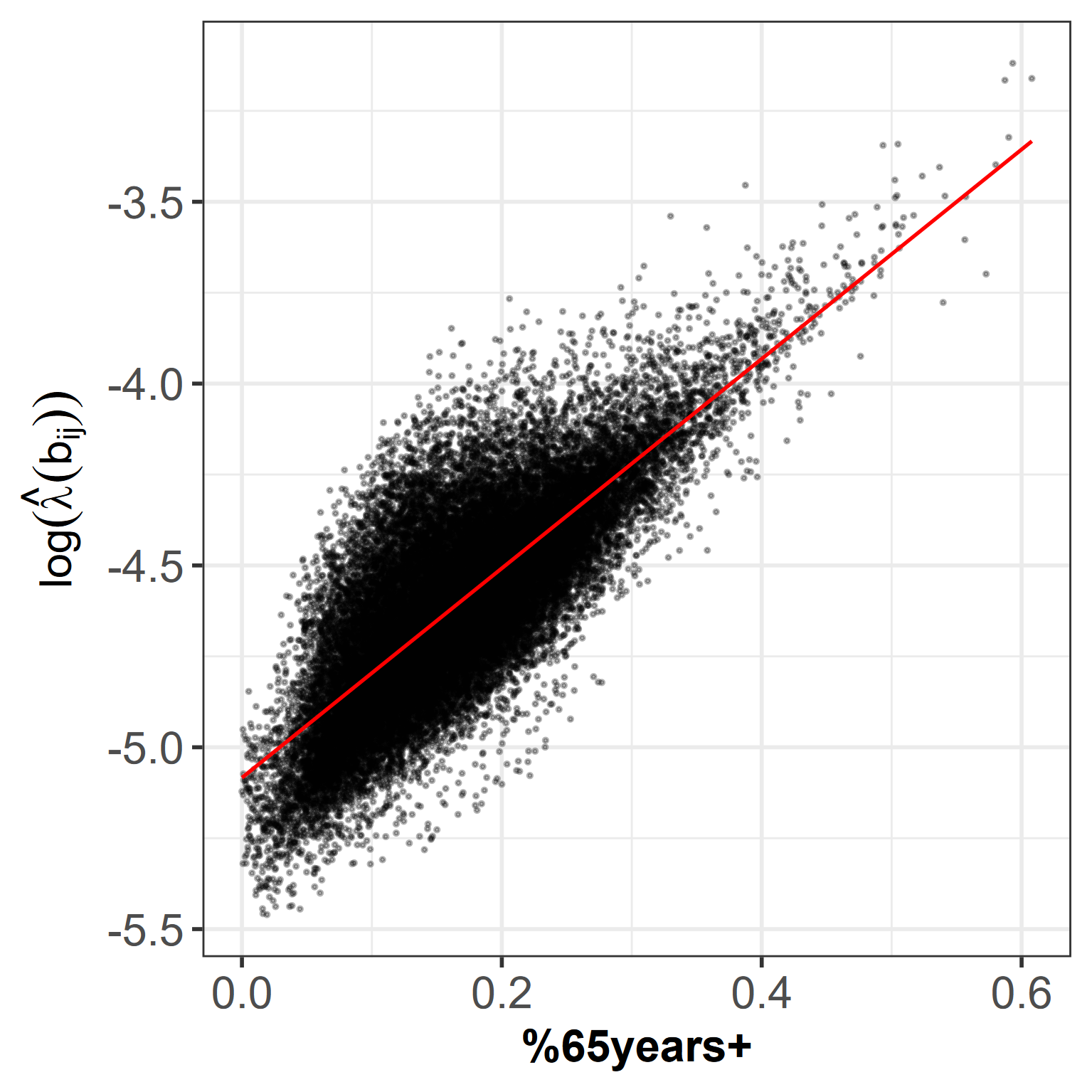}
        \vspace{0.6ex}
        \textbf{(a)}
        \label{fig:scatter_percent65plus}
    \end{minipage}
    \hspace{1cm}
    \begin{minipage}[b]{0.4\linewidth}
        \centering
        \includegraphics[width=\linewidth]{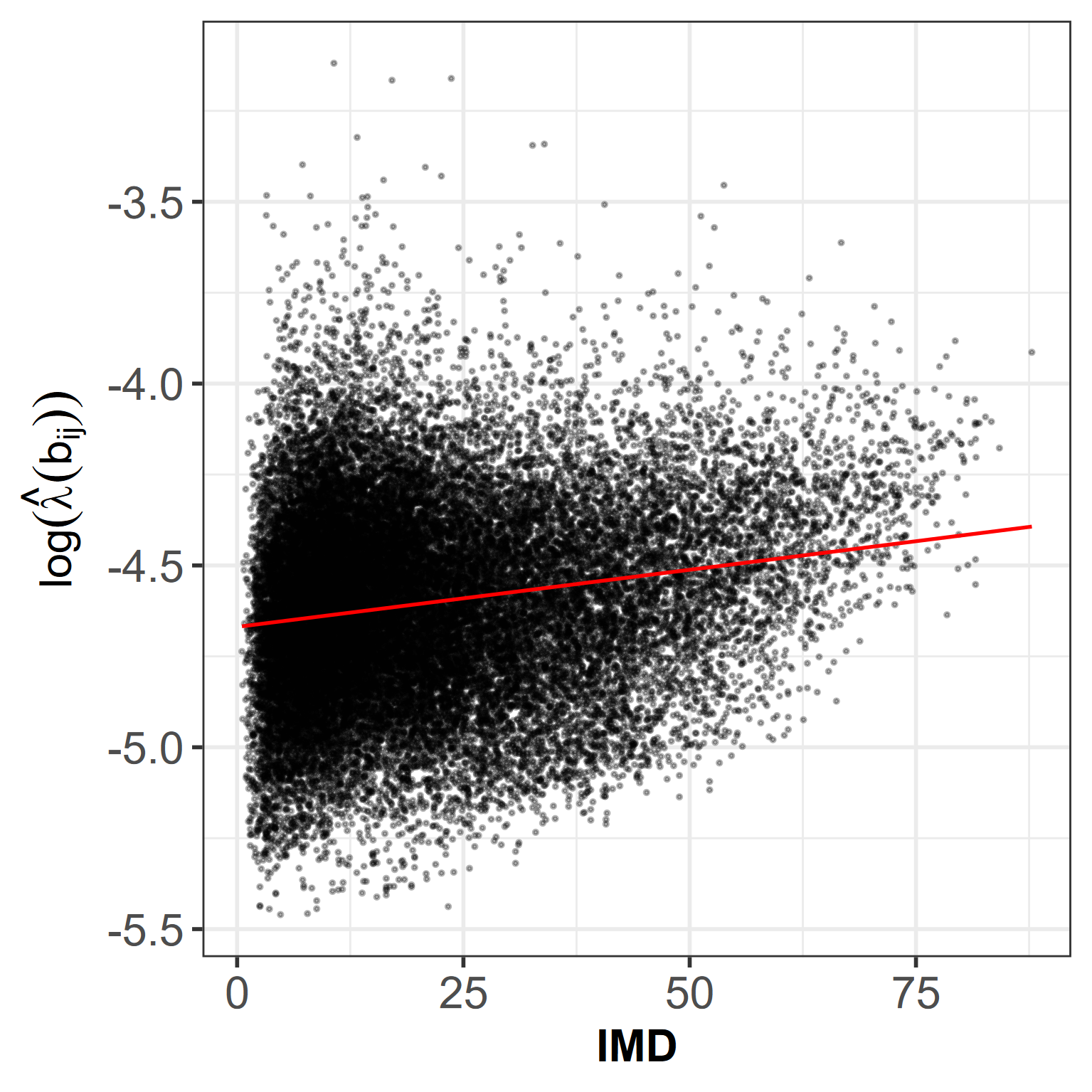}
        \vspace{0.6ex}
        \textbf{(b)}
        \label{fig:scatter_IMD}
    \end{minipage}

    \caption{Scatterplots comparing $\log\hat{\lambda}_{ij}$ with the covariates: 
    (a) $\texttt{OLD}_{ij}$, 
    (b) $\texttt{IMD}_{ij}$.}
    \label{fig:scatter_combined}
\end{figure}

\end{document}